\documentclass{article}

\usepackage{arxiv}

\usepackage[utf8]{inputenc} 
\usepackage[T1]{fontenc}    
\usepackage{hyperref}       
\usepackage{url}            
\usepackage{booktabs}       
\usepackage{amsfonts}       
\usepackage{nicefrac}       
\usepackage{microtype}      

\usepackage{amsmath}
\usepackage{amssymb}
\usepackage{longtable}
\usepackage[ruled]{algorithm2e} 
\usepackage{xcolor}
\usepackage{comment}

\providecommand{\shortcite}[1]{\cite{#1}}
\DeclareMathOperator{\Tr}{Tr}

\title{Graph Metrics for Internet Robustness -- A Survey}

\author{
  Milena Oehlers 
    \\
  Technische Universität Berlin\\
  \& Humboldt-Universität zu Berlin\\
  \texttt{milenaoehlers@gmail.com} \\
   \And
 Benjamin Fabian \\
    Technische Hochschule Wildau\\
    \& Humboldt-Universität zu Berlin\\
  \texttt{bfabian@wiwi.hu-berlin.de} \\
}

\begin{document}
\maketitle

\begin{abstract}
Research on the robustness of the Internet has gained critical importance in the last decades because more and more individuals, societies and firms rely on this global network infrastructure for communication, knowledge transfer, business processes and e-commerce. In particular, modeling the structure of  Internet has inspired several novel graph metrics for assessing important topological robustness features of large complex networks such as the Internet. 
This survey provides a comparative overview of these metrics, presents their strengths and limitations for analyzing the robustness of the Internet topology, and outlines a conceptual tool set in order to facilitate future analysis and adoption in research and practice.
\end{abstract}


\section{Introduction}\label{sec:int}

The Internet has become a part of our everyday lives and is used by more and more individuals, firms and organizations worldwide \cite{ho2007growth}. 
Internet disruptions can entail major risks and huge costs \cite{Doerr:2014}, in particular for the growing number of e-businesses. Being cut off from the Internet for even the shortest period of time affects customer confidence and could lead to a severe business losses. 
Motivated by these risks, a new field of research has evolved that investigates the robustness of the Internet with respect to random failures and targeted attacks, based on mathematical graph models \cite{albert2000barabasi, Magnien:2011, baumann2014robust}. 
This fundamental abstraction enables analysis and quantitative evaluation of the main function of large and complex networks such as the Internet: their ability to provide -- preferably short -- communication paths between pairs of entities. Assessing the robustness of the Internet structure and corresponding traffic flows has recently inspired a large number of novel robustness metrics. As of yet, however, 
there is no comprehensive review of this field. The present survey aims to fill this gap in order to foster future research. It presents important analysis strategies and metrics that are proposed in literature, along with major results. 

This article is structured as follows. 
First, \textit{Internet Robustness} is defined from a graph-theoretical perspective; then, layer and graph models of the Internet as well as network generators are presented, and challenges and aspects of robustness are introduced. The main categories of metrics are introduced, and the corresponding article structure is motivated. Then, important general features of metrics are compared and discussed, followed by a list of symbols, concepts and mathematical notations. 

In the main part of the article, the actual metrics and methods for robustness analysis are presented and discussed, using categories that reflect their main idea: \textit{Adjacency}, \textit{Clustering}, \textit{Connectivity}, \textit{Distance}, \textit{Throughput}, \textit{Spectral Methods}, and  \textit{Geographical Metrics}. 
This is followed by a discussion section that compares the suitability of the metrics for measuring major \textit{robustness aspects} and outlines a conceptual tool set of metrics for further research on Internet robustness. 
Finally, a summarizing table provides an overview of the main metrics and their features. 

\section{Methods and Notation}\label{sec:met}

\subsection{Internet Robustness}\label{sec:met:rob}

\textit{Internet Robustness} can be generally defined as the ability of the Internet to maintain its service of transferring data between node pairs in a timely manner under challenges. However, its interpretation and relationship to the more general concept of \textit{resilience} varies to some extent in different research communities \cite{albert2000barabasi, sterbenz2010resilience}. 

In our article, we focus on topological robustness and corresponding \textit{topological robustness metrics} or \emph{graph metrics}, which mainly abstract from technical or organizational details of the Internet by the use of mathematical graph theory and particular graph models

\subsection{Layer Models and Graph Models of the Internet}\label{sec:met:layers}

On the one hand, the well-known ISO/OSI and TCP/IP models focus on technical aspects of communication that are described by a hierarchy of layers, such as physical and local connections, Internet routing, transport control and application-layer protocol connections, including the Hypertext Transfer Protocol (HTTP) for the Web but also protocols for exchanging routing information (such as BGP, the Border Gateway Protocol, see RFC 4271), for email exchange, or peer-to-peer (P2P) overlay networks. For instance in \cite{sterbenz2010resilience}, a technically-layered model of \textit{Internet Resilience} is constituted by a \textit{Physical (P)}, \textit{Link+MAC (L)}, \textit{Topology (To)}, \textit{Routing (R)}, \textit{ Transport (Tr)}, \textit{Application (A)} and \textit{End User (U)} layer. Each layer provides the base for the next higher one and can be characterized by 
layer-specific metrics, including non-topological performance measures.

On the other hand, there are organizational and political perspectives on the exchange, forwarding, controlling and filtering of data on the Internet, for instance, models that distinguish the layers of Internet Protocol (IP) interfaces, routers, points-of-presence (PoP), autonomous systems (AS), Internet Service Providers (ISPs) or other organizations, and countries. As an example, in \cite{Fabian:2017a}, the authors construct an organizationally-layered family of graph models derived from empirical measurement data.

For both perspectives and most layers, there are historical examples of attacks and failures with impact on connectivity \cite{baumann2014robust,Fabian:2015}. 
Some attacks on higher layers, such as BGP misconfiguration, 
affect the robustness of AS-level and IP-level graphs \emph{without} damage to the underlying physical infrastructure and interconnections of the lower layers. Other failures and attacks take place on the \textit{Physical Layer} of the Internet, but can cause faults on higher abstraction levels, such as routing or AS layers. 

Correspondingly, it is useful to model and analyze entities and their interconnections as graphs for studying its topology and robustness at any particular layer. Traditionally, the complex-network community has studied \textit{Internet Robustness} at single layers \cite{Albert2002, PastorSatorras:2004,albert2000barabasi, Magnien:2011, baumann2014robust, Barabasi:2016}, and this is the main focus of our study; but there are several recent approaches to model the Internet and assess its \textit{Robustness} using layered hierarchies. Most graph metrics that are discussed in the following sections can be relevant for organizational as well as for technical graph models, and can be applied at every layer where an exchange or relation between different entities is of interest. 

Each layer can be modeled by several different graph types \cite{Barabasi:2016}. A simple graph is a strong abstraction as it is constituted solely of two entity types: nodes (e.g., routers or ASes) and undirected edges (e.g., links or routing connections), but it has the advantage that fewer information is needed for analysis. 
In edge-weighted graphs, transmission speed or capacity constraints are also modeled \cite{Costa2007}. A directed graph accounts for routing policies between the ASes, e.g., denoting a customer-provider relationship by an edge that points from the former to the latter; peer-to-peer ASes are connected through two links, one in either direction \cite{tangmunarunkit2001impact, Tangmunarunkit2002}. Nodes can only communicate if both have a directed path towards a common node, e.g., if they share a common upstream provider. A weighted directed graph can also be taken into consideration. Further important modeling aspects are the amount of traffic flow between nodes \cite{Doyle2005} and their Euclidean distance \cite{Crucitti2006}. 

Concerning properties of complex networks, Internet graphs on both router and AS-levels have been shown to feature scale-free (SF) degree distributions \cite{faloutsos1999power} and the small-world property \cite{Albert2002, Barabasi:2016}, in particular, shorter paths and higher clustering than expected within random networks with a given degree distribution \cite{Watts1998}.

Not all network-generating models exhibit these characteristics: The Erd\"os-R\'enyi (ER) model \cite{erdios1959r} exhibits none of them, the Watts-Strogatz (WS) model \cite{Watts1998} lacks the scale-free degree distribution, and the clustering coefficient of the Barab\'asi-Albert \cite{barabasi1999emergence} (BA) model is slightly too high. The Klemm-Egu\'iluz (KE) model \cite{klemm2002growing} and the HOT model \cite{li2004first} of the router level of the Internet \cite{Doyle2005} seem to be suitable in this sense. However, there is currently no final consensus on a most accurate model, in particular also with respect to the highly dynamic and evolutionary character of the Internet \cite{PastorSatorras:2004,Aggarwal:2014}. Therefore, in this article, metrics are generally discussed independently of these models.

\subsection{Challenges and Aspects of Robustness}\label{robasp}

Even with an abstract and topological approach, the question whether an Internet graph is robust should be considered with respect to the requirements of the particular service that is to be delivered; usually, this is a multidimensional problem, which requires several metrics measuring different \textit{robustness aspects} that should remain in an acceptable state even under severe challenges. 

Two main variants of \emph{topological robustness} are often discussed in complex network literature: The robustness of a network against random failures of entities, and its robustness against targeted attacks \cite{albert2000barabasi, Magnien:2011, baumann2014robust}. These challenges can be further motivated (\cite{sterbenz2010resilience}): \textit{Failures} of entities are random and can be caused by software, hardware or human errors. Furthermore, unexpected non-malicious \textit{traffic increases} in parts of the network can be caused by flash crowds. Not least, \textit{geographically extended faults} can occur due to earthquakes or power outages. 

Targeted and correlated \textit{attacks}, such as distributed denial of service (DDoS) and malware, aim at destroying crucial entities, e.g., those identified with graph metrics such as highest \textit{Degree} or \textit{Betweenness}. Not specifically investigated in 
our study is the topic of network epidemics, which is relevant for self-disseminating malware such as Internet worms and their propagation strategies, but only indirectly affects network robustness, e.g., by malicious payloads or network saturation. Interactions of malware with topological network properties could be the focus of a dedicated survey on network epidemics.

Many metrics can be used with both random failure and targeted attack strategies. Local or non-adaptive attack strategies calculate the metric values of all entities once and take them down in decreasing order. In contrast, global or adaptive strategies recalculate metrics after every deletion and tend to have a more severe impact while also needing real-time and global information access \cite{ccetinkaya2015multilevel}. In this article, if an \emph{attack} is analyzed without further specification of its strategy, it always refers to an adaptive attack on highest-degree nodes. 

Loosely corresponding to the challenges, there are major aspects of topological robustness. \textit{Disconnection Robustness} of a graph is measured by metrics that assess path redundancy, detect and evaluate topological bottlenecks, and analyze the severity of 
graph disconnections. It is defined as the ability of the network 
to maintain a connection between all (or most) of the node pairs when challenged. 
Another crucial feature to take into consideration 
is the small-world property 
of the Internet, which is associated with 
the inverse of the average minimum hop distance between node pairs. \textit{Transmission Speed Robustness} is the ability to keep the decay of this measure low, even under stress. 

Another aspect, \textit{{Traffic Robustness}} of a network,  
is defined as the ability to endure and transmit highly dynamic traffic demands without becoming overloaded. 

All of these aspects are important for the \textit{Internet Backbone}, which forms the core of the Internet, providing shortest paths through high-capacity links and routers. Corresponding distance concepts focus, for example, on the number of hops a data package has to traverse in order to reach the target node, implicitly assuming that the longer the path, the slower the transmission speed. Throughput takes into account that capacities for information forwarding are limited, and calculates the maximum workload of network entities or their likelihood to become overloaded when other entities fail. 


However, apart from the above mentioned exceptions, the study of layer interactions and their implications for Internet robustness is still an emerging research field, whose tools and methods are not yet broadly established. The same holds for the study of robustness of interacting networks \cite{Muro:2017} such as electricity and communication networks. For the broader and multi-facet concept of \emph{Resilience}, we refer the reader to related work \cite{sterbenz2010resilience}. 

As a baseline for future extensions, our study aims to survey the most relevant metrics for single-layer \emph{Internet Robustness} analyses, with emphasis on the \textit{Topology} and \textit{Routing Layer} and above. To what extent the surveyed metrics are suitable
for studying these mentioned aspects, is a complex research question in itself, which will be discussed in more detail in Section \ref{sec:com}.

\subsection{Metric Selection and Article Structure}\label{sec:met:res}

The selection of metrics is based on literature research, mainly conducted via Google Scholar using diverse topic-related keywords, and inverse search using the references in publications already identified. The selection of metrics for this survey, out of the vast number of general graph metrics in the literature \cite{Kamisinski:2015}, is conducted based on the following criteria: The metric has to be applicable to at least one graph type and either directly measure an aspect of network robustness or at least a clearly related characteristic. 

Exclusively assigning the extensive amount of metrics relevant for Internet robustness to non-overlapping groups would prove to be infeasible since some metrics are relevant for several aspects. Therefore, the article is structured as follows: the presentation of metrics and corresponding results for Internet graphs is conducted by dividing them into six major groups that indicate their respective main task. Three of them, \textit{Adjacency}, \textit{Clustering} and \textit{Connectivity} build on each other and describe the general structure of graphs. The \textit{Distance} and \textit{Throughput} categories focus on the concepts that are crucial for communication networks, approximating the concrete Internet routing processes via shortest-paths. \textit{Throughput} accounts for Internet-specific link capacity restrictions. Aspects of all these groups are also regarded in \textit{Spectral Methods} with the help of random walks. The more sophisticated matrix-calculation schemes involved justify their bundling into a dedicated chapter. Afterwards, a separate section assesses the suitability and interactions of the metrics with respect to the different \textit{Robustness Aspects}. 


\subsection{Features of Metrics}\label{sec:met:met}
Many metrics are used with different names in the literature. In order to ensure consistency of notation, only one name is selected and used throughout the article; the others are provided in brackets. At the outset of every section, general features of the metric are displayed.

First, the technical layers as defined above in Section \ref{sec:met:layers}, following \cite{sterbenz2010resilience}, 
on which the metric can be used particularly well (but often not exclusively), are indicated by their respective abbreviations given in Section \ref{sec:met:layers}. This categorization should be understood as a preliminary step to combine the traditional single-layer graph analysis of the Internet with a more detailed technical layer model. 

The next feature indicates the graph types to which the metric has been applied in the literature. 
Whenever an application to further graph types is straightforward, 
by substituting simple-graph elements by their directed or weighted counterparts,  
this will be listed in brackets. For example, the entry '(un)weighted (directed)' indicates that the metric is applied to simple and weighted graphs, but can be modified for directed ones. In order to highlight metrics that can be applied to every graph type, '\textit{all}' is used. 

The third entry states whether the metric is \emph{local} -- either measuring the individual node's \textit{impact} on network robustness or its \textit{liability} to faults of connectivity to the rest of network --
 or if it is \emph{global}, capturing features about the entire graph \cite{Baumann2013}. Local metrics often depend on a smaller information set, provide a more detailed insight and need less calculation time, while global metrics are more meaningful for assessing the state of the entire network and also allow for comparison of different graphs.

A metric is denoted \textit{static} if it is applied without simulating faults -- taking a snapshot of characteristics that influence robustness -- or \textit{dynamic} if it assesses network behavior under arbitrary 
removal strategies; the entries \textit{failures} or \textit{worst-case} indicate metrics only suitable for these scenarios. 


Next, the \emph{codomain} is provided, providing a quick impression of the value range of a metric -- for instance, if the metric ranges from zero to one. Its dependency on parameters such as the network size indicates whether a direct comparison of graphs that differ in these aspects is possible. 
If it is set to '-', the presented method is not a metric in the narrower sense, but a general algorithm, e.g., for finding cluster structures and minimum cuts. 

Finally, an \textit{efficiently computable} metric is denoted 
by a check mark, otherwise its exact calculation is \textit{NP-hard} or \textit{NP-complete}. If calculation orders $\mathcal{O}$ of (approximating) algorithms are provided in the literature, they are stated in brackets. A non-calculable metric is not practical for robustness analysis of large graphs, but still can provide interesting ideas and a base for further heuristics.

\subsection{List of Symbols}\label{sec:met:lis}
The following notations are used throughout the article: \vspace{-0.2cm}
\begin{table}[h!]
\begin{center}
\begin{tabular}{ll}
$G(V,E)$& Graph $G$ with a set of vertices $V$ and set of edges $E$.\\
$v=|V|$& Number of nodes in the graph.\\
$e=|E|$& Number of edges in the graph.\\
$V_i$& set of nodes that are neighbors of node $i$ in a simple graph\\
$\overline{x}$& Arithmetical mean of $x_y$ for all nodes/edges/etc. $y$.\vspace{-0.3cm}
\end{tabular}
\end{center}
\end{table}

By substituting the following elements of simple graphs with their weighted and/or directed equivalents, various metrics can be enhanced with the additional information contained in more advanced graph models.  
\vspace{-0.2cm}
\begin{table}[h!]
	\begin{center}
		\begin{tabular}{ll}
$\textbf{A}=[a_{ij}]$& Adjacency matrix; $a_{ij}=1$ if an edge from node $i$ to $j$ exists, $a_{ij}=0$ else.\\
\vspace{2mm}
$\textbf{A}^w=[w_{ij}]$& Weight matrix; $w_{ij}$ is the edge weight between nodes $i$ and $j$.\\
$k_i^{(in/out)}$& (In-/out-) degree of node $i$ (Section \ref{sec:con:nod}).\\
\vspace{2mm}
$s_i^{(in/out)}$& (In-/out-) strength of node $i$ (Section \ref{sec:con:str}).\\ 
$d_{(m,n)}$& Minimum hop distance from node $m$ to node $n$ (Section \ref{sec:dis}).\\
$d^w_{(m,n)}$& Minimum weighted distance from node $m$ to node $n$ (Section \ref{sec:dis}).
\end{tabular}
\end{center}
\end{table}
\vspace{-0.3cm}\\

\newpage 


\section{Adjacency}\label{sec:con}

The assessment of node adjacency is one of the first and easiest approaches for investigating network robustness. The intuition is that a vertex with many edges, i.e., a high-degree node, could be more important to the overall graph structure than one with a low degree. The average node degree is a first indicator for the overall network robustness, and the node-degree variance for the resilience against high-degree attacks. 
		
		\subsection{Node Degree {\scriptsize (Degree Centrality)}}\label{sec:con:nod}
\hrule \vspace{0.03cm}
\centerline{{L, R}\hfill simple, directed  \hfill local impact \& liability / global \hfill static \hfill $\left[0, v-1\right]$ \hfill \checkmark} 
\hrule 
\begin{eqnarray}
\label{eqn:01}
k_{i}&=&\sum_{j\in V}{a_{ij}}=\sum_{j\in V}{a_{ji}}\\
k^{out}_{i}&=& \sum_{j\in V}{a_{ij}} \neq\sum_{j\in V}{a_{ji}}= k^{in}_{i}\label{eqnodi}
\end{eqnarray}

The \textit{Node Degree} \cite{hakimi1962realizability} is used in most of the classical results on Internet robustness. 
The directed version of this metric (eq. \ref{eqnodi}) is given in \cite{Costa2007}. 

Notably, in \cite{Ghedini2011}, the simple-graph version $k_i$ is used to build an attack algorithm that first deletes the nodes with the highest degrees. The authors show that this attack algorithm affects ER, KE and WS networks more than random node failure in terms of \textit{Global Network Efficiency} $E_i$ (Section \ref{sec:dis:glo}) and \textit{Number of Nodes in the Largest Component} $|V_L|$ (Section \ref{sec:min:ana}). However, at each step, the removal of a different node than the one with the highest \textit{Node Degree} 
could cause even larger global damage. Therefore, the \textit{Degree} of a node has some, but also a limited, influence on global \textit{Disconnection} and \textit{Transmission Speed Robustness}, beyond assessing the local connectivity of a node to direct neighbors. 


Hence, relying on the degree alone for estimating the importance of a node could be misleading. In \cite{Xia2008}, a high correlation between \textit{Node Degree} and \textit{Betweenness Centrality} (Section \ref{sec:thr:bet}) was discovered at the AS-level of the Internet in particular for higher degree nodes, as well as in a WS, BA, ER, and clustered scale-free networks.

The global version of this metric is the \textit{Degree-Frequency Distribution}, $P(k)=\left|V_{k}\right|/ v$, 		
with $|V_k|$ defined as the number of nodes with degree $k$.
Its codomain is $[0,1]$. A classical result of \cite{faloutsos1999power} states that the Internet belongs to the class of scale-free networks with $P(k)\propto k^{-\gamma}$. 
 This distribution indicates that few highly connected nodes (hubs) exist that are very important for the network connectivity. More precisely, the authors found $\gamma\in[2.15,2.2]$ for the Internet on the AS-level between 1997 and 1998, whereas for the router level from 1995, $\gamma = 2.48$. 
However, in \cite{Doyle2005}, is is challenged that the Internet exhibits a hub-like core. The authors claim that a more suitable Internet topology with the same $P(k)$, the HOT network, exhibits a rather different topology with hubs on the periphery and hence not as a crucial \textit{Backbone}. Their removal would hence only have local effects.

		\subsection{Strength}\label{sec:con:str}
\hrule \vspace{0.03cm}
\centerline{{L, R}\hfill weighted (un)directed \hfill local impact / global \hfill static \hfill $\left[0, \infty\right[$ \hfill \checkmark} 
 \hrule
\begin{eqnarray}
s_{i}&=&\sum_{j\in V} w_{ij}=\sum_{j\in V} w_{ji}\\
s_{i}^{out}&=&\sum_{j\in V} w_{ij}\neq \sum_{j\in V} w_{ji}=s_{i}^{in}
\end{eqnarray}
\noindent 

\textit{Strength} is the adaptation of \textit{Node Degree} for weighted graphs \cite{DallAsta2006, Costa2007}. Quantifying the amount of information that would have to be redirected in the case of a failure of node $i$, it can be used as a first indicator for its impact on \textit{Transmission Speed} and \textit{Traffic Robustness}. Moreover, the largest-capacity  \textit{Backbone} nodes are indicated by highest \textit{Strength}. 
Its global version \cite{Barrat2004} ranges from zero to one: $P(s)=|V_s|/v$, with $V_s$ defined as the set of nodes with strength $s$. It gives an overview over the heterogeneity of local node importance. Similar to the \textit{Degree-Frequency Distribution}, a more uniform \textit{Strength Distribution} leads to a network that is more robust against attacks, but more vulnerable to random failures.




\subsection{Entropy}\label{sec:con:ent}
\hrule \vspace{0.03cm}
\centerline{{L, R}\hfill simple (\textit{all}) \hfill global \hfill static \hfill $[0,\log(v-1)]$ \hfill \checkmark} 
\hrule \vspace{0.1cm}
\textit{Entropy} $H$ measures the homogeneity of the \textit{Degree-Frequency Distribution} $P(k)$ (Section \ref{sec:con:nod})
\begin{equation}
H=-\sum_{k=1}^{v-1}P(k)\log P(k)
\end{equation}

The maximum of this metric is $H_{max}=\log (v-1)$.
Wang et al. \shortcite{wang2006entropy} find that optimizing an SF-network's \textit{Disconnection Robustness} to random failures in terms of \textit{Percolation Threshold} $p_c$ (Section \ref{sec:min:per}) is equivalent to maximizing its \textit{Entropy} -- and thus homogeneity of $P(k)$ -- while maintaining the SF-structure and keeping the average 
\textit{Node Degree} constant. 




\subsection{Skewness}\label{sec:con:ske}
\hrule \vspace{0.03cm}
\centerline{{L, R}\hfill simple (\textit{all}) \hfill global \hfill static \hfill $\left[0, 1\right]$ \hfill \checkmark} 
\hrule \vspace{0.1cm}
\begin{eqnarray}
Sk&=&\frac{\sum_{i} \left(r_{i}*k_{i}\right)}{Sk_{u}}\\
Sk_{u}&=&\sum_{i}r_{i}*\overline{k}=\overline{k}*\frac{v*\left(v+1\right)}{2}
\end{eqnarray}
\noindent \textit{Skewness} $Sk\approx 0$ indicates a very preferential or heterogeneous network, $Sk\approx 1$ a very uniform one. $Sk_u$ is the \textit{Skewness} of a network with uniform degree $\overline{k}$. The lowest node rank $r_i=1$ is given to the node with the highest degree, $r_i=2$ to the one with the second highest, and so on. Each rank is assigned only once; if there is more than one node with the same degree, the order in which they receive the respective ranks is randomly chosen \cite{park2004comparing}.  

Ghedini and Ribeiro \shortcite{Ghedini2011} show that a more homogeneous \textit{Degree-Frequency Distribution} $P(k)$ (Section \ref{sec:con:nod}), as measured by \textit{Skewness} $Sk$, contributes to higher \textit{Transmission Speed} and \textit{Disconnection Robustness} in terms of \textit{Global Network Efficiency} $E_i$ (Section \ref{sec:dis:glo}) and \textit{Number of Nodes in the Largest Component} $|V_L|$ (Section \ref{sec:min:ana}) against highest-\textit{Degree} attacks, but leaves the graph more vulnerable in this respect to random failures. 
	In the preceding and current sections, contrasting conclusions about the impact of $P(k)$ on \textit{Disconnection Robustness} to random failures are drawn. It must be noted, though, that $p_c$ measures a worst-case scenario at high failure rates and thus a different facet of \textit{Disconnection Robustness} than $|V_L|$, which evaluates graph behavior at lower failure rates. In this light, the apparent conflict between the metrics is solved.

\subsection{Vulnerability Function}\label{sec:con:vul}
\hrule \vspace{0.03cm}
\centerline{{L, R}\hfill simple graph \hfill global \hfill static \hfill $\left[0,1\right]$ \hfill \checkmark} 
\hrule
\begin{equation}
v_\sigma (G)=\exp\left[\frac{\sigma}{v}+v-e-2+\frac{2}{v}\right]
\end{equation}
\noindent The \textit{Vulnerability Function} $v_\sigma (G)$ \cite{Criado2005, yazdani2010note}  aims at measuring the vulnerability of a graph with an axiomatic approach, only based on information contained in the \textit{Node Degree Distribution} $D(k)=P(k)\cdot v$.  In the formula, $\sigma$ is the standard deviation of $D$, and 
$P(k)$ is the \textit{Degree-Frequency Distribution} (Section \ref{sec:con:nod}). A high value of $v_\sigma (G)$ indicates a vulnerable network, a low value a robust one. 

Similarly to \textit{Entropy} (Section \ref{sec:con:ent}) and \textit{Skewness} (Section \ref{sec:con:ske}), the Vulnerability Function assesses the impact of the homogeneity of $P(k)$ on robustness, additionally taking the relation of numbers of nodes to edges into account -- it is designed not to simply increase when edges are added. For graphs with the latter two parameters fixed, the \textit{Vulnerability Function} thus shares the implications of the other two metrics on \textit{Disconnection} and \textit{Transmission Speed Robustness}. 

\subsection{Assortative Coefficient}\label{sec:con:ass}
\hrule \vspace{0.03cm}
\centerline{{L, R}\hfill simple graph \hfill global \hfill static \hfill $\left[-1,1\right]$ \hfill \checkmark} 
\hrule
\begin{eqnarray}
r&=&\frac{e^{-1}*\sum_{(i,j)\in E}\left(k_{i}k_{j}\right)-\left[e^{-1}*\sum_{(i,j)\in E}\frac{1}{2}\left(k_{i}k_{j}\right)\right]^{2}}{e^{-1}*\sum_{(i,j)\in E}\left(k_{i}^{2}+k_{j}^{2}\right)-\left[e^{-1}*\sum_{(i,j)\in E}\frac{1}{2}\left(k_{i}k_{j}\right)\right]^{2}}\\
&=&\sum_{k_{1}\leq k_{2}} k_{1}k_{2}*\left(P(k_{1},k_{2}) - \frac{k_{1}k_{2}P\left(k_{1}\right)P\left(k_{2}\right)}{\overline{k^{2}}}\right)
\end{eqnarray}
The \textit{Assortative Coefficient} \cite{Mahadevan2005} calculates whether the network is \textit{assortative}, $r>0$, meaning that most nodes are connected to nodes with a similar degree, or \textit{disassortative}, $r<0$, as the Pearson correlation coefficient of node degrees at both ends of an edge. 
Similar to \textit{Average Neighbor Connectivity}, this metric summarizes of the $JDD$ by calculating the probability that a hub node is connected to another hub, but the information loss is even bigger. 
The advantage is the clear codomain, with $r>0$ indicating an assortative network, which is thus robust against failures and attacks, and $r<0$ a disassortative one. If $r=0$, there is no correlation between the node degrees of an edge. 

In \cite{Newman2002}, assortative networks are found to exhibit a higher \textit{Disconnection Robustness} concerning the \textit{Number of Nodes in the Largest Component} $|V_L|$ (Section \ref{sec:min:ana}) against both failures and attacks and the disassortativity of the Internet is explained by high-degree connectivity providers and low-degree clients. 
Furthermore, Liu et al. \shortcite{liu2017comparative} find assortative networks to be more \textit{Disconnection Robust} with respect to \textit{Natural Connectivity} $\overline{\lambda}$ (Section \ref{sec:spe:nat}) and stability of $|V_L|$ under attacks, 
but less so 
in terms of \textit{Algebraic Connectivity} $\lambda_2$ (Section \ref{sec:spe:alg}). 
The latter metric is thus in contrast with the former two. As it only assesses how easily any set of nodes can be disconnected, irrespective of its size, more weight should be put on $|V_L|$, which conveniently addresses this problem. 
Still, further analysis of these three metrics is necessary to finally decide if they indeed measure different facets of this robustness aspect. Only then, an unambiguous interpretation of the impact of $r$ on \textit{Disconnection Robustness} would be possible.

		\subsection{Average Neighbor Connectivity}\label{sec:con:ave}
\hrule \vspace{0.03cm}
\centerline{{L, R} \hfill (un)weighted (directed) \hfill global \hfill static \hfill $\left[0, v-1\right]$ \hfill \checkmark} 
\hrule
\begin{eqnarray}
k_{nn}(k)&=&\sum_{k'}k'*P(k'|k) \\
P(k'|k)&=&
\frac{\overline{k} P(k,k')}{kP(k)}\quad
\end{eqnarray}
\noindent \textit{Average Neighbor Connectivity} 
calculates the average neighbor degree of $k$-degree vertices \cite{Mahadevan2005}, 
 summarizing the \emph{Joint Degree Distribution}
$JDD(k,k')=P(k,k')$, which measures the probability that a randomly chosen edge points from a $k$- to a $k'$-degree node.
\cite{Mahadevan2007}. 
It thus assesses the level of network \textit{assortativity} and its \textit{Disconnection Robustness} as measured by \textit{Number of Nodes in the Largest Component} $|V_L|$ (Section \ref{sec:min:ana}) in a more detailed but also more complicated way than \textit{Assortative Coefficient} (Section \ref{sec:con:ass}). According to Costa et al. \shortcite{Costa2007}, it holds that $k_{nn}(k)\approx\overline{k^2}/\overline{k}$ for all $k$ if there is no correlation between $k_{nn}(k)$ and $k$. 
If $k_{nn}(k)$ is increasing in $k$, the investigated graph is assortative; if it is decreasing, the graph is disassortative. 

For weighted networks, Barrat et al. \shortcite{Barrat2004} propose the following formulae:
\begin{eqnarray}
k^w_{nn}(k)&=&\frac{1}{|V_k|}\sum_{k_i=k}{k^w_{nn,i}} \\
k^w_{nn,i}&=&\frac{1}{s_i}\sum^{v}_{j=1}{a_{ij}w_{ij}k_j}
\end{eqnarray}
\noindent The local weighted average of the nearest-neighbor degree is calculated according to the normalized weight of the connecting edges $w_{ij}/s_i$. Similarly, $k^w_{nn,i}$ measures the effective affinity to connect with high-degree or low-degree neighbors, considering the magnitude of interactions. The weighted version of \textit{Average Neighbor Connectivity}, $k^w_{nn}(k)$, measures the average weighted neighbor degree of a degree-$k$ node. When applying both the unweighted and weighted versions of this metric to a weighted graph, then $k^w_{nn,k}>k_{nn}(k)$ indicates that edges with larger weights point to neighbors with larger degree, while $k^w_{nn,k}<k_{nn}(k)$ shows that such edges are connected to lower-degree neighbors. This procedure can thus help to identify whether hubs are located at the high-capacity meaning high-strength Internet \textit{Backbone} core or at the periphery (Section \ref{sec:con:nod}).

 	\subsection{Rich-Club Connectivity {\scriptsize (Rich-Club Coefficient)}}\label{sec:con:ric}
\hrule \vspace{0.03cm}
\centerline{{L, R}\hfill \textit{-- all --} \hfill global \hfill static \hfill $[0, \infty[$ \hfill \checkmark} 
\hrule 
\begin{eqnarray}
\rho_{null}(k)&=&\phi(k)/\phi_{ran}(k)\\
\phi(k)&=&\frac{\sum_{i,j\in R(k)}a_{ij}}{\left|R(k)\right|(\left|R(k)\right|-1)}\\
R(k)&=&\left\{v\in V |k_{v}>k\right\}
\end{eqnarray}
\noindent 
The \textit{Rich-Club Coefficient} $\phi(k)$ measures the interconnectedness of hubs by calculating the ratio of existing edges between nodes with a degree higher than $k$ to the maximum possible number of edges \cite{Colizza2006}. Its codomain is $[0,1]$ with zero indicating no direct interconnections, and one for a fully connected cluster. 
Uncorrelated networks that are neither assortative nor disassortative usually have a non-zero value of $\phi$. Therefore, $\phi(k)$ has to be normalized by $\phi_{ran}(k)$ in order to assess with \textit{Rich-Club Connectivity} $\rho_{null}$ if there is a rich-club behavior ($\rho_{null}>1$), or not ($\rho_{null}<1$). 

The Internet graph investigated in \cite{Colizza2006} lacks rich-club ordering, which is explained by the \textit{Backbone} hubs gaining their importance through high bandwidths and traffic capacities between each other rather than by high numbers of interconnections. Along with \cite{Doyle2005}, it is suggested that hubs solely provide connectivity to local networks on the periphery without redundant connections between themselves. 
The weighted \textit{Rich-Club Coefficient} $\phi^{w}(s)$ thus measures the weighted interconnectedness of high-strength nodes,
\begin{eqnarray}
\rho^w_{null}(s)&=&\phi^w(s)/\phi^w_{null}(s)\\
\phi^{w}(s)&=&\frac{\sum_{i,j\in R^{w}(s)}{w_{ij}}}{\sum_{i\in R^{w}(s)}s_{i}}\\
R^{w}(s)&=&\left\{v\in V|s_{v}>s\right\}
\end{eqnarray}
with the same codomain as above and again $\rho^w_{null}>1$ indicating a rich club behavior as well as increased \textit{Transmission Speed} and \textit{Traffic Robustness}. 
For more details, see \cite{Opsahl2008}.

	\subsection{Recapitulation}\label{sec:con:dis}

The adjacency metrics presented in this chapter are descriptive and provide a static snapshot of basic characteristics of a graph. 
The \textit{Degree-Frequency Distribution}, its summarizing metrics \textit{Entropy}, \textit{Skewness}, and \textit{Vulnerability Function} and  \textit{Strength Distribution} 
indicate whether a network is vulnerable to \textit{Disconnection} through attacks or random failures and to \textit{Transmission Speed} reduction due to attacks. 
In order to attain the same goal, the easy-to-handle \textit{Assortative Coefficient} can be used, which also gives a first indicator whether the graph is assortative and thus more \textit{Disconnection} robust against both failures and attacks. 
The \textit{Average Neighbor Connectivity} metric uses a similar measuring approach, but 
provides more details about the network
and has a counterpart for weighted graphs, which makes it very suitable for Internet robustness analysis. 
Another very interesting metric is the weighted \textit{Rich-Club Connectivity}. It evaluates the \textit{Backbone} structures that provide many alternative transmission ways if one of the most important nodes fails. 
Since this is the only metric measuring this aspect, it should be part of the methodological repertoire for robustness analysis. 

\section{Clustering}\label{sec:clu}

Clustering metrics aim to provide a detailed overview of the community structure of the Internet in order to gain an understanding of its dynamical evolution and organization \cite{Costa2007}. 

		\subsection{Clustering Coefficient {\scriptsize (Transitivity)}}\label{sec:clu:clu}
\hrule \vspace{0.03cm}
\centerline{{L, R}\hfill\textit{-- all --} \hfill local impact / global \hfill static \hfill $\left[0,1\right]$ \hfill \checkmark} 
\hrule \vspace{0.1cm}

The \textit{Clustering Coefficient} \textit{(CC)} is commonly applied to measure the interconnectedness of nodes with the same neighbor. This metric has many different versions. In all of them, the coefficient ranges from zero to one, with a high value indicating high clustering.  The frequently used local version $C_i$ for simple graphs compares $e(V_i)$, defined as the number of edges among the neighbors $V_i$ of node $i$, to the maximum possible number \cite{Watts1998, Crucitti2003}:
\begin{equation}
C_{i}=\frac{e(V_{i})}{k_{i}(k_{i}-1)/2}
\end{equation}
\noindent 
A high $C_i$ denotes a node against whose removal the network is \textit{Disconnection} and \textit{Transmission Speed} robust, as many alternative routes between its neighbors exist \cite{Edwards2012}.

$C_i$ is not defined for nodes having degree of one, 
which is problematic 
as the global \textit{CC} is defined as $\hat{C}=\overline{C_{i}}$ \cite{Sato2009}. 
In \cite{Edwards2012} the Internet is evaluated from 
2002 to 
2010 using AS-level graphs based on snapshots of BGP routing-table dumps. 
The authors discover that most structural metrics 
are time-invariant. 
An exception is $\hat{C}$, which decreases while maintaining its distribution, possibly due to a flattening of net growth of transit providers in 2005 \cite{Oliveira2007} as they are substituted by an increasing number of unobservable peer-to-peer links. 

An enhanced version 
includes only nodes of degree higher than one \cite{Soffer2005}:
\begin{equation}
C=\frac{\sum_{i|k_i>1}{C_i}}{\sum_{i|k_i>1}{1}}
\end{equation}

Generally, many real-world networks exhibit a high global $C$ \cite{Soffer2005}. 

Another approach to deal with the bias problem uses the following definition: \cite{newman2001scientific}:
\begin{equation}
C_{\Delta} = \frac{3T\Delta}{T_3}=\frac{\sum_{i=1}^v e(V_i)}{\sum_{i=1}^v k_i(k_i-1)/2}=\frac{3\sum_{k>j>i}a_{ij}a_{jk}a_{ik}}{\sum_{k>j>i}(a_{ij}a_{ik}+a_{ji}a_{jk}+a_{ki}a_{kj})}\\
\end{equation}
\noindent 
As this approach normalizes by the sum of the possible edges between the neighbors of a node, the nodes of degree zero or one can be considered without further precautions. It gives the same weight to each triangle in the network. As they are usually involved in a larger number of triangles than low-degree vertices, hubs tend to have a higher weight in the calculation than with the other two global versions that weigh each node equally. 
A comparison thus reveals whether clustering occurs relatively more on hubs ($C<C_\Delta$), or on low-degree vertices ($C>C_\Delta$). 

In \cite{Soffer2005}, the average $C_i$ for nodes with degree $k$, 
$C(k)$, is found to be a decreasing function of vertex degree because of the disassortative nature of Internet graphs (Section \ref{sec:con:ass}), where high-degree vertices mainly interconnect subgraphs consisting of smaller-degree vertices with few inter-subgraph connections. It holds that $C(k)\propto k^{-\alpha}$. 
Ravasz and Barab\'asi \shortcite{Ravasz2003} find a hierarchical exponent $\alpha=0.75$ on the AS-level and a $C(k)$ independent of $k$ on the router-level.

In \cite{Soffer2005}, a \textit{CC} is introduced that is not correlated to the \textit{Assortative Coefficient}:
\begin{equation} 
\tilde{c_i}=e(V_i)/\omega_i
\end{equation}
where $\omega_i\leq \Omega_i=\left[\frac{1}{2}\sum_{j,l\in  V_i}[\min(k_j,k_l)-1]\right]\leq\frac{k_i(k_i-1)}{2}$, in contrast to $\frac{k_i(k_i-1)}{2}$,  takes into account that not all the $k_i-1$ excess edges are available at the neighbors of $i$. $\tilde{c_i}$ is obtained by a rewiring algorithm. 
This metric has the following characteristics: If all neighbors of a vertex have degree one, the \textit{CC} is undefined; $\tilde{c_i}\geq C_i$ holds for all $i$; and if all neighbors of $i$ have a degree larger than or equal to $k_i$, it follows that $\tilde{c_i}=C_i$. Also, two different global versions are proposed:
\begin{equation} 
\tilde{C}=\frac{\sum_{i|\omega_i>0}\tilde{c}_i}{\sum_{i|\omega_i>0}1}\quad\text{and}\quad
\tilde{C_{\Delta}}=\frac{\sum_ie(V_i)_i}{\sum_i\omega_i}
\end{equation}
For disassortative networks such as the Internet, $C$ and $C_{\Delta}$ give highly contrasting results, whereas $\tilde{C}$ and $\tilde{C_{\Delta}}$ are similar, and $\tilde{c_i}$ does not  depend as strongly on the node degrees but remains constant or decays logarithmically with increasing vertex degree.

Comparing these simple-graph \textit{CCs}, a high $C$ or $C_\Delta$ as such indicates a graph of greater \textit{Disconnection} and \textit{Transmission Speed Robustness} as more alternative routes with hardly lengthened paths exist around the failed node. An assessment whether clustering rather occurs on hubs $C<C_\Delta$ or on low-degree vertices $C>C_\Delta$, has direct implications for Internet resilience: Doyle et al. \shortcite{Doyle2005} (contrasting Albert et al. \shortcite{albert2000barabasi}) state that the Internet \textit{Backbone} is formed by a low-degree mesh graph, and the access networks by hub ring-, star- or tree-like graphs located at the network periphery \cite{jabbar2010framework}. This implies that $C>C_\Delta$ should hold. An  increased clustering of low-degree vertices ($C$) enhances global \textit{Transmission Speed} and \textit{Traffic Robustness} of the network as a whole; locally though, a higher clustering of hubs ($C_\Delta$) significantly enhances the \textit{Disconnection Robustness} of single peripheral clusters and end-users. Generally, a node of high $C_i$ has a low local impact in terms of these aspects, while one of very low $C_i$ is a local bottleneck.   If graphs of equal  \textit{Degree-Frequency Distribution} $P(k)$ (Section \ref{sec:con:nod}) and varying $C$ or $C_\Delta$ are investigated, though, the interpretation becomes more complex: The higher $C$, the more intra-connected the clusters and the less inter-connected. Thus, \textit{Disconnection} and \textit{Transmission Speed Robustness} against the removal of inner cluster entities is increased, but decreased against faults of community-peripheral bridge entities. To determine the position of a node in a cluster, the \textit{Participation Coefficient} (Section \ref{sec:clu:par}) can be used. 

As $\tilde{C}$ and $\tilde{C_\Delta}$ calculate how many of available neighbor ties are used for clustering, thus not as community-connecting bridges, its interpretation corresponds to the one of altered $C$ with constant distribution $P(k)$. Properly addressing this problem with one single metric, $\tilde{C}$ and $\tilde{C_\Delta}$ are very suitable to discover a too-pronounced community structure that bears the risk of fragmenting \textit{Disconnections}. It could also cause severe \textit{Transmission Speed} delays and \textit{Overload} increases in the whole network in worst-case scenarios of cutting crucial bridges as presented in Section \ref{sec:min}.

In \cite{Barrat2004}, the following local version is proposed for weighted networks:
\begin{equation}
\hat{C}_i^w=\frac{1}{s_i(k_i-1)}\sum_{(j,k)}{\frac{w_{ij}+w_{ik}}{2}a_{ij}a_{ik}a_{jk}}
\end{equation}
where the global version $\hat{C}^w$ is again obtained by averaging $\hat{C}_i^w$. Here, connections between nodes with high-weighted connections to node $i$ get a higher weight, regardless of their own weight. When comparing $C$ and $\hat{C}^w$, $\hat{C}^w>C$ indicates a network where interconnected triplets are more likely to be formed by edges with larger weights. For $\hat{C}^w<C$, the opposite holds: clustering is less important for the traffic flow and thus also for network organization \cite{Barthelemy2004}. 

This metric is only able to assess the individual impact of node $i$ on \textit{Disconnection Robustness} on a mere topological level. As the weight of neighbor-interconnecting edges is not taken into account, meaning the ability to redirect possibly large traffic flows through them when $i$ fails,
 it is not suitable for assessing local impacts on \textit{Transmission Speed} and \textit{Traffic Robustness}. 
 A better approach in this sense is introduced in \cite{Onnela2005}: 
\begin{equation}
\tilde{C}_i^w=\frac{2}{k_i(k_i-1)}\sum_{jk}(\hat{w}_{ij}\hat{w}_{jk}\hat{w}_{ik})^{1/3}\, ,\quad\text{with}\quad \hat{w}_{ij}=\frac{w_{ij}}{\max_{i,j}w_{ij}}
\end{equation}
$\tilde{C}_i^w$ decreases smoothly with the weight of the neighbor-interconnecting edge, as opposed to $\hat{C}_i^w$, which first stays constant, but abruptly drops when the weight turns zero.

In \cite{Opsahl2009}, a weighted equivalent to the global 
$C_{\Delta}$ is introduced: 
\begin{equation} \label{wasser}
C_{\tau}=\frac{\text{total value of closed triplets}}{\text{total value of triplets}}=\frac{\sum_{T\Delta}\tau}{\sum_{T_3}\tau}
\end{equation}
The triplet value $\tau$ can be defined as either the arithmetic mean of the weights of links that form the triplet -- which is severely influenced by extreme values of weights -- or as the less sensitive geometric mean, or the maximum respectively minimum value of edge weights. The latter two options, given two triplets with the same average weight, apply a higher/ lower value to the one with larger weight dispersion, thus making $C_{\tau}$ with the minimum definition of $\tau$ very useful for global \textit{Transmission Speed} and \textit{Traffic Robustness}.

Furthermore, the definition of Wassermann and Faust \shortcite{wasserman1994social} is used to make eq. \ref{wasser} applicable to directed graphs: the nodes $i$, $j$, $k$ form a triplet if an edge points from $i$ to $j$ and one from $j$ to $k$, and a closed triplet if, additionally, a direct link from $i$ to $k$ exists, which preserves the path if $j$ fails. This definition can thus be used to measure all three \textit{Robustness Aspects} on the AS level.

		\subsection{Edge Clustering Coefficient}\label{sec:clu:edg}
\hrule \vspace{0.03cm}
\centerline{{L, R}\hfill simple (\textit{all}) \hfill local impact \hfill static \hfill $\left[0,1+\left[\min{(k_i-1,k_j-1)}\right]^{-1}\right]$ \hfill \checkmark} 
\hrule 
\vspace{1mm}
\begin{equation}
C_{(i,j)}=\frac{T\Delta_{(i,j)}+1}{\min(k_i-1,k_j-1)}\label{edgclu}
\end{equation}
\noindent In this definition of the \textit{Edge Clustering Coefficient} \cite{radicchi2004defining}, $T\Delta_{(i,j)}$ denotes the number of triangles the edge $(i,j)$ belongs to. According to Newman \shortcite{Newman2004}, the $+1$ in the numerator is added in order to not overestimate the importance of edges that do not belong to triangles and connect to a low-degree vertex. $C_{(i,j)}$ is not defined for edges that connect degree-one nodes. 

More complex loops, such as squares, can be also taken into account. Then, low values identify edges that connect different communities 
and are bottlenecks not only on a local scale as in eq. \ref{edgclu}, with high impact on \textit{Disconnection} and \textit{Transmission Speed Robustness}. As stated in \cite{Costa2007}, this holds only for networks with a high average \textit{CC}, which is the case for the Internet. 


		\subsection{Modularity}\label{sec:clu:mod}
\hrule \vspace{0.03cm}
\centerline{{L, To, R}\hfill simple (weighted) \hfill global \hfill static \hfill $\left[0,1\right]$ \hfill \checkmark} 
\hrule 
\begin{equation}
Q=\sum_i \left(e_{ii}-\left(\textstyle\sum_j{e_{ij}}\right)^2\right)=\Tr \mathbf{E}-||\mathbf{E}^2||
\end{equation}
Determining the clusters or communities in large graphs is of particular significance for detecting vulnerable points and assessing the network structure. 
To this end, \textit{Modularity} \cite{Newman2004a} measures the quality of a given global network division into $k$ communities, where $\mathbf{E}$ is a $k\times k$ matrix whose elements $e_{ij}$ are defined as the fraction of all links in the graph that interconnect communities $i$ and $j$. Consequently, $\Tr \mathbf{E}$ gives the total fraction of inner-community edges. $\sum_j{e_{ij}}$ is the fraction of edges that connect to vertices in community $i$. In a graph where edges connect vertices randomly without considering their community, $e_{ij}=\sum_i{e_{ij}}\sum_j{e_{ij}}$ holds, thus $Q=0$. Therefore, a low $Q$ indicates a bad division into communities, and $Q=1$ a graph where every community represents one disconnected cluster. 

The following two algorithms are the most suitable out of the ones provided in \cite{Newman2004, Newman2006} for Internet graphs since they apply to real networks without homogeneous community sizes and knowledge of the number of communities, using minimal calculation time:

		\subsubsection{\textit{Modularity Matrix}}\label{sec:clu:mod:mod}
 This bisection algorithm  \cite{Newman2006} 
 maximizes the value of \textit{Modularity} \cite{Costa2007}. The \textit{Modularity Matrix} $\mathbf{B}$ is constructed with its elements defined as
\begin{equation}
b_{ij}^g=a_{ij}-\frac{k_i k_j}{2e}-\delta_{ij}\sum_{u\in V}\left[a_{(i,u)}-\frac{k_i k_u}{2e}\right]\quad 
\end{equation}
\noindent where $\delta_{ij}$ is the Kronecker Delta. 
The entries of the eigenvector corresponding to the largest positive eigenvalue divide the nodes into a group with positive eigenvector values, and one with negative values. 
Existing edges between them are not removed as this would change the degrees of the nodes and hence bias further calculations. The calculation is repeated within the groups until the \textit{Modularity Matrix} eigenvalues are all zero or negative, which indicates that no further useful division is possible. The advantage of this algorithm compared to other bisection methods is that the sizes of the communities, into which the network finally splits, do not have to be homogenous or known. The disadvantage is that at any step only a division into exactly two communities is possible. As stated in \cite{Newman2004}, dividing a network into two parts and then dividing one of those again does not yield the best possible result for a division into three communities.

		\subsubsection{Edge Betweenness Partitioning Algorithm}\label{sec:clu:mod:edg}
The \textit{Edge Betweenness Centrality} metric (Section \ref{sec:thr:bet}) used in this algorithm is the ratio of shortest paths between two nodes that pass through the considered edge, 
summed up over all node pairs. The idea behind it is that communities are characterized by having scarce inter-community edges that act as bottlenecks, obtaining high \textit{Edge Betweenness Centrality} values \cite{Newman2004, Girvan2002abc}.
The algorithm repeatedly deletes the highest valued links and recalculates the \textit{Modularity} and the \textit{Edge Betweenness Centrality} after each removal. Finally, the state with the highest \textit{Modularity} is chosen. 

The implementation of this algorithm does not require homogeneity of community sizes or knowledge about their number, and as opposed to the \textit{Modularity} Matrix, the graph can be split into any number of communities. 
Still, the whole process is very costly since after every edge removal, the \textit{Edge Betweenness Centrality} of each remaining edge must be calculated again, which takes $\mathcal{O}(ev)$ every time. In total, this can take $\mathcal{O}(v^3)$ on a sparse graph or $\mathcal{O}(e^2v)$ on a non-sparse graph in the worst case. 
In order to reduce the calculation time and to introduce a stochastic element, Tyler et al. \shortcite{tyler2003proceedings} propose to only sum over a random subset of vertices, giving the partial \textit{Edge Betweenness Centrality} scores for all edges as a Monte Carlo estimation, which provides good results with reasonably small sample sizes. 

		\subsection{Z-Score of the Within Module-Degree}\label{sec:clu:zsc}
\hrule \vspace{0.03cm}
\centerline{{L, R}\hfill simple (\textit{all}) \hfill local impact \hfill static \hfill $\left[-\infty,\infty\right]$ \hfill \checkmark} 
\hrule 
\begin{equation}
z_i=\frac{e_i(K_i)-\overline{e(K_i)}}{\sigma_{e(K_i)}}
\end{equation}
\noindent 
\textit{Z-Score of the Within Module-Degree} \cite{guimera2005functional} calculates the rank of a node within its community, which is higher for a larger $z_i$. $e_i(K_i)$ is the number of connections that node $i$ has within its community, $\overline{e(K_i)}$ is the average number of intra-community connections of the nodes in $K_i$,  and $\sigma_{e(K_i)}$ the corresponding standard deviation. This is a normalized version of the \textit{Node Degree} (Section \ref{sec:con:nod}) if just the community as a subgraph is considered. This approach of finding inner-community hubs is sufficient by itself for detecting vertices of high impact on inner-cluster \textit{Disconnection} and \textit{Transmission Speed Robustness}, as by definition inside of communities no bottlenecks exist. 

		\subsection{Participation Coefficient}\label{sec:clu:par}
\hrule \vspace{0.03cm}
\centerline{{L, To, R}\hfill simple (weighted) \hfill local impact \& liability \hfill static \hfill $\left[0,1\right]$ \hfill \checkmark} 
\hrule 
\begin{equation}
P_i=1-\sum_j \left(\frac{e_i(K_j)}{k_i}\right)^2
\end{equation}
\noindent 
$e_i(K_j)$ is the number of links from node $i$ to community $K_j$. The \textit{Participation Coefficient} \cite{guimera2005functional} measures how equally distributed the connections of $i$ are among all communities. $P_i=0$ indicates that $i$ is only connected to nodes in its own community, while $P_i=1$ shows that its edges are uniformly distributed among all communities. $P_i$ serves as an indicator of the impact of a node, as a bridge, on the \textit{Robustness} against community \textit{Disconnection} or increased \textit{Transmission Speed}. Furthermore, the removal of a small share of neighbors of a node with high $P_i$ can degrade the node's \textit{Transmission Speed} severely, if the removed neighbors belong to the same community.  

	\subsection{Recapitulation}\label{sec:clu:dis}
Knowledge about the community structure of a graph provides insights into its hierarchical characteristics and gives first indicators of bottlenecks. 
The algorithms presented with \textit{Modularity} are useful tools for community detection, and a resulting high value of the metric itself or of the \textit{CC} of available neighbor edges, $\tilde{C}$, serves as an indicator of a too pronounced clustering structure that is prone to severe \textit{Disconnections}, \textit{Transmission Speed} decreases and \textit{Overloads}. 
In the presence of such a structure, the role of a node inside its communities as determined by \textit{Z-Score of the Within Module-Degree} and \textit{Participation Coefficient} is crucial; especially the latter detects vulnerable points. 

In contrast to $\tilde{c}_i$, the local $C_i$ by itself measures the ability to locally redirect information flows if node $i$ fails, thus its local impact on \textit{Disconnection} and \textit{Transmission Speed Robustness}.

The importance of clustering for network flows can be estimated by the comparison of 
$\hat{C}^w$
and $C$. 
As indicators for \textit{Transmission Speed} and \textit{Traffic Robustness}, in a weighted graph $\tilde{C}_i^w$ serves best by accounting for weights of neighbor-interconnecting edges; in a weighted directed network, $C_{\tau}$ with the minimum-value triplet definition can be used.

\section{Connectivity}\label{sec:min}
\textit{Clustering} in a network contributes to its local robustness but entails the presence of fewer bridges between these clusters, and hence a reduced global \textit{Disconnection Robustness}. In the subsequent sections, the fundamental and NP-complete task \cite{Mohar1989} of finding a graph-separating \textit{Minimum Cut Set} and approximating bi-partitioning algorithms are discussed. After this worst-case approach, metrics to assess \textit{Connectivity} under multiple random failures are presented. 

	\subsection{Vertex-, Edge- \& Conditional Connectivity {\scriptsize (Cohesion,  Adhesion \& P-Connectivity)} }\label{sec:min:con}
\hrule \vspace{0.03cm}
\centerline{{To, R}\hfill simple graph \hfill global \hfill worst-case \hfill $\left[0,v-1\right]$ \hfill NP-complete} 
\hrule 
\hspace{1pt}

The \textit{Vertex-} $\kappa (G)$ and \textit{Edge-Connectivity} $\mu (G)$ of a graph are defined as the minimum number of nodes and edges whose removal disconnects the graph, respectively \cite{chartrand1996graphs, yazdani2010note}. A severe flaw is that the resulting component sizes are not accounted for, making it unsuitable for robustness analysis: On sparse graphs such as an Internet graph containing stub nodes, the return will equal the minimum \textit{Node Degree}. On networks with a high minimum \textit{Node Degree} and severe bottlenecks, this problem is NP-complete \cite{npcomplete}.

A more useful, though also NP-complete concept is the \textit{Conditional Connectivity} $\kappa(G,P)$. It is the smallest number of vertices (links) whose removal disconnects the graph, with every component having property $P$ \cite{harary1983conditional}. In order to address the above-mentioned flaws, $P$ can be set to the minimum of the resulting component sizes. Further metrics that address this problem are presented in the following sections.


		\subsection{Sparsity}\label{sec:min:spa}
\hrule \vspace{0.03cm}
\centerline{{To, R}\hfill simple graph \hfill global \hfill worst-case\hfill $\left[4/(v-1)^2,v-2\right]$ \hfill \checkmark} 
\hrule
\begin{equation}
Q_X=\frac{|X|}{|A||A'|}
\end{equation}
\noindent Here, $A$, $X$ and $A'$ are three non-empty sets of nodes that together form the entire graph G. $X$ is the set of vertices that,  if deleted, causes a separation from $A$ and $A'$. Therefore, 
a low value for $Q_X$ indicates a network of low \textit{Disconnection Robustness}, and a high value a robust one. 

Calculating the minimum of this metric 
is NP-hard \cite{bui1992finding}. In \cite{Sreenivasan2007}, this problem is addressed by considering every bi-partition of the graph into $\widehat{A}$ and $\widehat{A'}$, defining $c(\widehat{A})$ as node subset in $\widehat{A}$ adjacent to $\widehat{A'}$, and $c(\widehat{A'})$ analogously, and taking 
\begin{equation}
\min_X Q_X= \min_{\hat{A},\hat{A'}\subset V} \left( \frac{|c(\widehat{A})|}{|\widehat{A}-c(\widehat{A})||\widehat{A'}|}, \frac{|c(\widehat{A'})|}{|\widehat{A'}-c(\widehat{A'})||\widehat{A}|}\right)
\end{equation}
Interestingly, the inverse of this metric is the minimum possible average \textit{Betweenness Centrality} (Section \ref{sec:thr:bet}) of the nodes in $X$, as every shortest path from $A$ to $A'$ must pass through $X$. 

		\subsection{Cheeger Constant {\scriptsize (Isoperimetric Number)} 
		}\label{sec:min:che}
\hrule \vspace{0.03cm}
\centerline{{To, R}\hfill simple graph \hfill global \hfill worst-case\hfill $\left[0, v/2\right]$ \hfill NP-complete} 
\hrule
\begin{eqnarray}
h(G)&=&\min\left\{\frac{|\partial A|}{|A|}:A\subseteq V(G), 0\leq |A|\leq \frac{v}{2}\right\}\\
\partial A&=&\{(x,y)\in E(G): x\in A, y \in V(G)\backslash A\}
\end{eqnarray}
This metric \cite{Mohar1989} is similar to the \textit{Sparsity} bisection approach, but focuses on edges instead of nodes. It 
calculates the minimum ratio of edge cuts necessary for graph bisection to nodes in the resulting smaller cluster. Small cut sets and large impacts in terms of \textit{Reachability} (Section \ref{sec:min:ana}) are rewarded. $h(G)=0$ holds for a disconnected graph, and $h(G)=v/2$ for a fully connected one.

In \cite{Merris1994}, bounds for the NP-complete \textit{Cheeger Constant} are derived using the calculable \textit{Algebraic Connectivity} $\lambda_2$ (Section \ref{sec:spe:alg}): $\frac{\lambda_2}{2}\leq h(G)\leq \sqrt{\lambda_2(2k_{max}-\lambda_2)}$. However, they are not very tight and thus of little use due to the scale-free nature of the Internet. With the subsequently presented \textit{Network Partitioning Algorithm}, the \textit{Cheeger Constant} 
can be approximated by setting the balancing criterion to $m/v=0.05$, repeatedly running the algorithm for $a=|A|/v=\{0.05, 0.15, ..., 0,45\}$, respectively, and taking the minimum resulting $h(G)$.

		\subsection{
			Minimum m-Degree}\label{sec:min:minm}
\hrule \vspace{0.03cm}
\centerline{{To, R}\hfill simple graph \hfill global \hfill worst-case\hfill $\left[1,m(v-m)\right]$ \hfill NP-C. \& $\mathcal{O}\left((\substack{v\\m})e\right)$} 
\hrule 
\begin{eqnarray}
\xi(m)&=&\min\left\{\partial A: A\subseteq V(G), |A|=m\right\}\\
\partial A&=&\{(x,y)\in E(G): x\in A, y \in V(G)\backslash A\}
\end{eqnarray}
Very similar to the \textit{Cheeger Constant} (Section \ref{sec:min:che}), the \textit{Minimum m-Degree} denotes the smallest number of edges that must be removed to split the network into two components, one of which contains exactly $m$ nodes \cite{boesch1970graphs}. 
 Since for estimating \textit{Disconnection Robustness} the exact component size is not crucial, this restriction can be relaxed as follows.
\subsubsection{Network Partitioning Algorithm}\label{sec:min:net}
This 
algorithm
finds a graph bi-partitioning \textit{Minimum Edge Cut Set} $E_{min}(m/v)$ of size $\xi(m)=|E_{min}(m/v)|$ with splitting ratios of approximately $m/v$. 
First, a balancing criterion of $c$ nodes is set, by which each partition may deviate at most from the splitting ratio $m/v$.
A separation then bisects the graph, denoting the resulting inter-partition links as cuts. Afterward, the following pass 
is executed: The gain of all nodes, initially denoted as unlocked, is calculated as the decrease in the number of cuts if the node is moved to its opposite partition. The largest (possibly negative) gain node is moved, as long as the balancing criterion is not violated, and denoted as locked. 
All node gains are recalculated, and the process is repeated until either all nodes are locked or the balancing criterion prevents further moves. The split with the fewest cuts  is 
 executed. Then, all nodes are unlocked again, and the procedure is repeated until the cut set cannot be further reduced.
 The computation takes $\mathcal{O}(e)$ per pass and the convergence occurs quite fast. In the worst case, $\mathcal{O}(e)$ passes are needed  \cite{fiduccia1988linear}.

Wang et al. \shortcite{Wang2008a} studied Internet models both on AS and router levels and find that approximately 
6.5\% of the links have to be removed in order to obtain a splitting ratio of 50\%. 
For a splitting ratio up to 30\%, the disconnected parts are highly fragmented pieces in the router model, whereas in the AS model a large cluster 
contains 20-40\% of disconnected nodes. Afterward, in both models, 
the vertices glue together, which enables them to communicate. 
The largest cluster in a scale-free network separated this way remains scale-free and functions well. 

Furthermore, it is analyzed if the \textit{Minimum Edge Cut Set} for randomly distributed nodes, which can be a connected cluster, can be significantly reduced by this algorithm compared to just cutting off all the links adjacent to the target nodes. The results show that this is only possible if the target nodes are connected and only at the expense of cutting off many non-target nodes with them.

\subsection{Ratio of Disruption}\label{sec:min:rat}
	\hrule \vspace{0.03cm}
	\centerline{{To, R}\hfill simple graph \hfill global \hfill worst-case\hfill $\left[0, v/2\right]$ \hfill NP-complete} 
\hrule

Another metric that assesses edge cuts is the \textit{Ratio of Disruption} where $[.]$ denotes the floor function \cite{lipman1985toward}: 
\begin{equation}
\rho=\max \left\{\frac{|A|}{|\partial A||V-A|}:1\leq|A|\leq \left[\frac{v}{2}\right]\right\}
\end{equation}
Since this metric takes the size of the other (in most cases larger) component $V-A$ into account, it is more likely to result in more equally-sized graph cuts, but at the expense of bigger cut-set sizes $|\partial A|$ than the \textit{Cheeger Constant}. This metric is thus designed to find cut-sets which have considerably more severe impacts on a graph's \textit{Reachability} (Section \ref{sec:min:ana}). 
Unfortunately, it is NP-complete and cannot be conveniently approximated 
by the \textit{Network Partitioning Algorithm} (Section \ref{sec:min:net}). Therefore, this important concept is currently not applicable to large Internet graphs. 
\\

		\subsection{Local Delay Resilience {\scriptsize (Resilience)}}\label{sec:min:res}
\hrule \vspace{0.03cm}
\centerline{{To, R, Tr}\hfill simple, directed \hfill local liability / global \hfill worst-case\hfill $\left[1, v(h)^2/4\right]$ \hfill NP-hard ($\mathcal{O}(e)$)} 
\hrule

\vspace{0.03cm}

\begin{alignat}{2}
R_i(h)&=|E_{min}\left(0.5,G_i(h)\right)|, \quad&\text{with}\quad& G_i(h)=\{m\in V|d_{(i,m)}\leq h, h\in\mathbb{N}\}\\
R(v(h))&=\overline{R_i(h)}, \quad&\text{with}\quad& v(h)=\overline{|G_i(h)|}\quad\quad\quad
\end{alignat}
$G_i(h)$ is the subgraph induced by nodes within the $h$-hop environment of node $i$. The local \textit{Local Delay Resilience} $R_i(h)$ measures the size of its \textit{Minimum Cut Set} $E_{min}$ of splitting ratio $0.5$, assessing its proneness to local \textit{Disconnection} or severe \textit{Transmission} delays when entities around it fail. 
The global \textit{Local Delay Resilience} $R$ 
is a function of the hop-radius $h$, but as 
the number of vertices within it is higher 
in graphs with a high \textit{Expansion} (Section \ref{sec:dis:exp}), $R$ is 
presented as a function of 
$v(h)$, 
denoting the average number of vertices within $h$ hops. 
Karypis and Kumar \shortcite{karypis1998fast} present a partitioning heuristics for this NP-hard problem.

In \cite{Tangmunarunkit2002}, the investigated AS and router-level graphs are stated to have a high \textit{Local Delay Resilience} for any $h$. 
Their directed adaptation, which 
takes policy-based routing into account, yields substantially lower values than the simple one. Now, $G_i(h)$ is induced by nodes of policy-compliant paths no longer than $h$ hops. 
Thus, in AS-level graphs, only paths that do not violate provider-customer relationships are considered; in order to determine a router-level policy path, first the corresponding AS-level path is computed and then shortest paths within the ASes.

The name of this metric given above is an adapted version of the one found in the literature. It needs to be emphasized that it only measures one aspect of \textit{Resilience}, and could by no means aggregate all facets of this complex concept (Section \ref{robasp}) into one single-valued number.

\subsection{Toughness, Integrity, Scattering Number} \label{sec:min:toughness} 
\hrule \vspace{0.03cm}
\centerline{To, R\hfill simple graph \hfill global \hfill worst-case\hfill N/A \hfill NP-complete} 
\hrule 
The previously presented \textit{Connectivity} metrics concentrate on detecting worst-case bisections, but do not conveniently assess the damage of partitioning the network into more than two components. \textit{Toughness} $T_{\overline{S}}$ \cite{CHVATAL1973215}, \textit{Integrity} $I$ \cite{barefoot1987vulnerability} and \textit{Scattering Number} $s$ \cite{Jung1978} aim to fill this gap by finding cut sets $\overline{S}$ (either the set of removed nodes $\overline{V}$ or edges $\overline{E}$), which minimize or maximize certain parameters.
\begin{eqnarray}
T_{\overline{S}}&=&\min_{\overline{S}} |\overline{S}|/\omega_{G-\overline{S}}\\
I_{\overline{S}}&=&\min_{\overline{S}}|\overline{S}|+|V_L|_{G-\overline{S}}\\
s(G)&=& \max \omega_{G-\overline{V}} - |\overline{V}|
\end{eqnarray}
They account for important characteristics of the disconnected graph, namely cut set size $\overline{S}$, number of nodes in the largest component $|V_L|$ and the number of resulting components $\omega_{G-\overline{S}}$, but every metric omits one of these. 
An inner similarity of these metrics can be ascertained by the results they provide: Knowing the \textit{Scattering Number} and some basic information about the graph, a lower or upper bound for the other metrics can be derived \cite{paperw}.

\subsection{Tenacity}\label{sec:min:ten}
\hrule \vspace{0.03cm}
\centerline{To, R\hfill simple graph \hfill global \hfill worst-case\hfill N/A \hfill NP-complete} 
\hrule 
As opposed to the metrics presented in the previous section, \textit{Tenacity} $T(G)$, \textit{Edge Tenacity} $T_e(G)$ and \textit{Mixed Tenacity} $T_m(G)$ \cite{Alon1997} take all of the three  important parameters into account for evaluating the \textit{Disconnection Robustness} of a graph.
\begin{eqnarray}
T(G)&=&\min_{\overline{V}\subset V(G)}{\left[|\overline{V}|+|V_L|_{G-\overline{V}}\right]/\omega_{G-\overline{V}}}\\
T_e (G)&=&\min_{\overline{E}\subset E(G)}{\left[|\overline{E}|+|E_L|_{G-\overline{E}}\right]/\omega_{G-\overline{E}}}\\
T_m&=&\min_{\overline{E}\subset E(G)}{\left[|\overline{E}|+|V_L|_{G-\overline{E}}\right]/\omega_{G-\overline{E}}}
\end{eqnarray}
\noindent Here, $\overline{V}$ and $\overline{E}$ represent the sets of removed nodes and edges, respectively, 
$|V_L|_X$ and $|E_L|_X$ the resulting number of nodes and edges in the largest component of graph $X$, 
and $\omega_X$ its number of components. 

A study on the behavior of \textit{Tenacity}, or an approximating algorithm, applied to Internet graphs is currently still due. As stated in \cite{Piazza1995}, a large set of graph classes exists that are edge-tenacious, meaning that the minimum of the term measured by $T_e(G)$ is obtained by cutting the entire set of edges in the graph. 
As the same presumably holds for Internet-representing graphs, this metric would become useless. 

\color{black}

\subsection{Percolation Threshold}\label{sec:min:per}
\hrule \vspace{0.03cm}
\centerline{{To, R}\hfill simple graph \hfill global \hfill failures \hfill $\left[0,1\right]$ \hfill \checkmark} 
\hrule
\begin{equation}
1-p_c=\left(\overline{k^2}/\overline{k}-1\right)^{-1}
\end{equation} 
\textit{Percolation Threshold} $p_c$ measures the threshold of uniform node failure probability after which the network disintegrates, meaning that no giant component of graph size order exists \cite{cohen2000resilience}. For SF-networks with exponent $\gamma\approx 2.5$ like the Internet, $\displaystyle\lim_{v\rightarrow \infty}p_c=1$ for infinitely large graphs. For finite Internet networks, the \textit{Largest Component Size}  (Section \ref{sec:min:ana}) decreases with increasing node failure probability $p$, but the largest component persists for a $p$ of nearly 100\%.

\subsection{Reliability Polynomial}\label{sec:min:rel}
\hrule \vspace{0.03cm}
\centerline{{To, R}\hfill simple graph \hfill global \hfill failures \hfill $\left[0,1\right]$ \hfill \checkmark 
} 
\hrule
\begin{equation}
Rel(K)=\sum_j a_j p^j (1-p)^{e-j}
\end{equation} 
The \textit{K-Reliability Polynomial} $Rel(K)$ measures the probability that, given a uniform edge-reliability $p$ and thus edge failure probability $1-p$, all nodes of a subset $K$ are connected. The number of $K$-connecting subgraphs with $j$ edges is denoted by $a_j$. This is the most general definition, denoted \textit{K-Reliability}. 
If for all $p\in(0,1)$, $Rel_{G-E_i}(K)\geq Rel_{G-E_h}(K)$ and $Rel_{G^*E_i}(K)\leq Rel_{G^*E_h}(K)$ holds, Page and Perry \shortcite{page1994reliability} rank the importance of edge $i$ higher than the one of edge $h$ based on purely topological characteristics. Here, $G^*E_i$ corresponds to graph $G$, where the nodes adjacent to edge $i$ are contracted to being one vertex. The \textit{2-Terminal Reliability}, $|K|=2$, calculates the probability that a message can be transmitted between both members of $K$. 

By calculating the \textit{All-Terminal Reliability}, $K=G$, before and after removing single entities, their importance to the topological structure can be ranked for a given $p$ \cite{page1994reliability}.  The \textit{All-Terminal Reliability} contains the size of minimum edge cutsets, which disconnect the graph (corresponding to \textit{Edge-Connectivity}, Section \ref{sec:min:con}) as well as their number. According to the authors, the \textit{Number of Spanning Trees} (Section \ref{sec:spe:num}) is  correlated to  \textit{All-Terminal Reliability}. The algorithm provided in \cite{page1988practical} is not practical for large Internet graphs, but can be used to measure the \textit{Disconnection Robustness} of its \textit{Backbone}. 

Interestingly, the coefficients of the polynomial indicate the \textit{Number of Spanning Trees}, \textit{Size of Smallest Edge-Cutset} and \textit{Number of Smallest Edge-Cutsets}. The latter two metrics could be useful for analysing the \textit{Backbone}, if the \textit{Size} is not determined by the node with minimum degree. An algorithm for \textit{Reliability Polynomial} calculation is presented in \cite{page1988practical}. 

\subsection{Partition Resilience Factor {\scriptsize (Resilience Factor)}  
}\label{sec:min:resfac}
\hrule \vspace{0.03cm}
\centerline{{To, R}\hfill simple graph \hfill global \hfill failures \hfill $\left[0,1\right]$ \hfill \checkmark 
} \hrule 
\begin{equation}
R_F=\frac{\sum^{v-1}_{i=2}k(i)}{n-2}
\end{equation} 
Again, an adapted metric name is used here to distinguish it from the much broader \textit{Resilience} concept. The \textit{Partial $i$-Connectivity} $k(i)$ denotes the ratio of node failure sets of size $i$, which disconnect the graph. 
A low \textit{Partition Resilience Factor} $R_F$ thus indicates a well-connected network \cite{salles2011strategies}. It attempts to capture subtle \textit{Disconnection Robustness} differences, but it does not consider the severity of disconnection  measured by \textit{Reachability} (Section \ref{sec:min:ana}), and is computationally very costly. 
Only an application to formerly identified \textit{Backbone} graphs seems appropriate due to much smaller network sizes and the high importance of every single vertex.

\subsection{Analysis of Disconnected Components \& Reachability \emph{{\scriptsize (Flow Robustness)}}}\label{sec:min:ana}
\hrule \vspace{0.03cm}
\centerline{{To, R}\hfill simple graph \hfill global \hfill dynamic \hfill different codomains \hfill \checkmark} 
\hrule \vspace{0.1cm}
In order to assess the impact of an attack or failure, it is necessary to take some characteristics of the disconnected components into account. According to Sun et al. \shortcite{Sun2007}, these could be the \emph{Total Number of Isolated Components} $N\in[1,v]$, the \emph{Fraction of Nodes in the Largest Component} $|V_L|/v\in[0,1]$ and the \emph{Average Size of Isolated Components} $\overline{|V_C|}\in [1,v]$. In \cite{Sacks2009} it is stated that the latter is not suitable for an application to the Internet as the disconnected component size seems to be heavy-tailed, always resulting in a low \emph{Average Size of Isolated Components}. 
In \cite{Xiao2010}, investigating an AS-level network measured 
in 2000, 
less than 3\% of the highest-degree nodes have to be removed to reach  $|V_L|/v\leq 0.05$. 

For large graphs,  
the \emph{Distribution of Component Class Frequency} \cite{Sacks2009} assigns disconnected components of similar size to disjunct classes $c\in [1,2,...,C]$, whose numbers indicate the decimal logarithm of the upper bound size of the class, and counts the nodes $|V_c|$ in each of them. 
Another related metric  is the \emph{Distribution of the Relative Number of Nodes per Class} \cite{Sacks2009}, which 
is calculated as $f(c,r)=|V_c|/v\in[0,1]$. It gives the probability that a node can communicate with at most a given number of nodes for a given percentage $r$ of removed nodes.

All these aspects are covered by \emph{Reachability} (or \emph{Flow Robustness}) $R$, 
being the reason why \textit{Disconnection Robustness} is best evaluated in terms of the resulting decrease in $R$. It is defined as $R=\frac{1}{v(v-1)}\sum_{i\neq j\in G} R_{ij}$, with $R_{ij}=1$ if a path exists between node $i$ and $j$ and $R_{ij}=0$ otherwise \cite{Sato2009}. 
It calculates the fraction of node pairs which can still communicate, thus  
$R\in[0,1]$. 
Confirming the findings in (Section \ref{sec:con:nod}), a positive correlation is found between the exponent $\gamma$ of an SF network and its \textit{Disconnection Robustness} in terms of $R$ 
 under node attacks \cite{Sato2009}.

In \cite{cetinkaya2013flow} the $R$ of the highest level of a multilevel AS graph after faults on lower levels is investigated. The multilevel graph consists of $L$ levels of simple graphs; for any pair of them, the set of all nodes in the higher level is a subset of the node set in the lower level. A connecting path between two nodes on a higher level can only exist if such a path also exists on the lower level. A partition of the tier-1 ISPs is found to be unlikely when the logical link layer is attacked. An increase in $L$, a switch from adaptive to non-adaptive attacks, as well as from static routing on the original shortest-path to perfect dynamic routing resulted in a significantly higher \textit{Reachability}. 

		\subsection{Recapitulation}\label{sec:min:dis}
The fundamental task of calculating exact \textit{Minimum Cuts} is NP-complete. \textit{Cohesion} and \textit{Adhesion} are not useful, as resulting component sizes are not considered (contrary to the other worst-case cut finders). A fundamental drawback of those is that suitable approximating algorithms, exist so far only for bisections: \textit{Sparsity} and \textit{Network Partitioning Algorithm} are both apt for global \textit{Disconnection Robustness} analysis. Yet, 
their calculation requires predetermined sizes of the resulting components, 
contrary to the NP-complete \textit{Cheeger Constant}, 
which could be conventiently approximated by the repeated execution of the \textit{Network Partitioning Algorithm}. 
For evaluating general cuts, \textit{Reachability} in combination with the fraction of removed entities is a convenient tool, 
as it implicitly accounts for number and sizes of all components. 
Nodes close to local bottlenecks are found by \textit{Local Delay Resilience}, 
which needs little calculation time relying only on 
subgraph $G_i(h)$.

Of the measures presented in Sections \ref{sec:min:toughness} and \ref{sec:min:ten}, only the \textit{Tenacity} metrics consider all features of graph partitions of more than two components. But as they probably yield trivial results Internet graphs, and as all metrics in these sections are NP-complete, none is suitable for robustness assessment.

\textit{Percolation Threshold} suitably assesses minimum \textit{Connectivity} requirements after multiple failures, namely, at least a large part of the network still is connected and functioning. \textit{Reliability Polynomial} and \textit{Partition Resilience Factor} are more precise failure evaluation measures, but only useful for application to the \textit{Backbone} due to their computational cost. 

\section{Distance}\label{sec:dis}
Distance metrics consider 
the path length between node pairs.  
Its simple or directed version $d_{(m,n)}$ is defined by the minimum number of hops it takes from node $m$ to $n$. 
Its weighted counterpart, $d^w_{(m,n)}=\min_{a,b,...,z\in V} \{w^{-1}_{ma}+w^{-1}_{ab}+...+w^{-1}_{zn}\}$ \cite{DallAsta2006},
further takes the speed and capacities of the connections into account. 

\subsection{Average Shortest Path Length \\
	{\scriptsize (Closeness Centrality/ Hop Count/ Average Diameter/ Network Diameter/ Geodesic Distance)}}\label{sec:dis:ave}
\hrule \vspace{0.03cm}
\centerline{{Tr}\hfill simple (\textit{all}) \hfill local impact / global \hfill static \hfill $\left[1, v/2\right]$ / $\left[1, \frac{v+1}{3}\right]$ \hfill \checkmark ($\mathcal{O}(v+e)$ / $\mathcal{O}(ve)$)}
\hrule
\begin{eqnarray}
\overline{d}_{i}&=&\frac{\sum_{m\in V}{d_{(i,m)}}}{v-1}\\
\overline{d}&=&\overline{\left\{\overline{d_i}\right\}}
\end{eqnarray}
The local \emph{Average Shortest Path Length (ASPL)} is useful for finding central nodes that can play a significant role in forwarding information. 
The global \emph{ASPL} $\overline{d}$ characterizes the average \textit{Transmission Speed} with which node pairs communicate \cite{Sun2007}. Its degradation due to \textit{Challenges} hence serves as an indicator for comparing \textit{Transmission Speed Robustness} of networks that display differing characteristics as measured by other metrics. 
An implementation of Dijkstra's algorithm is used in \cite{brandes2005centrality}.

In SF networks, \emph{ASPL} tends to remain unchanged even when up to 5\% of the nodes fail \cite{albert2000barabasi}. 
However, after a certain threshold of attacks, \textit{ASPL} is no longer suitable as it approaches $\infty$ when a node is disconnected. 
To address this issue, only $d_{(m,n)}\neq \infty$ or just nodes in the giant connected component 
\cite{Sun2007} could be taken into account. 
These versions, however, 
can obtain low values for disintegrated graphs, suggesting one with redundant connections. 

\emph{Diameter-Inverse-K (DIK)}  addresses this problem with the infinite paths excluding $\overline{d}$ \cite{park2003static}:
\begin{equation}
DIK=\overline{d}/K\quad \text{with}\quad K=\left|\Pi\right|/\left|\Psi\right|
\end{equation}
where $|\Psi|$ is the number of node pairs and $|\Pi|$ the number of pairs connected by a path. It increases continuously as the graph disconnects. 

As \emph{Distance Distribution Width}, the standard deviation of path lengths $d_{(m,n)}$ \cite{Mahadevan2005}, 
is small, $\overline{d}$ is only slightly smaller than the \emph{Diameter} $D$ \cite{Edwards2012, ng2006structural}, whose codomain is $[1,\infty[$ if infinite paths are taken into account, and $[1,v-1]$ otherwise.
\begin{equation}
D=\max_{m,n\in V} d_{(m,n)}
\end{equation}
Due to these findings, $\overline{d}$, $DIK$ and $D$ overlap considerably and can be used interchangeably in connected graphs, while $DIK$ is preferable in disconnected ones.

The formulae above refer to simple networks. For weighted networks, the following adaptation 
 proposed in \cite{DallAsta2006} 
is applicable to 
every metric containing $d_{(m,n)}$:
\begin{equation}\label{eq:dist}
d^w_{(m,n)}=\min_{a,b,...,z\in V} \{w^{-1}_{ma}+w^{-1}_{ab}+...+w^{-1}_{zn}\}
\end{equation}
The weighted shortest path between two nodes is the path for which the sum of the inverted weights of the traversed edges is minimal, and is independent of the number of traversed edges, which can be very useful as the way that a data package takes depends highly on the speed and capacities of the connections. 

\subsection{Global Network Efficiency {\scriptsize (Harmonic Centrality / Closeness Centrality)}}\label{sec:dis:glo}
\hrule \vspace{0.03cm}
\centerline{{Tr}\hfill simple (\textit{all})\hfill local impact / global \hfill static \hfill $[0,1]$ \hfill \checkmark ($\mathcal{O}(v+e)$ / $\mathcal{O}(ve)$)} 
\hrule
\begin{eqnarray}
E_i&=&\frac{\sum_{j\neq i}{d_{(i,j)}^{-1}}}{v-1}\\
E_{global}&=&\overline{E_i}
\end{eqnarray}
Similar as with the metrics presented in the previous section, the degradation of \emph{Global Network Efficiency} indicates the \textit{Transmission Speed Robustness} of a network under \textit{Challenges}. By summing up the reciprocal of the distances, the problem of disconnected graphs is solved 
\cite{Rochat2009, latora2001efficient}: For the local version, $E_i=0$ only holds for a totally disconnected node $i$, while $E_i=1$ indicates that node $i$ is directly connected to every other node.  
Analogously, $E_{global}$ only becomes zero for a totally disconnected network and one for a network where every node is directly connected to all others. 

As opposed to \textit{ASPL} and \textit{DIK} (Section \ref{sec:dis:ave}), which weigh a one-hop increase equally for any path length, the \textit{Global Network Efficiency} is considerably more affected by one-hop increases in short than in long paths. This is interesting if, as assumed in the work of Gkantsidis et al. \shortcite{Gkantsidis2003a}, traffic is not homogeneously distributed between all node pairs, but is to a certain extent concentrated on topologically close nodes. In this case, \textit{Global Network Efficiency} would be more suitable for assessing \textit{Transmission Speed} than \textit{ASPL}.
	
However, for connected SF-graphs that do not exhibit highly improbable structures, there is a high correlation between the reciprocal \textit{ASPL} \cite{Opsahl2010} and this metric according to Rochat \shortcite{Rochat2009}. 
Since it most conveniently assesses the \textit{Speed}, with which a node or entire graph can communicate, and also works in disconnected graphs, \textit{Transmission Speed Robustness} against \textit{Challenges} is best evaluated, of all metrics presented so far, in terms of \textit{Global Network Efficiency} decrease. 

In \cite{cetinkaya2013flow}, multilevel Internet networks as described in Section \ref{sec:min:ana} are analyzed simulating non-adaptive attacks against highest-$E_i$ nodes. After a certain number of deletions, the resulting \textit{Reachability} (Section \ref{sec:min:ana}) and thus \textit{Disconnection Robustness} is higher than for the case of random failures. The reason is suggested to be the non-adaptive strategy, but it could also be due to mere unsuitability for detecting nodes with a high impact on this aspect as well-interconnected \textit{Backbone} vertices tend to exhibit the largest $E_i$. 

The Internet graph investigated by Latora and Marchiori \shortcite{latora2001efficient} showed a relatively high $E_{global}$.

Another version is the \textit{Harmonic Mean of Geodesic Distances} \cite{Costa2007}:
\begin{equation}
h=E_{global}^{-1}
\end{equation}
Its codomain is $[1,\infty[$, a higher $h$ indicating a less efficient network. Both $E_{global}$ and $h$ allow for graph comparison, but the former is more useful due to its codomain.


\subsection{Local Network Efficiency {\scriptsize (Cyclic Coefficent)}}\label{sec:dis:loc}
\hrule \vspace{0.03cm}
\centerline{{Tr}\hfill simple (\textit{all}) \hfill local impact \& liability / global \hfill static \hfill $\left[0,1\right]$ \hfill \checkmark} 
\hrule 
\begin{eqnarray}
E(V_i)&=&\frac{1}{k_i(k_i-1)}\sum_{m\neq n\in V_i}{\frac{1}{d_{(m,n)}}}\\
E_{local}&=&\overline{E(V_i)}
\end{eqnarray}
\noindent The \emph{Local Network Efficiency}  \cite{latora2001efficient} 
calculates the average of inverse shortest-path lengths between the neighbors $V_i$ of a node $i$ after it failed. The global version of this metric is found to show relatively high values for the assessed Internet graph. $E(V_{i})=0$ indicates a node in a tree branch, and $E(V_{i})=1$ a node that is connected to all others. It follows that bottleneck vertices, which are the only connectivity providers for their branches to the giant cluster (thus with crucial impact on \textit{Disconnection Robustness}), always exhibit $E(V_i)\approx 0$. At the same time, this indicates that only a few of $i$'s edges can lead to the giant cluster, making itself prone to \textit{Disconnection} when these links fail. For slightly larger but still low $E_i$ values, the detected node acts as a bridge between communities, thus having a high impact on the \textit{Transmission Speed} between them. 

In \cite{Kim2005}, a very similar metric called \textit{Cyclic Coefficient} is introduced, which in its local version calculates the average of the smallest loop sizes of node $i$ with two of its neighbors: 
\begin{align}
	\hat{E}(V_i)&=\frac{1}{k_i(k_i-1)}\sum_{m\neq n\in V_i}{\frac{1}{d_{(m,n)}+2}}\\
	\hat{E}_{local}&=\overline{\hat{E}(V_i)}
\end{align}
In the analysis of the Internet  AS-level topology 
in 1999 by Kim and Kim \shortcite{Kim2005}, 
tree-structure nodes $\hat{E}(V_i)=0$ dominate, while most of the other nodes' neighbors have a small average distance to each other. 

A disadvantage of this metric is the codomain, $[0,1/3]$, and a less intuitive interpretation. Thus, to measure this aspect, the \textit{Local Network Efficiency} is more appropriate. 

\subsection{Characteristic Path Length}\label{sec:dis:cha}
\hrule \vspace{0.03cm}
\centerline{{Tr}\hfill labeled (\textit{all}) \hfill local impact \hfill static \hfill $\left[1,\infty\right[ / \left[1,\frac{v+1}{3}\right]$ \hfill \checkmark} 
\hrule 
\begin{eqnarray}
L^{Q_{i}Q_{i}}(G)&=&\frac{1}{|Q_{i}|(|Q_{i}|-1)}\sum_{m\neq n\in Q_{i}}{d_{(m,n)}}\label{eq:cha1}\\
L^{Q_{i}Q_{j}}(G)&=&\frac{1}{|Q_{i}||Q_{j}|}\sum_{m\in Q_{i}, n\in Q_{j}}{d_{(m,n)}}\label{eq:cha2}
\end{eqnarray}
\textit{Characteristic Path Length (CPL)} \cite{Trpevski2010} is an adaption of the global \textit{ASPL} (Section \ref{sec:dis:ave}) for labeled graphs. Eq. \ref{eq:cha1} is used for calculations between nodes of the same label $Q_i$, eq. \ref{eq:cha2} for different labels $Q_i$ and $Q_j$. 
This allows detailed investigations, e.g., inspecting the \textit{Transmission Speed Robustness} between customer AS to failures of higher hierarchy nodes. 

Trpevski et al. \shortcite{Trpevski2010} show that large ISP exhibit the lowest \textit{CPL} of $1.5$ between themselves, confirming that the \textit{Backbone} is almost a full-mesh. Its value between IXPs and large or small ISPs, respectively, is also small, while the one between customer ASes and between universities is
substantially larger ($L=4$). 
Removal of all small ISPs results in a remaining of all large ISPs in the giant component, while 20 to 30\% of nodes of other labels are disconnected. It is shown though, that attacks based on highest-\textit{Betweenness} (Section \ref{sec:thr:bet}) and even more on highest-\textit{Degree} (Section \ref{sec:con:nod}), irrespective of their label, are substantially more damaging in terms of \textit{Fraction of Nodes in the Largest Component} (Section \ref{sec:min:ana}) than the same strategies, when only nodes of a certain type are removed. These metrics are thus more suitable for assessing a node's impact on \textit{Disconnection Robustness} than its label.


\subsection{Expansion}\label{sec:dis:exp}
\hrule \vspace{0.03cm}
\centerline{{R, Tr}\hfill (un)directed (weighted) \hfill local impact / global \hfill static \hfill $\left[0,1\right]$ \hfill \checkmark} 
\hrule 
\begin{eqnarray}
E(h)_i&=&|\{j\in V|d_{(i,j)}\leq h, i\neq j\}|/v\\
E(h)&=&\overline{E(h)_i}
\end{eqnarray}
\textit{Expansion} calculates the fraction of vertices reachable within $h$ hops. 
In a tree-like and in a router-level Internet graph, local and global \textit{Expansion} follows a power-law $E(h)_{(i)}\propto h^{p_{(i)}}$ \cite{Tangmunarunkit2002, Palmer2001}. 
Palmer et al. \shortcite{Palmer2001} find a strong negative correlation between $p_i$ and \textit{Effective Eccentricity} (Section \ref{sec:dis:eff}), suggesting that highest-$p_i$ nodes are likely to form the \textit{Backbone}. Furthermore, in a router-level graph of 1999, the vulnerability to \textit{Disconnection} and \textit{Transmission Speed} decreases in terms of \textit{Reachability} (Section \ref{sec:min:ana}); $p$ degradation is reported to be highest for largest-\textit{Degree} attacks, followed by highest-$p_i$ and then random failures. This supports the findings of Section \ref{sec:dis:ave}, that \textit{Transmission Speed} providing \textit{Backbone} vertices might not have the largest impact on network \textit{Transmission Speed Robustness} since, when removed, traffic is easily redirected through their neighbors.

The directed version inspects only 
customer-provider compliant AS paths.
A minimum cut set for halving the local node data \textit{Transmission Speed} in terms of $E(h)_i$ is calculated by \textit{Local Delay Resilience} (Section \ref{sec:min:res}).

\subsection{Effective Eccentricity \& Effective Diameter}\label{sec:dis:eff}
\hrule \vspace{0.03cm}
\centerline{{R, Tr}\hfill simple (\textit{all}) \hfill local impact / global \hfill static \hfill $\left[1,v-1\right]$ \hfill \checkmark} 
\hrule 
\vspace{5pt}
\begin{equation}
Ecc_{\textit{eff}}(i,r)=\min_h\left\{h\in \mathbb{N}|N(i,h)\geq rN(i,\infty)\right\}
\end{equation}
\begin{equation}
D_{\textit{eff}}(r)=\min_h\{h\in\mathbb{N}|\textstyle\sum_{v\in V}N(v,h)\geq r\sum_{v\in V}N(v,\infty)\}
\end{equation}
\textit{Effective Eccentricity} and \textit{Diameter} 
measure exactly the same as the previously presented \textit{Expansion}, but from another perspective: They count the minimum hop number $h$ required to reach a certain percentage $r$ of the nodes reachable from node $i$, and on average in the graph, respectively \cite{Palmer2001}. $N(i, h)$ is the number of nodes within the $h$-hop environment of node $i$. 
\textit{Effective Diameter} is a global metric that calculates the minimum number of hops required to enable communication between a fraction $r$ of node pairs that are connected through a path. 

As described in the previous section, $Ecc_{\textit{eff}}(i,r)$ finds \textit{Backbone} nodes with some impact on \textit{Disconnection} and \textit{Transmission Speed Robustness}.
Palmer et al. \shortcite{Palmer2001} analyze an Internet router-level graph from 1999 with an approximating algorithm, obtaining $D_{eff}(0.9)=10$, and a skewed distribution of $Ecc_{eff}(0.9)\in[5,19]$, with most routers having an $Ecc_{eff}(0.9)=8$.

\subsection{Recapitulation}\label{sec:dis:dis}
Though less intuitive to interpret than \textit{ASPL}, due to its applicability to disconnected graphs, \textit{Global Network Efficiency} is most appropriate for measuring and comparing the average \textit{Transmission Speed} of graphs and for finding distance-central \textit{Backbone} vertices. Those are the main providers of \textit{Transmission Speed}, but they do not have the largest impact on this aspect and on \textit{Disconnection} when removed, as they are embedded in a mesh-like neighborhood that can resume the traffic-forwarding tasks. 

In contrast, very low \textit{Local Network Efficiency} values indicate bottlenecks with high local impact on and liability to \textit{Disconnection Robustness}, and low values suggest a local impact on \textit{Transmission Speed}.
Distance metrics that address the difficulty of giving unbiased results even in partially-disconnected networks can be divided into two groups: One group uses adaptations of the \textit{ASPL}, the other the \textit{Global Network Efficiency}. In the first group, the \textit{Diameter Inverse $K$} metric seems most appropriate to take disconnected areas into account. The advantage of the metrics in the second group is their codomain $[0,1]$, which allows the expression of dynamic changes after failures as percentage rates. 
A disadvantage is the less intuitive interpretation of the results as they do not correspond to a certain number of hops.
Therefore, if a quick and comparable abstract evaluation of distances in a network is conducted, metrics of the second group are preferable. For a concrete transmission cost or speed analysis, the \textit{ASPL}-based group provides more suitable measurement tools.
The local version of the \textit{Local Network Efficiency} can detect local bottleneck nodes.
Their \textit{Expansion} or, equivalently, \textit{Effective Eccentricity} further indicates whether they are peripheral or constitute critical bottlenecks closer to or in the \textit{Backbone}. 
\textit{Characteristic Path Length} is useful for assessing the path structure and \textit{Transmission Speed Robustness} of different node types, such as ISPs or customer ASes.  

\section{Throughput}\label{sec:thr}
Evaluating path redundancy and distance increases is not enough for assessing robustness: traffic loads induced by flows along shortest paths and their capacity constraints must also be considered. 

		\hypertarget{abc}{\subsection{Betweenness Centrality \& AS Hegemony 
		}}\label{sec:thr:bet}
\hrule \vspace{0.03cm}
\centerline{{Tr}\hfill (un)weighted (directed) \hfill local impact\hfill static \hfill $\left[0, \frac{v-1}{2}(v-2)\right] / \left[1, (\frac{v}{2})^2\right]$ \hfill \checkmark ($\mathcal{O}(ve+v^2\log{v})$)} 
\hrule 
\begin{equation}
B_u=\sum_{\substack{(i,j)\in V^2 \\ i\neq j\neq u}}\frac{\sigma(i,u,j)}{\sigma(i,j)}
\end{equation}
\noindent \textit{Betweenness Centrality} $B_u$ sums up over all node pairs $(i,j)$, the fractions of shortest paths between $i$ and $j$ that pass through $u$, $\sigma(i,u,j)$, relative to the total number of shortest paths $\sigma(i,j)$ between $i$ and $j$ \cite{freeman1977set}. 
$B_u$ quantifies the number of flows which have to be redirected if $u$ fails, thus how much control $u$ has over network traffic \cite{newman2005measure}. 
The first codomain holds for $u$ as a vertex, the second for $u$ as a link.  Brandes and Fleischer \shortcite{brandes2005centrality} present a corresponding algorithm. 
For the average \textit{Edge Betweenness Centrality} $B^e(G)=\frac{1}{|E|}\sum_{u\in E}{B_u}$, it holds that $B^e(G)=\frac{v(v-1)}{2e}\overline{d}$, with global \textit{ASPL}  $\overline{d}$ \cite{boccaletti2007multiscale}. 
Furthermore, in \cite{DallAsta2006}, a weighted \textit{Betweenness Centrality} version is presented where $\sigma$ counts the number of shortest weighted paths as defined in Section \ref{sec:dis}.

Ghedini and Ribeiro \shortcite{Ghedini2011} find that highest \textit{Betweenness} entities have a huge impact on network \textit{Transmission Speed} and \textit{Disconnection Robustness} in terms of \textit{Global Network Efficiency} $E_{global}$ (Section \ref{sec:dis:glo}) and \textit{Fraction of Nodes in the Largest Component} $|V_L|/v$ (Section \ref{sec:min:ana}). 
These metrics decrease faster for highest-\textit{Betweenness} than highest-\textit{Degree} attacks in WS and KE models, while the difference in $E_{global}$ is moderate and the one in $|V_L|/v$ crucial. The average \textit{Local Network Efficiency} $E_{local}$ (Section \ref{sec:dis:loc}) decreases faster for highest-\textit{Degree} removals, thus leading to a more bottleneck-dependent graph. In a WS network, largest-\textit{Betweenness} attacks on up to 40\% of the nodes even increase $E_{local}$ of the graph, as the removed nodes often constitute community-interconnecting bridges with low \textit{Local Network Efficiency}.

In contrast, the investigation of an AS-level representing network in \cite{holme2002attack}, 
finds that both attack strategies are similarly harmful for the Internet with respect to $E_{global}$ and $|V_L|/v$.
This is explained by their high correlation, which according to Holme et al. \shortcite{holme2002attack} especially holds for high-\textit{Degree} nodes. Also, a weaker correlation between the \textit{Edge Degree}, which can be defined as $k_{(i,j)}$=$k_ik_j$ or $k_{(i,j)}=\min(k_ik_j)$, and the \textit{Edge Betweenness Centrality} is observed, which suggests 
that the existing edges between hubs play a crucial role as \textit{Transmission Speed} providers. Supporting the argumentation in Section \ref{sec:clu:clu}, for a network with fixed number of edges a higher \textit{CC} $C_i$ (Section \ref{sec:clu:clu}) is found to increase the vulnerability to bottleneck-removing \textit{Betweenness} attacks. 

These findings about correlations contrast those of Doyle et al. \shortcite{Doyle2005}, who state that ISP \textit{Backbones} are never represented by hubs as predicted in common SF-graph approaches. According to their study, high-\textit{Degree} nodes are mainly found in local networks at the  periphery of the Internet, and thus have low \textit{Betweenness} values. 
However, they investigate the Internet on the router level but not the AS level, which could be a reason for the substantial differences in the findings.

As \textit{Betweenness Centrality} can be used as an efficient attack strategy, an increase in its variance and maximal value substantially decrease network \textit{Robustness} in all of the three \textit{Aspects}.


Concerning further developments, Mahadevan et al. \shortcite{Mahadevan2005} investigate the AS-level topology and state that \textit{Edge Betweenness} is not a link-centrality measure, as the edges adjacent to peripheral one-degree nodes obtain medium-ranked, thus disproportionally high values. 
Fontunge et al. \shortcite{Fontugne:2017} attribute these findings to the non-random sampling method used to collect BGP data rather than the metric itself. Due to the selection of only a few viewpoints, the \textit{Betweenness} values of ASes close to them are disproportionally high and the ones of those further away too low. To obtain an unbiased measure in this respect, they propose \textit{AS Hegemony},
\begin{eqnarray}
\mathcal{H}(u,\alpha)=\frac{1}{n-2[\alpha n]}\sum_{i=\alpha n+1}^{n-\alpha n} BC_{(i)}(u)\quad\text{with}\quad BC_{(i)}(u)=\frac{1}{S}\sum_{j\in V}\sigma(i,u,j)
\end{eqnarray}
$BC_{(j)}(i)$ is the \textit{Betweenness} value computed with paths from only one viewpoint $j$, arranged in ascending order and $[.]$ is the floor function. $\sigma(i,u,j)$ denotes the number of paths from $i$ to $j$ that pass through $u$, and $S$ is the total number of paths. It uses only unbiased viewpoints for calculation, disregarding the $2 \alpha$ viewpoints with the highest / lowest number of paths passing through $u$, respectively. 
With this approach, they obtain centrality values for the ASes from only a few viewpoints, which are consistent with the ones calculated with all viewpoints. They find that the \textit{Hegemony} distribution does not to change over time and the one in the US seems most balanced.


\textcolor{white}{hidden}

		\subsection{Central Point Dominance}\label{sec:thr:cen}
\hrule \vspace{0.03cm}
\centerline{{Tr}\hfill simple (\textit{all}) \hfill global \hfill static \hfill $\left[0,1\right]$ \hfill \checkmark} 
\hrule \vspace{0.1cm}
The \textit{Central Point Dominance} \textit{(CPD)} \cite{freeman1977set} calculates the average \textit{Betweenness Centrality} difference of the most central node to all the other vertices in the network:
\begin{equation}
\textit{CPD}=\frac{\sum_u (B_{max}-B_u)}{v-1}
\end{equation}
\noindent  CPD quantifies the extent to which a network is decentralized ($\textit{CPD}\rightarrow 0$) or concentrated, indicating low \textit{Disconnection} and \textit{Transmission Speed Robustness} against their removal. 
Along with the correlation of \textit{Betweenness Centrality} to \textit{Node Degree}, it provides an indicator for determining if high-\textit{Betweenness} attacks should be investigated as an alternative to high-\textit{Degree} attacks. 

		\subsection{Effective Load}\label{sec:thr:eff}
\hrule \vspace{0.03cm}
\centerline{{Tr}\hfill simple (\textit{all}) \hfill local impact \& liability \hfill dynamic \hfill $\left[0, A(v/2)^2\right]$ \hfill \checkmark} 
\hrule 
\begin{equation}
\langle\lambda(u)\rangle_\Omega=\frac{1}{|\Omega|}\sum_{\Lambda\in \Omega}\sigma_{\Lambda}(u)\quad\text{with}\quad
\Omega=\left\{\Lambda:|\Lambda|=A v (v-1)\right\}
\end{equation}
\noindent The \textit{Effective Load} $\langle\lambda(u)\rangle_\Omega$ calculates the expected number of shortest paths passing through $u$ when a communicating node set $\Lambda$, containing a fraction of $A$ nodes, is chosen randomly \cite{holme2002vertex}.
Contrary to \textit{Betweenness Centrality}, it aims at modeling real flows and measuring their loads, accounting for the fact that not all node pairs communicate at the same time.

In \cite{holme2002vertex}, an edge is overloaded and removed, if its capacity  $\lambda_{max}(u)$ is exceeded, which leads to the possibility of congestion after an arbitrary \textit{Challenge}. After each removal, the \textit{Effective Load} is recalculated, allowing for breakdown avalanches during one time step. 

By defining $\Omega$ to be the ensemble of a small number of randomly chosen $\Lambda$, a stochastic element can be introduced. For repeated calculations with different $\Omega$, a load distribution for a given edge can be estimated, and hence the probability of a congestion-induced failure. 

For an appropriate modeling of the router-capacities vector $C=[C_u]$ considering their economic costs, Motter and Lai \shortcite{motter2002cascade} propose to measure node \textit{Betweenness Centrality} $B_u$ (Section \ref{sec:thr:bet}) in the unrestricted network and 
set $C_u=(1+\alpha)B_u$. 
Consequently, it depends on the choice of parameter $\alpha$ if avalanches of failures are limited or take down the entire network. In \cite{Lee2005}, 
the critical threshold for a BA graph is $\alpha_c=0.15$ and in an Internet graph the number of node failures during an avalanche process $s$ follow a power law: $p_{\alpha_c=0.15}\sim s^{-\tau}$ with $\tau\approx1.8$.

		\subsection{Performance}\label{sec:thr:per}
\hrule \vspace{0.03cm}
\centerline{{Tr}\hfill node-weighted (\textit{all}) \hfill global \hfill static \hfill $\left[0,\infty\right[$ \hfill \checkmark} 
\hrule
\begin{equation}
\label{eqper}
P(G)=\max_\rho \sum_{ij} x_{ij} ,\quad x_{ij}=\rho y_iy_j\quad s.t. \quad \textbf{R}\cdot\textbf{x}\leq \textbf{b}
\end{equation}
\textit{Performance} \cite{Doyle2005} measures the maximum aggregate network throughput of 'gravity flows' \cite{Zhang2003}. Such a two-way-traffic flow $x_{ij}$ is exchanged between all end vertex pairs $(i,j)$ with bandwidth demands $y_i$ and $y_j$. The global constant $\rho$ is to be maximized, given the restrictions on the right-hand side of eq. \ref{eqper}, and thus describes the worst capacities to demand share of a pair: 
The shortest-path routing matrix $\textbf{R}=[R_{rf}]$ entries are set to $R_{rf}=1$ if flow $f$ passes through router $r$, and $R_{rf}=0$ otherwise, and the vector $\textbf{x}$ of all flows $x_{ij}$ is indexed to match $\textbf{R}$. Their product must be consistent with the router bandwidth capacities vector $\textbf{b}$. 

A high \textit{Performance}-constant $\rho$ value indicates a network that is robust against flow congestions. In particular, a high $\rho>1$, which maximizes the term in eq. \ref{eqper}, indicates that in every node there are unused capacities which can at least carry $\rho-1$ percent more flows. $\rho<1$ indicates that an overloaded node exists, which is requested to carry $1-\rho$ percent more flows than it is capable of. In contrast to \textit{Effective Load} (Section \ref{sec:thr:eff}), breakdowns caused by overloads are not modeled. 

The router-level model used in  \cite{Doyle2005} is the HOT model, and the graph used aims to represent a small segment of the Internet, while taking  all hierarchical layers into account. 
In comparison to SF router-level models of the same \textit{Degree-Frequency Distribution} (Section \ref{sec:con:nod}), HOT performs significantly better with respect to \textit{Traffic Robustness} in terms of \textit{Performance}, with a difference of more than two orders of magnitude. This is explained by the hubs in SF-networks becoming saturated and thus severe bottlenecks. In the HOT topology, a modestly loaded mesh-like \textit{Backbone} provides enough alternative high-capacity paths if core nodes fail. 



\subsection{Elasticity}\label{sec:thr:ela}
\hrule \vspace{0.03cm}
\centerline{{Tr}\hfill simple (\textit{all}) \hfill global \hfill worst-case \hfill $\left[0,1\right]$
	\hfill \checkmark} 
\hrule 
\begin{equation}
E(n)=\int_0^n T_G(\hat{n}) d\hat{n}
 ,\quad
T_G(\hat{n})=\frac{1}{\alpha}\max_{\rho}\sum_{i,j}x_{ij}(\hat{n})\quad s.t.\quad \textbf{R}(\hat{n})\cdot \textbf{x}(\hat{n})\leq\textbf{b}
\end{equation}
\textit{Elasticity} $E(n)$ is a dynamic adaptation of \textit{Performance} (Section \ref{sec:thr:per}) with link- instead of router-bandwidth restrictions \cite{Sydney2008, Sydney2010}. It measures the gradual degradation of graph throughput during the successive removal of nodes up to a fraction $n$, which does not disconnect the graph.

This normalized graph throughput $T_G(\hat{n})$ is calculated by the normalized maximum sum of homogeneous flows $x_{ij}(\hat{n})$ between all node pairs $(i,j)$. They are restricted by the requirement that the product of shortest-path $(e\times v^2)$ routing matrix $\textbf{R}(\hat{n})=[R_{lf}]$ and flow vector $\textbf{x}(\hat{n})=[x(\hat{n})_{ij}]$ is not to exceed link bandwidth capacities vector $\textbf{b}$, which is set to be the unit vector; thus an unweighted graph is observed. The constant $\rho$ varies flow proportions in the network and the normalization parameter $\alpha$ is set so that $T_G(0)=1$. 
A benefit of \textit{Elasticity} is that it takes the \textit{Traffic Robustness} during the entire node-removal process into account by calculating the integral of $T_G(\hat{n})$. 
Its flaws are that it does not model congestion failures, and that unrealistically all flows are reduced equally if one capacity restriction is reached, thus assessing a worst-case scenario.

In \cite{Sydney2008}, \textit{Elasticity} results under attacks on high-\textit{Degree} nodes in a HOT and an SF topology are compared to those of \textit{Performance} of Doyle et al. \shortcite{Doyle2005}. They are found to be consistent, validating the simplifying assumptions of homogeneous flows and capacities. Again, HOT is more resilient than the SF model to \textit{Overloads} in terms of \textit{Elasticity} for a node-removal fraction of 80\%. Also, other Internet topologies 
with mesh-like \textit{Backbone} and negative \textit{Assortative Coefficient} $r$ (Section \ref{sec:con:ass}) are shown to feature a high $E$. Still, 
the relationship between $E$ and $r$ is not found to be clear-cut.

The upper bound for $E(1)$ is $1/3$ in a full-mesh network \cite{Sydney2010}. Also, on average, highest-\textit{Betweenness} vertices are reported to have the most impact on \textit{Traffic Robustness} in terms of removal induced \textit{Elasticity} degradation, followed by highest-\textit{Degree} and then random failures.
\textit{Elasticity} is found to be negatively correlated to \textit{Heterogeneity} \cite{dong2007understanding}, defined as the fraction of standard deviation of node degrees to the average node degree, for highest-degree and highest-betweenness attacks, and also to
the global \textit{ASPL} (Section \ref{sec:dis:ave}) for all attack types. 




\subsection{Vulnerability Impact Factors}\label{sec:thr:vuln}
\hrule \vspace{0.03cm}
\centerline{{Tr}\hfill labeled weighted (directed)\hfill local impact \& liability / global \hfill dynamic \hfill $\left[0,\infty\right] / [0,1]$ \hfill \checkmark} 
\hrule 

Hariri et al. \shortcite{hariri2003impact} present an agent-based framework for monitoring attacks designed to consume all system resources. Examples include TCP SYN, Smurf, ICMP flood, Ping of Death for server attacks  and  Distributed denial-of-service (DDoS) for router attacks. 

The following \textit{Vulnerability Impact Factors} are calculated by all agents $a\in \{1...A\}$ 
in real time for different fault scenarios $FS_k$ and agent-specific measures $M$: 
\begin{eqnarray}
CIF(a,FS_k)&=&\frac{|M_{norm}-M_{fault}|}{|M_{norm}-M_{min}|}\\
SIF_a(FS_k)&=&\frac{\sum_j\Theta(CIF_j-d)}{A}
\end{eqnarray}
The local \textit{Component Impact Factor} $CIF(a,FS_k)$ calculates the share of a fault-induced change $|M_{norm}-M_{fault}|$ in $M$ to the change that makes the agent operate in an abnormal state $|M_{norm}-M_{min}|$. Its global counterpart, \textit{System Impact Factor} $SIF_a(FS_k)$, indicates the share of agents in this abnormal state, thus those whose  $CIF_j$ surpassed the normal operation threshold $d$, with $\Theta$ denoting the Heaviside step-function. Example measures $M$ include data transfer rates for clients, buffer utilization, number of flows open or in process, request-processing rates for routers and connection queue lengths for servers as agents. 

Hariri et al. \shortcite{hariri2003impact} simulate single and multiple-router failures on a topology of clients, servers and routers, with determined bandwidth, using file transfer (TCP/IP) and open shortest path first routing protocol. The former induce drastic increases of $CIF_j$ of other routers into abnormal states, and breakdown avalanches with failure rates of up to $SIF_{Clients}=25\%$. 
In a similar way, agents with highest local impact on and liability to \textit{Traffic Robustness} can be determined.

\subsection{Survivability Function, Failures}\label{survfail}
\hrule \vspace{0.05cm}
\centerline{{A}\hfill weighted directed \hfill global \hfill failures \hfill $[0,1]$ \hfill \checkmark} 
\hrule \hspace{-1mm}

Molisz \shortcite{molisz2004survivability} introduces a measure for \textit{Traffic Robustness} of \textit{Backbone} networks of limited size for random occurrences of node or link failure (ROF), taking into account traffic demands, capacity constraints modeled by link weights, and effects of routing protocols: The \textit{Survivability Function} $S(\zeta)$ is defined as the probability function of the \textit{Share of Delivered Data Flow} $X(\zeta)$ after a (set of) $v_D$ node or $e_D$ link failure(s) $\zeta$ occurred. 
\begin{eqnarray}
S(x)&=&\sum_{\zeta:X(\zeta)=x}P(\zeta)\\
P(\zeta)&=&\prod_{l=1}^{e_D}e^{-1}\prod^{e-e_D}_{k=1}\frac{e-1}{e}
\end{eqnarray}

Investigating a 
broad-band network of 12 nodes and 34 edges, the author finds an \textit{Expected Survivability} $E[X]=\sum_x xS(x)$ of 89.60\% for a total load $l=90\%$ and uniform link failure probability $e^{-1}$, given a routing based on minimum number of hops. In this network, an increase of $l$ from $60\%$ to $90\%$ does not have a large impact on $S(x)$ for node failures. For $l=200\%$, though, congestion causes a dramatic degradation to $E(x)=62\%$. 


	\subsection{Recapitulation}\label{sec:thr:dis}

\textit{Betweenness Centrality}, measures the traffic load of an entity when all node pairs communicate. Its highest values indicate bottlenecks that render the network vulnerable in all three \textit{Robustness Aspects}. The ongoing discussion about the amount and nature of its correlation with \textit{Node Degree} provides crucial insights into the Internet graph structure. 
Furthermore, \textit{Betweenness Centrality} can be used to define capacity restrictions for nodes and edges. In AS-level graphs, \textit{AS Hegemony} is independent of the viewpoints adapted, and thus a more convenient metric. 

\textit{Effective Load} can assess the probability of nodes and edges to become overloaded, and also the whole network behavior during possible breakdown avalanches. The \textit{Performance} constant measures how much flow increase in percent could be carried or by how much it would have to decrease to match capacity restrictions, in the relatively most loaded node of the graph. 
It does not model congestion-induced failures, just like \textit{Elasticity}, which provides a worst-case flow-reduction scenario in the absence of congestion: It assumes that \textit{all} flows are reduced equally if the capacity of one single edge is reached, and measures its gradual degradation. 

The \textit{Vulnerability Impact Factors} are technical metrics, which allow the detailed investigation of the exact cause of congestion-induced avalanche breakdowns after a fault scenario. The \textit{Survivability Function} for failures conveniently assesses the \textit{Traffic Robustness} on the highest level, the \textit{Application Layer} against random failures on the \textit{Physical} or \textit{Routing Layer}, given that the occurrence probabilities are known.
For capturing the important aspect of data flow, \textit{Betweenness Centrality} is the most prevalent metric and has been studied extensively. 

\section{Spectral Methods}\label{sec:spe}
The following metrics cover aspects of all previous chapters. Their calculation relies on random walks, which provide a useful tool to estimate entity importance as their role in alternative, non-shortest backup-paths is accounted for. 
The assumption of shortest-path routing seems in general more suitable for approximating entity loads and distances in the majority of the classical Internet \textit{Backbone} \cite{medhi2017network}. However, the potential of mobile adhoc networks (MANETs) in particular at the Internet edges has increased, and with it the possibility of a more random character of routing \cite{yang2011transportation}. For this network type, the presented loads and distance metrics can be particularly convenient.

Since all sections in this chapter rely on matrix algebra, its basic concepts and notation are introduced subsequently, based on \cite{Seary2002, Gkantsidis2003a}.

The adjacency matrix $\textbf{A}$, its derivatives and corresponding $v$ eigenpairs $(\lambda_i, \textbf{u}_i)$ are key for all calculations. 
The eigenvalues are sorted by size: $\lambda_1\leq\lambda_2\leq...\leq\lambda_v$. 
In undirected graphs, 
the largest eigenpairs of the symmetric adjacency matrix are most important, and the \textit{Spectral Gap} is defined as 
$\Delta \lambda=\lambda_v-\lambda_{v-1}$. For any $k\in\mathbb{N}$, the entries of $\textbf{A}^k$ stand for the number of paths between the corresponding nodes that have length $k$, where revisiting the same nodes and edges in the same path is allowed. This feature is useful to estimate path redundancy, both locally and globally. 
For the eigenvalues of $\textbf{A}^k$, $\lambda_{i}(\textbf{A}^k)=[\lambda_i(\textbf{A})]^k$ holds with the same corresponding eigenvectors. 
For a directed graph, $\textbf{A}$ is not symmetric; 
the entries of $SIM(\textbf{A})=\textbf{A}\textbf{A}^T$ count the number of nodes to which both nodes, $i$ and $j$, point, hence on AS-level their common communication providers which are one step away, and $SIM^{T}(\textbf{A})=\textbf{A}^T\textbf{A}$ the one from which they are both pointed to, thus denoting their common customers.
The eigenvector values of the adjacency matrix are very concentrated on high-degree nodes. Two approaches that try to avoid this concentration are presented:

 The first is the Laplacian matrix $\textbf{L}=\textbf{D}-\textbf{A}$ with $\textbf{D}=diag\{k_1, k_2,..., k_v\}$. Here, the lowest eigenvalues are crucial, and the \textit{Algebraic Connectivity} equals to the \textit{Laplacian Spectral Gap}:  
$\Delta \lambda=\lambda_2-\lambda_1=\lambda_2$, as $\lambda_1=0$ always holds. 
The second approach, the stochastic normalization $N(\textbf{A})$ or $N(SIM(\textbf{A}))$, is calculated by dividing each entry of $\textbf{A}$ or $SIM(\textbf{A})$ by $2\sum_j a_{ij}$ and then replacing the diagonal entries by $1/2$. The resulting eigenvalues are bounded by $\lambda\in[0,1]$, and the largest ones are most important, with the \textit{Spectral Gap} defined as $\Delta \lambda=\lambda_v-\lambda_{v-1}$.  
Modified like this, the eigenvalues of $N(\textbf{A})$ are linear functions of the ones of the random walk matrix 
 $\textbf{R}(\textbf{A})=\textbf{D}^{-1}\textbf{A}=[r_{ij}]$, which calculates the probability of  traversing a node on a random walk, with the same eigenvectors. 

Golnari et al. \shortcite{golnari2018random} use the random walk matrix $\textbf{R}(\textbf{A})$ for simple and directed \textit{connected} graphs; in the latter case, $\textbf{A}$ is not symmetric and $\textbf{D}$ is the diagonal matrix of out-degrees. The expected number of visits at a transit node $m$ of a random walk with  source node $s$ and a target node $t$ is counted by the fundamental $t$-matrix  $\textbf{F}^t=[F_{sm}^t]=(\textbf{I}-\textbf{R}(\textbf{A}_{\backslash tt}))^{-1}$. $\textbf{A}_{\backslash tt}$ denotes the adjacency matrix without line and column $t$ and $\textbf{I}$ is the $(v-1)\times (v-1)$ identity matrix. From this, a fundamental tensor $\textbf{F}=[F_{smt}]$ is defined, whose entries equal $F_{sm}^t$ if $s,m\neq t$, and $0$ otherwise.

The metrics presented in the last three sections of this chapter aim at measuring distance and load aspects, thus implicitly assuming that random walks can be used to describe routing behavior. The majority of routing on the router-level, though, relies on shortest paths \cite{medhi2017network}. On the AS-level, restricting routing policies can be explicitly modeled by directed graphs, and the different costs of forwarding information through different neighbors can be conveniently accounted for by link weights. As cost minimization is the main parameter with which routing decisions are determined, finding a shortest path on this weighted graph is most suitable to model routing behavior.  




\subsection{Eigenvector Centrality {\scriptsize (Significance)}}\label{sec:con:sig}
\hrule \vspace{0.03cm}
\centerline{{To, R}\hfill simple (\textit{all}) \hfill local impact \& liability \hfill static \hfill $\left[0,1\right]$ \hfill \checkmark} 
\hrule 
\begin{eqnarray}
\textbf{A}\cdot\textbf{u}_v&=&\lambda_v\textbf{u}_v\\
u_v(i)&=&\max_t\frac{\sum_{j\in V_i}{u_{v,t-1}(j)}}{\sum_{k\in V} {u_{v,t}(k)}}, \quad\text{with}\quad u_{v,0}(i)=1
\end{eqnarray}
The \textit{Eigenvector Centrality} \cite{bonacich1972factoring} of a node $i$ is given by the $i$th entry of the eigenvector of the adjacency matrix $\textbf{A}$ corresponding to the largest eigenvalue. It can be calculated using the following algorithm provided by \cite{Tauro1995}. Initially, all nodes have equal \textit{Centrality} values. At each step $t$, the value of each node is set to the sum of the values of its neighbors $V_i$. Then all values are normalized so that their sum is one and the process is repeated until it converges. 

The vector entries reflect how many and how important the neighbors of a given node are, 
implicitly accounting for the entire topology. Unsurprisingly,  \textit{Eigenvector Centrality} highly positively correlates with \textit{Degree} (Section \ref{sec:con:nod}) \cite{Gkantsidis2003a} and shortest path stochastic traffic loads \cite{Gkantsidis2003a}, hence possibly also with \textit{Betweenness Centrality} (Section \ref{sec:thr:bet}), and negatively with \textit{Effective Eccentricity} (Section \ref{sec:dis:eff}) \cite{Tauro1995}, thus conceivably marking \textit{Backbone} entities. 

\subsection{Symmetry Ratio}\label{sec:spe:sym}
\hrule \vspace{0.03cm}
\centerline{To, R\hfill simple graph \hfill global \hfill static \hfill $\left[1, v/3\right]$ \hfill \checkmark} 
\hrule\hspace{1pt}
\begin{equation}
SR=\#_{\lambda}/(D+1)
\end{equation}
In this equation for the \textit{Symmetry Ratio} $SR$ \cite{Dekker2005}, $\#_{\lambda}$ is the number of distinct eigenvalues of the adjacency matrix, and $D=\max_{m,n\in V} d_{(m,n)}$ the \textit{Diameter} (Section \ref{sec:dis:ave}). For low-symmetry networks, $SR\in [8,15]$ holds and for high-symmetry networks, $SR\in [1,3]$. In the latter case, all nodes are similarly important, and the network is more \textit{Traffic Robust} against attacks as measured by the \textit{Share of Delivered Data Flow} $X(\zeta)$ (\ref{survfail}), which then have a similar impact as failures. 

An analysis of different graphs \cite{Manzano2011} finds that networks with smaller average \textit{Node Degree} (Section \ref{sec:con:nod}) tend to be more symmetric, but also more vulnerable to 
attacks. It has to be noted, though, that in \cite{Dekker2005} $SR$ is related to robustness after the average \textit{Node Degree} has been controlled for. 


\subsection{Spectral Cluster Identification}\label{sec:spe:spe}
\hrule \vspace{0.03cm}
\centerline{{L, To, R}\hfill simple, directed \hfill both \hfill static \hfill - \hfill \checkmark} 
\hrule \vspace{0.1cm}
The main goal of \textit{Spectral Cluster Identification} is the detection of cluster sets $S$ of nodes that have a large ratio of links between themselves to the total sum of their edges, or of clusters of edges that connect two topologically or geographically remote regions. 
In \cite{Gkantsidis2003a}, these clusters are found by sorting the nodes according to their weights in the eigenvectors of the largest eigenvalues of $\textbf{A}$, and bisecting the node set with special preference to sharp weight jumps. According to \cite{mihail2002eigenvalue}, the largest eigenvalues of this non-normalized adjacency matrix follow a power law as a result of the power-law  degree distribution, with an exponent between 0.45 and 0.5 in inter-domain and 0.18 in router graphs \cite{faloutsos1999power}. 
This method is thus considered to mainly show trivial clusters concentrating on hubs such as ISPs. 

Non-trivial groupings of ASes with clear semantic proximity, such as geography and business interests, are obtained for a directed AS graph where edges point from customers to providers by constructing 
$N(\textbf{A})$ or $N(SIM(\textbf{A}))$ is used.
Generally, high eigenvalues of the normalized matrices indicate the existence of clusters in the underlying topology, which 
are found not to be located in the core. 
Both in the core and the entire topology, North America exhibits less clustering than Europe and Asia. Also,  from 1997 to 2001 Internet graph spectra 
did not change substantially.
 
In order to discover edge clusters of high importance, a similar strategy can be applied: In an undirected graph, implement shortest paths between all node pairs $v^2$
and construct the $v^2 \times |E|$ traffic matrix 
$\textbf{T}$, where each row corresponds to a node pair that has a $1$ in the column if the corresponding link is used by the path. Applying the SVD method to gain eigenvalues and eigenvectors in non-square matrices, the link importance and cluster membership can be derived. In \cite{Gkantsidis2003a}, the resulting crucial link clusters are very intuitive, one example being cross-Atlantic and east-coast to west-coast connections in the US. 
 
Spectral methods are also used in \cite{Gkantsidis2003a} to model data traffic more realistically: Instead of assuming uniform traffic between every node pair, a certain percentage is switched to run between nodes in the same cluster. 
The resulting change in the maximum link load is crucial. For the extreme case of $100\%$ 'intra-cluster' traffic, the maximum load measured by the \textit{Edge Betweenness} (Section \ref{sec:thr:bet}) is reduced by more than $40\%$, which can only partially be explained by the reduction in the global \textit{ASPL} of approximately $15\%$. 
If, in fact, traffic distribution on the Internet turns out to exhibit this behavior, \textit{Throughput} metrics 
should be adapted accordingly as the importance of formerly identified bottleneck entities could turn out to be substantially lower. 

\subsection{Algebraic Connectivity {\scriptsize (Laplacian Spectral Gap)}}\label{sec:spe:alg}
\hrule \vspace{0.05cm}
\centerline{{To, R}\hfill  simple graph \hfill global \hfill static \hfill $[0, \frac{v}{v-1}\min\{k_i|i\in V\}]$ \hfill \checkmark} 
\hrule
\vspace{0.1cm}

The larger the \textit{Algebraic Connectivity}, which in \cite{Fiedler1973} is defined as the second-smallest eigenvalue of the Laplacian matrix $\lambda_2$ (and thus the \textit{Laplacian Spectral Gap}), the more difficult it is to cut a graph into independent components \cite{Wu2008}. 
Its codomain \cite{Jamakovic2007} implies that, for a complete graph, $\lambda_{2}=v$ holds and for a disconnected one $\lambda_{2}=0$. The multiplicity of eigenvalues that are zero corresponds to the number of connected components. The metrics \textit{Cohesion} $\kappa(G)$ and \textit{Adhesion} $\mu(G)$ (Section \ref{sec:min:con}) follow Fiedler's inequality $\lambda_{2}\leq \kappa(G)\leq\mu(G)$.

In \cite{Mohar1991} it is shown that 
the minimum edge cut set $\partial A$ between two disjoint node sets $A$ and $V-A$ 
is bounded by the Laplacian eigenvalues:
$
\lambda_2|A||V-A|/v\leq|\partial A|\leq\lambda_{v}|A||V-A|/v
$.
Merris \shortcite{Merris1994} derives bounds for the NP-complete \textit{Cheeger Constant} (Section \ref{sec:min:che}): $\frac{\lambda_2}{2}\leq h(G)\leq \sqrt{\lambda_2(2k_{max}-\lambda_2)}$ . 

As discussed in \cite{Wu2008}, a disadvantage of \textit{Algebraic Connectivity} is that it becomes zero for disconnected networks. Furthermore, in contrast to \textit{Natural Connectivity} (Section \ref{sec:spe:nat}), 
it does not always distinguish between networks with intuitively different robustness values, for instance, given a graph and the same graph with a removed edge. Furthermore, 
it cannot distinguish between trivial cuts that separate single nodes, and those which separate large components.

\subsection{Good Expansion}\label{sec:spe:goo}
\hrule \vspace{0.03cm}
\centerline{{To, R}\hfill simple, weighted \hfill global \hfill static \hfill $\left[0,1\right]$ \hfill NP-hard} 
\hrule
\vspace{0.1cm}
\begin{equation}
G(V,E)\in GE(\alpha) \;\Leftrightarrow\; \forall S ( S\subset V \land |S|<0.5v )\; \exists\; N_S (|N_S| >\alpha |S|)
\end{equation}
A graph is a \textit{Good Expansion (GE)} network of factor $\alpha$, if and only if any subset of nodes $S$ with less then half of the nodes of the graph in it has a neighborhood $|N_S|$ that is larger than $|S|$ multiplied by $\alpha$. $N_S$ is defined as the set of vertices with one endpoint in $S$ and the other in $\overline{S}=V\backslash S$. 
The global exponent $\alpha$ is known as the \textit{Expansion Parameter}, its computation is NP-hard.  \textit{GE} networks are desirable 
because the minimum cut set of vertices necessary to disconnect a node set exceeds a certain fixed lower bound, which means that there are no crucial bottlenecks \cite{Estrada2007}.

A large \textit{Spectral Gap} $\Delta\lambda=\lambda_v-\lambda_{v-1}$ of $A$ is a necessary, but not sufficient condition for \textit{GE} \cite{Estrada2007}, 
as in some cases it 
fails to distinguish between networks with obvious bottleneck structures and others that are relatively well-connected. 

The author also presents a method to determine whether a network is \textit{GE} -- in that case, 
\begin{equation}
\log[\textbf{u}_v(i)]=\log\left(\sinh[\lambda_v]^{-0.5}\right)+0.5 \log [SC_{odd}(i)]\label{ge}
\end{equation}
holds. $\textbf{u}_v(i)$ denotes the $i$-th component of the $v$-th eigenvector of the adjacency matrix $\textbf{A}$. The local odd \textit{Subgraph Centrality} $SC_{odd}(i)=\sum_{k=0}^{\infty}\frac{\mu_{2k+1}(i)}{(2k+1)!}$ stands for the weighted sum of closed walks $\mu_l(i)=(A^l)_{ii}$ of different but odd lengths $l=2k+1$ starting and ending at node $i$, where re-visiting the same edges and nodes is allowed. 
Thus, for a GE-network, a linear regression of $\log[\textbf{u}_v(i)]$ on $\log[SC_{odd}(i)]$ must approximately exhibit the form of eq. \ref{ge}. 
Regarding the resulting  plot for all nodes, tightly connected clusters will be close to a linear trend while nodes that are connected to the main node cluster through a bottleneck lie substantially below the curve, whose number reveals the importance of the bottleneck. 
Nodes that are more remote from the main cluster exhibit a larger negative bias from the curve. 
Although this method cannot estimate the exact parameter $\alpha$, it is a very helpful tool for both global and local \textit{Disconnection Robustness} analysis. 
A weighted graph can also be analyzed, which is particularly interesting for detecting capacity bottlenecks. 

In \cite{Estrada2007}, two Internet graphs at the AS-level from 1997 and 1998 are 
found to be \textit{GE}.



\subsection{Number of Spanning Trees }
\label{sec:spe:num}
\hrule \vspace{0.03cm}
\centerline{{To, R}\hfill simple graph \hfill global \hfill static \hfill $[1,\prod_{i=1}^{v-1}k_i]$ \hfill \checkmark} 
\hrule
\vspace{1pt}
\begin{equation}
N_{ST}= det(\tilde{\textbf{L}})
\end{equation}
$\tilde{\textbf{L}}$ is the Laplacian Matrix $\textbf{L}$ with one row and column deleted; its dimension is thus $(v-1)\times (v-1)$  \cite{szabo2003geometry}. A \textit{Spanning Tree} is defined as a subgraph without loops where all nodes are connected. Thus, an increased \textit{Disconnection Robustness} to random failures would be expected if the \textit{Number of Spanning Trees} $N_{ST}$ rises because it indicates enhanced \textit{Path Redundancy}. However, Zhang et al. \shortcite{zhang2010enumeration} show that scale-free networks exhibit  significantly lower $N_{SF}$ than regular lattices with the same number of nodes and edges while being more robust against failures. This metric is also not suitable to measure robustness against attacks, as it does not account for bottlenecks. 
Further investigations are therefore necessary before an unambiguous conclusion about the effect of an increased \textit{Number of Spanning Trees} on network robustness can be drawn.

%

\subsection{Natural Connectivity {\scriptsize (Natural Eigenvalue)}}\label{sec:spe:nat}
\hrule \vspace{0.03cm}
\centerline{{To, R}\hfill simple graph \hfill global \hfill static \hfill  $\left[0, \ln ((v-1)e^{-1}+e^{v-1})-\ln v\right]$ \hfill \checkmark} 
\hrule
\hspace{-5mm}
\begin{eqnarray}
\overline{\lambda}&=&\ln\left(\frac{SC(G)}{v}\right)=\ln\left(\frac{\sum_{i=1}^v e^{\lambda_i}}{v}\right)\\
SC(G)&=&\sum_{l=0}^{\infty}\frac{\mu_l}{l!}=\sum_{l=0}^v\sum_{l=0}^\infty\frac{\lambda_i^l}{l!}=\sum_{i=1}^v e^{\lambda_i}
\end{eqnarray}
Here, the global \textit{Subgraph Centrality} $SC(G)$ \cite{estrada2000characterization} stands for the weighted sum of numbers of closed walks $\mu_l$ of length $l$ in the graph, where higher weights are assigned to shorter walks. 
In contrast to the \textit{Algebraic Connectivity} (Section \ref{sec:spe:alg}),  $\overline{\lambda}$  does not become zero for disconnected graphs. Furthermore, it yields more intuitive results when applied to a BA-network: The \textit{Natural Connectivity} $\overline{\lambda}$ \cite{Wu2008} increases strictly monotonically with the addition of edges, capturing subtle differences in path redundancy. This contributes to a higher \textit{Disconnection Robustness}. However, the same values can be obtained for graphs with severe bottlenecks which are else well-connected, and a medium-densely connected network without this flaw. 
According to Wu et al. \shortcite{Wu2011}, for SF networks with sufficiently large \textit{Spectral Gaps}, $\overline{\lambda}_{SF}\approx \lambda_1-\ln v$ holds. 

\subsection{Random Walk ASPL}\label{sec:spe:rwaspl}
\hrule \vspace{0.05cm}
\centerline{Tr\hfill (un)weighted (directed)  \hfill local impact / global \hfill static \hfill $[0,\infty]$\hfill \checkmark} 
\hrule \vspace{1pt}

The random walk (rw) equivalents for the distance $d_{(s,t)}$ between a source $s$ and target node $t$ and the local \textit{Random Walk ASPL} $d_t^{rw}$ are provided in \cite{golnari2018random} as follows:
\begin{eqnarray}
d^{rw}_t&=&\sum_m d^{rw}_{(s,t)}\quad\text{and}\;\quad d^{rw}_{(s,t)}=\sum_{s}F_{smt}
\end{eqnarray}
They count the expected number of hops between a node pair or towards a target node on a random walk. Note that $d^{rw}_{(s,t)}$ is not symmetric: A stub node has a random-walk distance $d_{(s,t)}=1$ to its neighbor with a larger degree, while when starting from that neighbor, wandering off into the network is possible, thus increasing the expected distance $d_{(t,s)}>1$. 

In the presence of random routing, the degradation of this metric under \textit{Challenges} indicates the \textit{Transmission Speed Robustness} of the network.

\subsection{Current-Flow Closeness {\scriptsize (Information Centrality)}}\label{sec:spe:curcl}
\hrule \vspace{0.05cm}
\centerline{Tr\hfill (un)weighted (directed)  \hfill local impact / global \hfill static \hfill  N/A \hfill \checkmark} 
\hrule \vspace{1pt}
Brandes and Fleischer \shortcite{brandes2005centrality} present the \textit{Current-Flow Closeness} 
\begin{eqnarray}
C^c_s 
&=&v\cdot \{(\textbf{L}+\textbf{J})^{-1}\}_{ss}+\text{trace}\{(\textbf{L}+\textbf{J})^{-1}\}-2/v
\end{eqnarray}
whose degradation under \textit{Challenges}, just like the one of \textit{Random Walk ASPL} (Section \ref{sec:spe:rwaspl}), indicates the \textit{Transmission Speed Robustness} between source and target nodes $s$ and $t$.

In electrical networks, it measures the inverse of the average resistance of a source node $s$ to all others. 
In Internet graphs, it can be interpreted as \textit{Information Centrality}, with $\textbf{J}=\textbf{1}\textbf{1}^T$ where $\textbf{1}=[1... 1... 1]^T$. The source-target vector is defined as $\textbf{u}_{st}=[0 ... 1 ... -1 ... 0]$ with $1$ and $-1$ in the $s^{th}$ and $t^{th}$ position.
Only connected graphs are considered. 
An algorithm with a total running time of $\mathcal{O}(ev^{1.5})$ on sparse matrices is presented in \cite{brandes2005centrality}.

In \cite{cetinkaya2013flow}, multilevel Internet networks as described in Section \ref{sec:min:ana} are analyzed under attack strategies targeted at nodes of highest \textit{Current-Flow Closeness} or local \textit{Average Shortest Path Length} (Section \ref{sec:dis:ave}). They find that after a certain number of deletions, \textit{Disconnection Robustness} in terms of \textit{Reachability} (Section \ref{sec:min:ana}) is \textit{higher} than for random failures. This is ascribed to the fact that only non-adaptive attacks are performed, but could also be due to a unsuitability of these metrics for detecting nodes with a large impact on \textit{Disconnection Robustness}. 




\subsection{Random Walk Betweenness }\label{sec:spe:rwbe}
\hrule \vspace{0.05cm}
\centerline{Tr\hfill \textit{-- all --} \hfill local impact \hfill static \hfill $[k_{min}v/a(G), k_{max}v(v-1)/a(G)]$ \hfill \checkmark} 
\hrule

\begin{equation}
B^{rw}_i=\sum_{s,t}F_{smt}=e\pi_i\tau
\end{equation}
The \textit{Random Walk Betweenness} $B^{rw}_i$ counts the times a flow passes through a node on any random walk in a connected network \cite{tizghadam2011robust,golnari2018random}, where at each time step the transition probability through the outgoing links is proportional to their weights. It is proportional to the \textit{Network Criticality} $\tau$ (Section \ref{sec:spe:net}), and, in undirected networks with a stationary probability $\pi_i=\frac{s_i}{2e}$, to \textit{Strength} $s_i$ (Section \ref{sec:con:str}). Thus, in undirected networks it does not provide additional information about the network. In a directed network, however, it shares the implications of its shortest-path equivalent \textit{Betweenness Centrality} (Section \ref{sec:thr:bet}) on robustness for MANETs.

\subsection{Current-Flow Betweenness}\label{sec:spe:curbe}
\hrule \vspace{0.05cm}
\centerline{Tr\hfill (un)weighted (directed) \hfill local impact \hfill static \hfill N/A \hfill \checkmark} 
\hrule
The \textit{Current Flow-Betweenness} $B^c_i$ is introduced by \cite{newman2005measure}:
\begin{eqnarray}
B^c_i&=&\sum_{s,t}\sum_k\frac{1}{2}|F_{smt}r_{mk}-F_{skt}r_{km}| 
\end{eqnarray}
The running time to calculate this metric is $\mathcal{O}(v^3)$ on sparse graphs \cite{newman2005measure}, but a faster approximating algorithm is presented in \cite{brandes2005centrality}. The \textit{Current Flow-Betweenness} has two interpretations: In an electrical network, it calculates the amount of current that flows through node $i$, averaged over all sources $s$ and sinks $t$. Regarding random walks in general networks from any $s$ to any $t$, their averaged net flow through an edge is calculated as the difference between the number of times a flow passes in one direction,  to the one they pass the other way round. 
Only connected graphs are regarded. 

This variant of the metric thus differs from \textit{Random Walk Betweenness Centrality} $B^{rw}_i$ in that moving back and forth through a vertex does not increase its $B^c_i$. Still, its random walk nature has the advantage of assigning  a non-zero value to nodes that lie on a path between an $st$-pair, which is not the shortest one. Thus, the importance of alternative paths is accounted for. 

\textit{Current Flow-Betweenness} is shown to highly correlate with its shortest-path  equivalent (Section \ref{sec:thr:bet}) , and also with \textit{Node Degree} (Section \ref{sec:con:nod}) \cite{newman2005measure}. Due to these correlations, it is unsurprising that {\c{C}}etinkaya et al. \shortcite{cetinkaya2013flow} find that adaptive and non-adaptive attacks targeted at high $B^c_i$ nodes have a greater impact on the \textit{Disconnection Robustness} in terms of \textit{Reachability} (Section \ref{sec:min:ana}) than random failures.

Generally, like the previously presented metric, it shares the implications of its shortest-path equivalent \textit{Betweenness Centrality} (Section \ref{sec:thr:bet}) on robustness for MANETs.


\subsection{Network Criticality {\scriptsize (Effective Resistance)}}\label{sec:spe:net}
\hrule \vspace{0.05cm}
\centerline{Tr \hfill \textit{-- all --} \hfill global \hfill static \hfill $[v/a(G),v(v-1)/a(G)]$ \hfill \checkmark} 
\hrule
\begin{eqnarray}
\tau&=&2K(G)=\frac{2}{e}\sum_{s,m,t}F_{smt}=\sum_{s,t}\tau_{st}\quad\text{being}\quad\tau_{st}=\textbf{u}_{st}^t\textbf{L}^+\textbf{u}_{st}\\
n(k)&=& B^{rw}_k/s_k=\tau/2\quad \text{and}\quad n_{ij} =B^{rw}_{ij}/w_{ij}=\tau
\end{eqnarray}
$\textbf{L}^+$ is the Moore-Penrose inverse of the Laplacian $\textbf{L}$ of a connected networks, and $\textbf{u}_{st}=[0 ... 1 ... -1 ... 0]$, with $1$ and $-1$ in the $s^{th}$ and $t^{th}$ position, and $K(G)$ the \textit{Kirchhoff Index} \cite{golnari2018random}. The random walk begins at source node $s$ and ends at destination node $t$. At each step, at node $i$ it chooses to travel to the adjacent node $j$ with probability $p=\frac{w_{ij}}{s_i}$. If the weights are interpreted as capacities, the \textit{Network Criticality} $\tau$ globally quantifies the relative entity load, and thus the risk of using them: 
\textit{Node} and \textit{Edge Criticality}, $n(k)$ and $n_{ij}$, are defined as the share of the entity's \textit{Random Walk Betweenness Centrality} $B^{rw}$ to the \textit{Strength} or \textit{Weight}, respectively, and are constant for all nodes and edges and proportional to $\tau$. 


In \cite{tizghadam2011robust}, its suitability to capture \textit{Disconnection Robustness} against topological changes as node failures and \textit{Traffic Robustness} against variation in link capacities, changes in active sources and sinks, and traffic demand shifts is expounded: As its bounds are determined by \textit{Algebraic Connectivity} $a(G)$ and vice versa, a lower $\tau$ leads to a lower upper bound for $a(G)$, which makes it a (weak) indicator for \textit{Disconnection Robustness}. Furthermore, the authors show that the maximum of average input rate $\lambda$ and thus critical load, which a network with link capacities  $w_{ij}$ can transmit, is achieved by minimizing $\tau$ by altering the capacity matrix $W=[w_{ij}]$,  $\max_W\lambda =2v(v-1)/\min_W\tau$. Thus, for a given topology, the effect of the capacity distribution on \textit{Traffic Robustness} can be assessed by $\tau$. Additionally, it is proportional to average path cost $\overline{\varphi}=1/2\tau\sum_{ij}w_{ij}z_{ij}$ to traverse a network with link weights $w_{ij}$ and traverse costs $z_{ij}$ on random walks. A possible interpretation of the traversal costs is the message delay; thus, the degree of increase of this metric under \textit{Challenges} in special networks governed by random-walk routing indicates their \textit{Transmission Speed Robustness}.

With the \textit{Traffic-Aware Network Criticality (TANC)},  another version of this metric is provided in \cite{tizghadam2011robust}, which also relies on random walks and shares its above-mentioned implications about all \textit{Robustness Aspects}: 
	\begin{equation}
	\tilde{\tau}=\sum_{s,t}\alpha_{st}\tau_{st}\quad\text{where}\quad \alpha_{st}=1+\frac{\gamma_{st}+\gamma_{ts}}{2\gamma}+\frac{\gamma_{*s}-\gamma_{s*}}{v\gamma}
	\end{equation}
	where $\gamma$ and is the total traffic and $\gamma_{st}$ the traffic between source $s$ and destination $t$. 

	The results of Tizghadam and Leon-Garcia \shortcite{tizghadam2011robust} are not applicable to directed graphs.

	\subsection{Recapitulation}\label{sec:spe:dis}
Spectral metrics aim at globally assessing a graph's topology with random walks. 
Although relatively resource-consuming, they promise to be highly accurate and to provide solutions for the difficult detection of communities and minimum-cut sets. 

The \textit{Eigenvector Centrality} is more suitable for capturing the importance of a single node than the local \textit{Node Degree} (Section \ref{sec:con:nod})
as it can discriminate between hubs that are mostly connected to one-degree-nodes and hubs that are neighbors to other hubs close to the network \textit{Backbone}. The \textit{Symmetry Ratio} assesses how much the \textit{Traffic Robustness} against failures and attacks diverges. 
The methods \textit{Spectral Cluster Identification} and \textit{Good Expansion} detect community structures, 
similar to the algorithms in \textit{Modularity} (Section \ref{sec:clu:mod}), 
and estimate the severity of community-interconnecting cut sets, thus having similar goals as the \textit{Minimum Cut} algorithms. 
However, they are more practical 
as no parameter prefixing is necessary. 
For simple graphs, the \textit{GE} procedure should be used as it provides a very detailed insight into the network topology; 
for directed networks, the \textit{Spectral Cluster Identification} is more suitable. 
\textit{Natural Connectivity} and \textit{Number of Spanning Trees} can capture subtle differences in path redundancy, but not necessarily graph \textit{Disconnection Robustness} if applied by themselves. 

\textit{Random Walk ASPL} and \textit{Current-Flow Closeness} assume random-walk routing and thus mainly serve for analyzing mobile adhoc networks. In contrast, the \textit{Current-Flow Betweenness} can also be used in networks relying on shortest-path routing to assess the importance of entities as their role in crucial backup-paths is accounted for. As for \textit{Network Criticality}, its traffic-aware version \textit{TANC} is most convenient for assessing \textit{Transmission Speed} and \textit{Traffic Robustness} 
in mobile networks.


\section{Geographical Metrics}
This section discusses metrics that take geographical aspects into account. 

\subsection{Distance Strength \& Outreach}\label{sec:oth:out}
\hrule \vspace{0.03cm}
\centerline{{P}\hfill geographical weighted (directed) \hfill local impact \hfill static \hfill $\left[0,\infty\right]$ \hfill \checkmark} 
\hrule 
\begin{equation}
D_i=\sum_{j\in V_i} g_{ij}\qquad O_i=\sum_{j\in V_i} w_{ij}g_{ij}
\end{equation}
\textit{Distance Strength} $D_i$ and \textit{Outreach} $O_i$ \cite{DallAsta2006} are variations of \textit{Node Degree} and \textit{Strength} (Sections \ref{sec:con:nod}, \ref{sec:con:str}) that account for the geographical distance $g_{ij}$ between node $i$ and its neighbors $V_i$. High values indicate crucial bridges 
such as submarine cables or links connecting large ISPs.

\subsection{Survivability Function, Geographical}\label{survgeo}
\hrule \vspace{0.05cm}
\centerline{{P}\hfill geographical \hfill global \hfill dynamic \hfill $[0,1]$ \hfill \checkmark} 
\hrule \hspace{-1mm}
In \cite{liew1992framework, jabbar2010framework}, a measure for \textit{Disconnection Robustness} against \textit{geographically extended failures} is presented with the \textit{Survivability Function}:
\begin{equation}
P[|V_L|/v=x]=\sum_{e:V_L(e)/v=x}P(e)
\end{equation}
It measures the probability that the \textit{Fraction of Nodes in the Largest Component} $|V_L|/v$ (Section \ref{sec:min:ana}) is $x$, given the probabilities $P(e)$ of geographically extended failure events $e$ that result in this outcome. Due to the limited size of possible sets $e$, this metric can also be calculated in large graphs. From this, the \textit{Expected Survivability} $E[|V_L|/v]=\sum_x xP[|V_L|/v=x]$ and \textit{Worst-Case Survivability} $s^0=\min_{P[|V_L|/v=x]>0}s$ can be derived.

		\subsection{Pointwise Vulnerability {\scriptsize (Information Centrality)}}\label{sec:dis:poi}
		\hrule \vspace{0.03cm}
		\centerline{{P}\hfill geographical (\textit{all}) \hfill local impact \hfill dynamic \hfill $\left[0,1\right]$ \hfill \checkmark}
		\hrule 
		\begin{eqnarray}
		U_i&=&\frac{\Delta E_{eucl}}{E_{eucl}}=\frac{E_{eucl}-E_{eucl}(i)}{E_{eucl}}\\
		E_{eucl}&=&\frac{1}{v(v-1)}\sum_{
			m\neq n\in V} \left[\frac{g_{ij}}{d_{(m,n)}}\right]
		\end{eqnarray}
		\noindent \textit{Pointwise Vulnerability} is a dynamic metric that calculates the relative drop in the \textit{Euclidean Network Efficiency} $E_{eucl}$ caused by removing node $i$ 
		\cite{Crucitti2006, gol2004vulnerability}. It assigns higher weights to connections between node pairs of high geographical distance $g_{ij}$, 
		 which often constitute scarce bottleneck edges such as submarine cables. 
		$U_i=1$ indicates a very vulnerable highest-position vertex, $U_i=0$ a low-position vertex of little importance. 
		
		The global metrics \textit{Global Vulnerability} $U^{g}=\max_{i}U_i$ and the \textit{Relative Variance of Pointwise Vulnerability} $h=\overline{\left[\Delta U_i\right]^{2}}/\overline{U_i}^{2}$, being $\Delta U_i=U_i- \overline{U_i}$, make different graphs comparable \cite{gol2004vulnerability}.
		Its codomain is $[0,\infty[$, a higher $h$ indicating a more heterogeneous network in terms of the nodes' importance for the overall \textit{Euclidean Network Efficiency}, thus a higher vulnerability to attacks that select targets based on this feature.

		\subsection{Effective Geographical Path Diversity}\label{sec:oth:eff}
\hrule \vspace{0.03cm}
\centerline{{P}\hfill geographical (directed) \hfill local liability / global \hfill static \hfill $\left[0,1\right]$ \hfill \checkmark} 
\hrule \vspace{0.1cm}
\textit{Effective Geographical Path Diversity (EGPD)}  \cite{Cheng2013} 
estimates the liability of the communication of node $s$ with $d$ to geographically extended failures,
e.g.,
caused by natural disasters. 
\begin{eqnarray}
EGPD_{sd}&=&1-e^{-\lambda k_{sd}}\\
k_{sd}&=&\sum^k_{i=1}D_{min}(P_i)\\
D_{min}(P_i)&=&\min(D_g(P_i,P_0))\\
D_g(P_b,P_a)&=&\omega D^2_{min}+(1-\omega)A
\end{eqnarray}

$P$ is a path vector that contains all links and intermediate nodes from source $s$ to destination $d$, $P_a$ and $P_b$ paths 
with the same $s$ and $d$, and 
$D_{min}$ the minimum geographical distance between any member node of $P_a$ and any node of $P_b$. The area of the polygon, whose borders are formed by the paths $P_a$ and $P_b$, is denoted by $A$, and $\omega\in [0,1]$ is a weighting factor, which is set to $0.5$. The shortest path between nodes $s$ and $d$ is denoted by $P_0$, and $\lambda$ is an experimentally determined constant, which scales the impact of $k_{sd}$. For a node pair that is not geographically path-diversified, $EGPD$ converges to zero and for a highly diversified node pair, to one. 

The \textit{Total Geographical Graph Diversity} $TGGD$ is calculated as the average of $EGPD$ over all node pairs. $cTGGD$ is the corresponding normalized global metric, called the \textit{Compensated} $TGGD$:
\begin{equation}
cTGGD=e^{TGGD-1}\times |E|^{-\rho}\quad 
\end{equation}
The normalization by $\rho\approx 0.05$ 
eliminates the penalty to a dense network for a given size of the physical region. 
The codomain is again $[0,1]$, with a high value indicating a well-diversified network. 
In \cite{Cheng2013}, first the sets of geographically close nodes are identified that are not well path-diversified. Each of them is then successively taken down in a simulated natural catastrophe, and the flow robustness of the remaining network is calculated with an alternative routing algorithm. 


\subsection{Recapitulation}
\textit{Outreach} and \textit{Pointwise Vulnerability} use different approaches to detect crucial bottleneck nodes that connect geographically remote regions, for instance landing stations of submarine cables. The latter metric is more sophisticated and more likely to yield better results, especially when a robustness assessment instead of a mere topological analysis is conducted, but also has a substantially higher computational cost. \textit{Effective Geographical Path Diversity} measures the vulnerability to natural catastrophes in an elaborative way. For robustness analyses of networks in areas prone to such disasters, this metric is very useful. If the extent and probability of possible future incidents can be estimated, the geographical \textit{Survivability Function} yields the most reliable and exact results for a \textit{Disconnection Robustness} assessment of geographically extended failures.

\section{Comparison and Discussion}\label{sec:com}

Considering the diversity of goals and methods of the presented metrics, it becomes clear that an isolated application of a single metric or metrics of the same family would not be sufficient for properly studying \textit{Internet Robustness}. 
In the following, we outline a conceptual tool set that accounts for \textit{Backbone} identification and all previously defined \textit{Robustness Aspects}. 
Besides providing methodological insights, such a tool set can also be used to improve practical software for graph-based analysis of connectivity risks, such as the CORIA framework \cite{Fabian:2017}.


\subsection{Backbone Identification and Evaluation}\label{sec:com:bac}

In order to 
identify the \textit{Backbone}, 
the metrics \textit{Node Degree}, \textit{Strength}, weighted \textit{In-Degree}, \textit{Eigenvector Centrality}, \textit{Outreach}, \textit{Betweenness Centrality} and \textit{AS Hegemony} (Sectiosn \ref{sec:con:nod}, \ref{sec:con:str}, \ref{sec:con:sig}, \ref{sec:oth:out} and \ref{sec:thr:bet}) and their respective value distributions can be applied. 
An alternative metric to \textit{Node Degree}, which yet highly correlates with it, is \textit{Eigenvector Centrality}. It does not only take the direct neighbors of a node into account, but also the number of all nodes that can be reached with a random walk and hence can be considered as a better measure for node centrality.
Depending on the selected network-generating model, the \textit{Node Degree} may highly correlate \cite{holme2002attack} or not correlate at all \cite{Doyle2005} with 
\textit{Betweenness Centrality} and the similar \textit{AS Hegemony}, which is better suited for AS-level graphs. 
If the \textit{Backbone} is thought of as a well-interconnected set of high-capacity nodes, the findings of Colizza et al. \shortcite{Colizza2006}
, indicating that hubs are usually not very well-interconnected (but high-\textit{Strength} nodes are), reinforce the claims of Doyle et al. \shortcite{Doyle2005} and cast doubt on the idea that \textit{Node Degree} is a suitable tool for detecting \textit{Backbone} nodes. 
Contrary to this, \textit{Strength}, 
 weighted \textit{In-Degree} and \textit{Outreach} are appropriate for \textit{Backbone} detection, regardless whether the weights are defined as loads or capacities.  
Low \textit{ASPL} and high \textit{Effective Eccentricity} or \textit{Effective Diameter} (Sections \ref{sec:dis:ave}, \ref{sec:dis:eff}) can also serve as indicators for \textit{Backbone} nodes. 
They should not be applied in isolation since a one-degree node adjacent to a central node would be wrongly perceived as a \textit{Backbone} node.

After conducting community detection (Section \ref{sec:clu}), 
local \textit{Backbones} can be found using the \textit{Z-Score of the Within Module-Degree} and  \textit{Participation Coefficient} (Sections \ref{sec:clu:zsc}, \ref{sec:clu:par}). 
Note that both metrics are subgraph equivalents of the \textit{Node Degree}, but they do not share its drawbacks as they are applied to communities where, per definition, no internal bottlenecks should exist. 

To evaluate the \textit{Backbone} robustness, weighted \textit{Rich-Club Connectivity} (Section \ref{sec:con:ric}) can be applied to measure how easily data can be redirected through other high-capacity nodes. 
Average distances between \textit{Backbone} node types such as large ISPs or IXPs are measured by \textit{CPL} (Section \ref{sec:dis:cha}). 
The \textit{Reliability Polynomial} and \textit{Partition Resilience Factor} (Sections \ref{sec:min:rel}, \ref{sec:min:resfac}) are computationally too costly for large Internet graphs, but in the smaller \textit{Backbone}, they could conveniently assess the damage of random failures. The same holds for the \textit{Survivability Function} (\ref{survfail}), which is an important indicator for \textit{Traffic Robustness} on the highest level, the \textit{Application Layer}, against random failures on the \textit{Physical} or \textit{Routing Layer}, given that the occurrence probabilities are known.

Both, the entire network and the \textit{Backbone} can be analyzed with respect to the different aspects of robustness as expounded in the following three sections.

\subsection{Disconnection Robustness}\label{sec:com:dis}
The scale-free nature of the Internet, which entails a heterogeneous \textit{Degree Distribution} (Section \ref{sec:con:nod}), as measured by \textit{Entropy}, \textit{Skewness} or \textit{Vulnerability Function} (Sections \ref{sec:con:ent},\ref{sec:con:ske} and \ref{sec:con:vul}), makes it vulnerable to \textit{Disconnection} by high-\textit{Degree} attacks, 
but less so by failures. Its \textit{disassortativity} (Section \ref{sec:con:ass}), though, contributes to a larger vulnerability to both in terms of a fast decline in giant component size and reduced path redundancy.
 
Path redundancy is a crucial prerequisite for \textit{Disconnection Robustness}.
Spectral methods for quantifying this aspect by a single number are \textit{Number of Spanning Trees} (Section \ref{sec:spe:nat}) and \textit{Natural Connectivity} (Section \ref{sec:spe:nat}), which is computationally more costly but for which an approximation method exists. 

However, path redundancy can also be displayed in networks that feature serious bottlenecks. A high \textit{CC} $\tilde{C}$ or $\tilde{C_\Delta}$ (Section \ref{sec:clu:clu}) can indicate that a network is prone to such weak points because of its pronounced clustering structure. In this case, most available edges of a vertex are "used up" for connections inside of its cluster, so that no links remain to interconnect communities. As opposed to this metric, whose high values indicate a structural or \textit{global} vulnerability to disconnection, large values of their counterparts $C$ or $C_\Delta$ indicate that node removals \textit{locally} have no large impact due to the high degree of interconnections between neighbors. $C_\Delta$ is more critical for \textit{Disconnection Robustness} of peripheral structures than $C$ as it focuses more on the clustering of hubs, which, according to \cite{Doyle2005}, are more involved in ring-, star- or tree structures at the network periphery. They are thus more prone to partitions than the mesh-like \textit{Backbone}, which is dominated by low-degree nodes.

Due to its small-world property, the clustering in Internet graphs is relatively high. The first step is thus to find local topological bottlenecks and to determine whether they are central and important or peripheral. The calculation of the local \textit{CC} (Section \ref{sec:clu:clu}) is fast since it only relies on a local information set. In simple graphs, $C_i$ is most appropriate, and for a weighted network $\tilde{C}_i^w$, as the quality of the alternative path is taken into account with the weight of the neighbor-interconnecting edge. 
For a (weighted) directed network, $C_{\tau}$ with the triplet definition by Wassermann and Faust \shortcite{wasserman1994social} is most suitable.
Better results can be obtained with the 
\textit{Local Network Efficiency} (Section \ref{sec:dis:loc}) since in addition to direct connections, it measures shortest paths of all lengths between the neighbors of the concerned node and weighs them accordingly. 
For links, the 
\textit{Edge Clustering Coefficient} (Section \ref{sec:clu:edg}) 
can be used  for different loop sizes. 
If information about the geographical position of the nodes is available, \textit{Outreach} and \textit{Pointwise Vulnerability} (Sections \ref{sec:oth:out}, \ref{sec:dis:poi}) can be applied to detect important bridge nodes such as landing stations of submarine cables.
 
In order to estimate the importance of the detected bottlenecks, this can be combined with the local version of the \textit{Expansion} (Section \ref{sec:dis:exp}) that measures the fraction of nodes that can be reached within a certain number of hops. 
This approach, however, entails the risk of detecting only locally important or peripheral $k$-degree nodes that connect $k$-1 stub nodes to the rest of the network and that are close to well-interconnected clusters. 
Another possible combination 
is with \textit{Betweenness Centrality} (Section \ref{sec:thr:bet}), 
on router-level or \textit{AS Hegemony} on AS-level graphs. Its random-walk equivalent \textit{Current-Flow Betweenness} (Section \ref{sec:spe:curbe}) might be even more suitable, as it assigns nodes in crucial, but non-shortest backup paths a non-zero value. 
Since it relies on a global information set, the substitution of  \textit{Edge Betweenness} by the \textit{Edge Clustering Coefficient} as applied by Radicci et al. \shortcite{radicchi2004defining} could be suitable, if their strong negative correlation also holds for disassortative networks.
The nodes of highest \textit{Strength} (Section \ref{sec:con:str}) 
coincide with the most central ones, and the detected bottlenecks can be expected to be non-trivial. 
The simple \textit{CC}, $C_i$, (Section \ref{sec:clu:clu}) is a further important detector for local bottlenecks. 

Despite their drawbacks, combinations of these metrics are most convenient for local bottleneck detection as they are applicable to all graph types and can be normalized to the codomain $[0,1]$. The combination of \textit{Local Network Efficiency} and \textit{Betweenness Centrality}, respective \textit{AS Hegemony} seems most promising.

The next step could be finding topologically-close minimum link sets, the removal of which locally disconnects subgraphs, with the local 
\textit{Local Delay Resilience} (Section \ref{sec:min:res}). By using it on small to medium-size overlapping subgraphs in the whole graph or around locally detected bottlenecks, a cut set that globally disconnects the graph could emerge in less calculation time than with global disconnection algorithms. 

Other substantially more time-consuming approaches are global network partitioning or cluster-identifying algorithms. \textit{Sparsity} and the \textit{Network Partitioning Algorithm} (Sections \ref{sec:min:spa}, \ref{sec:min:net}) do not rely on spectral methods and use bisections that require a predetermination of the resulting network sizes and are only applicable to simple graphs. 
Therefore, 
spectral methods, such as \textit{Modularity Matrix}, \textit{Spectral Cluster Identification} and \textit{Good Expansion} procedure (Sections \ref{sec:clu:mod}, \ref{sec:spe:spe}, \ref{sec:spe:goo}) are more suitable. The advantage of the \textit{Modularity Matrix} is that 
it maximizes the value of \textit{Modularity}, a well-defined measuring tool for the quality of a graph division. A major disadvantage is that it is only applicable to simple graphs, as opposed to the \textit{Spectral Cluster Identification}, which is also applicable to directed graphs, and the \textit{Good Expansion} approach, which can be used in weighted networks. 
The latter is particularly appealing since it is the only metric that does not necessarily create a bisection and  
indicates which nodes are in the giant cluster and which are affected by bottlenecks. 

Further analysis of communities by the \textit{Participation Coefficient} (Section \ref{sec:clu:par}) finds the most important nodes that interconnect communities. This strongly depends on the quality of the division into communities, but offers an easy-to-calculate approach for bottleneck detection. For the analysis of a disconnected network, the metrics in Section \ref{sec:min:ana} can be applied, especially \textit{Reachability} summarizes the network state accordingly. 
\textit{Effective Geographical Path Diversity} (Section \ref{sec:oth:eff}) detects geographical bottlenecks of non-diversified areas that are particularly vulnerable to natural catastrophes.

\subsection{Transmission Speed Robustness}\label{sec:com:tra}
A very convenient definition of the network-transmission speed for a simple graph is 
based on the inverse of the number of hops that a flow has to take on its path from one node to another. In a weighted network, the sum of the inverse weights of the edges that constitute this path is used (Section \ref{sec:dis}). 
These definitions match exactly the \textit{Global Network Efficiency} (Section \ref{sec:dis:glo}), which, 
in contrast to the global \textit{ASPL} (Section \ref{sec:dis:ave}), can also be applied in disconnected networks. 
It seems likely to correlate with \textit{Effective Eccentricity} (Section \ref{sec:dis:eff}) and \textit{Expansion} (Section \ref{sec:dis:exp}), which calculates the fraction of nodes a vertex can reach in a certain amount of time, and its \textit{Expansion Exponent} $p_i$. Although more damaging than failures, attacks on nodes of high \textit{Expansion Exponent}, counter-intuitively, may not yield the highest decreases of global \textit{Expansion Exponent}. This indicates that vertices of high \textit{Transmission Speed} are not the most vulnerable points in terms of \textit{Transmission Speed Robustness}. The reason for this can be their position in the \textit{Backbone}, as suggested by Palmer et al. \shortcite{Palmer2001}, and its mesh-like structure \cite{Doyle2005} that can redirect traffic easily. This argument is supported by the very short \textit{CPL} for large ISPs (Section \ref{sec:dis:cha}). 

In MANETs, the degradation of \textit{Random Walk ASPL}, \textit{Current-Flow Closeness} and, most suitably, the \textit{Average Path Cost} derived with the \textit{Traffic-Aware Network Criticality} (Sections \ref{sec:spe:rwaspl}, \ref{sec:spe:curcl}  and \ref{sec:spe:net}) under \textit{Challenges} measure their \textit{Transmission Speed Robustness}. They are only applicable in connected networks.

Vulnerable points with respect to \textit{Transmission Speed Robustness} can be discovered with \textit{Node Degree} (Section \ref{sec:con:nod}) and \textit{Betweenness Centrality} on router- or \textit{AS Hegemony} on AS-level graphs (Section \ref{sec:thr:bet}). A more uniform distribution of \textit{Degree} \cite{Ghedini2011}, as measured by \textit{Entropy}, \textit{Skewness} or \textit{Vulnerability Function} (Sections \ref{sec:con:ent},\ref{sec:con:ske} and \ref{sec:con:vul}), leads to an increase in \textit{Transmission Robustness} against the respective attacks, but also to a higher vulnerability to failures. 
The removal of bottlenecks does not only involve a threat of \textit{Disconnection}, but also of severe \textit{Transmission Speed} decrease. 
Inside communities, nodes with a high \textit{Z-Score of the Within Module-Degree} (Section \ref{sec:clu:zsc}), and between them, those of large \textit{Participation Coefficient} (Section \ref{sec:clu:par}) are relevant providers of short-cuts. The most convenient metrics for estimating the hop increase of local traffic redirections are \textit{Local Network Efficiency} (Section \ref{sec:dis:loc}) in case of a node removal, and the generalized \textit{Edge Clustering Coefficient} (Section \ref{sec:clu:edg}) in case of a link removal. 

If the focus is on \textit{Transmission Speed Robustness} between geographically remote nodes, which are often vulnerable in this sense, \textit{Pointwise Vulnerability} (Section \ref{sec:dis:poi}) can be applied. Its relative variance indicates whether particularly weak spots do exist. 
\textit{Outreach} (Section \ref{sec:oth:out}) identifies crucial bridge nodes for \textit{Transmission Speed} by taking into account their weights (thus bandwidths or capacities) and the geographical distance of their neighbors. 

The local version of \textit{Local Delay Resilience} (Section \ref{sec:min:res}) calculates the 
minimum edge cut set in the vicinity of a node for halving the number vertices reachable 
within a certain amount of time, and thus its \textit{Transmission Speed} liability to other network failures.

Special attention must be given to the routing policies and the general distribution of flows \cite{Doyle2005, Gkantsidis2003a}. Bandwidth demands of single nodes can vary highly, and thus also the number of data transmissions between node pairs. Geographically close vertices in the same cluster may communicate more intensively with one another than with others. This needs to be considered by modeling non-uniform communication flows between nodes since it influences the measured \textit{Global Network Efficiency}.
Furthermore, the routing behavior is not only determined by shortest paths, but by routing policy constraints, which can be accounted for by applying a directed graph.

\subsection{Traffic Robustness}\label{sec:com:dat}
In this section, finite capacity and its implication of a maximum data flow are considered. As routing policies and flow distribution crucially influence \textit{Traffic Robustness} \cite{Doyle2005}, they must be modeled properly.

The worst ratio of capacity to traffic-load demand is calculated by the \textit{Performance} constant (Section \ref{sec:thr:per}). Its advantage is that it allows for modeling different bandwidth demands. However, it does not reflect the fact that in practice vertices usually do not communicate equally with all all other vertices, but often with geographically close nodes.  
Similar benefits and drawbacks hold for  \textit{Elasticity} (Section \ref{sec:thr:ela}), which dynamically  
measures the uniform reduction of data flow in the entire graph when only one node becomes overloaded. Here, it is unrealistically 
assumed that every node has access to global information about the load state of the other nodes. In fact, it only serves as a worst-case scenario assessment in the absence of congestions, but captures the gradual performance degradation in one number. 

Possible cascading overloads can be assessed by explicitly modeling capacity constraints and calculating loads with \textit{Betweenness Centrality}, respectively \textit{AS Hegemony}, or its counterpart metric \textit{Effective Load} 
(Sections \ref{sec:thr:bet}, \ref{sec:thr:eff}). 
While the assumption of a uniform flow distribution with \textit{Betweenness Centrality} and \textit{AS Hegemony} seems unrealistic, \textit{Effective Load} at least accounts for the fact that not every pair of nodes communicates at the same time. 
If only a small subset of possible node-pair communications are repeatedly averaged, a load distribution can
be calculated, which together with the edge capacity determines its risk of
failure. After attacking the nodes with highest loads,
the risk increase of congestion failure in the entities passed by the redirected
flows can be calculated.
The more technical \textit{Vulnerability Impact Factors} (Section \ref{sec:thr:vuln}) pursue the same goal and allow the detailed investigation of the exact cause of congestion-induced avalanche breakdowns after a fault scenario.

For networks of fixed average \textit{Node Degree} (Section \ref{sec:con:nod}), a low value of the \textit{Symmetry Ratio} (Section \ref{sec:spe:sym}) indicates a network whose \textit{Share of Delivered Data Flow} is more robust against attacks.

In the presence of random routing, the \textit{Traffic-Aware Network Criticality} (Section \ref{sec:spe:net}) is a suitable global parameter which assesses the load to capacity relation, and thus its proneness to overloads. Furthermore, it serves to find an optimal capacity allocation for a given topology and traffic demand and compare it to the one in the investigated network.


The weighted \textit{Good Expansion} procedure (Section \ref{sec:spe:goo}) finds clusters that share ties of higher weights between themselves than with 
the rest of the graph. Based on this,
the identification of the largest-strength bridge tie
is easy. Such bridges represent heavily loaded bottlenecks. Their deletion and successive redistribution of flows to other, lower-capacity links could make those prone to overload.


\subsection{Ambiguity and Correlations of Metrics}

The interpretation of metrics can be ambiguous and their implications for network robustness may be inconclusive without further investigation. Three main challenges present themselves most prominently in this context.

Firstly, 
the individual impact of a characteristic as measured by a metric cannot be estimated without controlling for other metrics that correlate with it.
One example for this statement is the \textit{Clustering Coefficient} $C$ (Section \ref{sec:clu:clu}). If estimated alone, a graph with a higher value would be considered more densely connected and featuring fewer bottlenecks than a graph with a lower value. If both networks have the same number of links, however, it becomes apparent that the graph with a higher $C$ exhibits a pronounced community structure, leading to local \textit{Transmission Speed} and \textit{Disconnection Robustness} \emph{inside} of the communities, yet a more vulnerable global graph structure \emph{between} communities. These ambiguous conclusions are related to the positive correlation of the \textit{CC} with the number of links in a graph; if more links exist in total, some will interconnect neighbors and increase the \textit{CC}. Therefore, it is of high importance to further study metric interactions and in particular correlations on the various graph types used for modeling the Internet. With the help of this information, the individual impact of the assessed graph features on robustness can be evaluated correctly. 

Secondly, many of the presented metrics overlap as they measure similar features that indicate or influence one or several \textit{Robustness Aspects}. The investigation of these groups of overlapping metrics shows that some may lead to conclusions about robustness that seem to be incompatible with each other. An example is \textit{Entropy} and \textit{Skewness} (Section \ref{sec:con:ske}): Both measure the same feature, namely the heterogeneity of the \textit{Degree Distribution}, which however is found to have different effects on \textit{Disconnection Robustness} against random failures, measured by \textit{Percolation Threshold} $p_c$ (Section \ref{sec:min:per}) and \textit{Number of Nodes in the Largest Component} $|V_L|$ (Section \ref{sec:min:ana}), respectively. In the former case, robustness increases for more homogeneous networks, and in the latter case, it decreases. It must be noted though, that $p_c$ measures the threshold failure probability at which the worst-case scenario, a disintegration without giant component, occurs, while $|V_L|$ measures the size of just this giant component at lower failure rates. They thus assess different facets of the same aspect. Another example is provided by Liu et al. \shortcite{liu2017comparative} who report that the \textit{Assortative Coefficient} $r$ is positively correlated with \textit{Natural Connectivity} and the stability of the \textit{Largest Component Size} under attacks, 
and indicates a more \textit{Disconnection Robust} graph. 
At the same time, though, it is negatively correlated to \textit{Algebraic Connectivity}, and thus to another measure of \textit{Disconnection Robustness}.

Thirdly, there could be possible conflicts between robustness goals in certain graphs. For example, as recently shown by Liu et al. \shortcite{liu2017comparative}, the optimization of a BA network with respect to \textit{Natural Connectivity} and limiting the decrease of the \textit{Largest Component Size} caused by removals, can lead to a significant increase in \textit{ASPL}. Hence, given certain constraints such as a fixed edge number, path redundancy and \textit{Disconnection Robustness} are in conflict with \textit{Transmission Speed}. 


The last two problems cannot be solved generally. The question if a graph is robust or not must always be answered depending on the adapted point of view -- is the global graph robustness of interest or the one of a certain set of nodes -- and the services that are to be delivered and their special demands. These demands, expressed by certain metric values which are to be maintained at an acceptable level, may stand in direct conflict to each other. 
In this case, a careful evaluation of their relative importance is to be conducted.

These and many similar problems need to be addressed in future research on the way to a profound and general understanding of 
Internet robustness. 

\section{Conclusion}\label{sec:sum}

This survey discusses an extensive set of robustness metrics in six major categories, their advantages, drawbacks and the main results obtained so far on Internet graphs. Furthermore, a comparison and assessment of their suitability for detecting crucial backbone structures and for measuring important robustness aspects is presented. Thus, we provided an outline for the conceptual tool set that will support future research in this field. Before selecting metrics for an individual robustness assessment, it should be very carefully analyzed which graph type is investigated and which network-generating model is applied. For every research goal, the accuracy-cost trade-off should be evaluated and the limitations of the metrics selected should be taken into account. 
To enable even more accurate and efficient robustness assessments, an investigation of metric interactions and correlations, extending for example the research of \cite{Costa2007, DallAsta2006} on the various graph types representing the Internet would constitute a fundamental future milestone in this research area. 

\newpage

\section{Summarizing Table}

The following table summarizes all discussed metrics and their main characteristics. The notation of graph types is as follows: \textit{simple}: \textit{s}, \textit{directed}: \textit{d}, \textit{(edge)-weighted}: \textit{w}, \textit{node-weighted}: $\tilde{w}$, \textit{labeled}: \textit{l}, and \textit{geographical}: \textit{g}.
Letters without brackets stand for required graph types, letters within brackets for further possible types. 
In order to highlight metrics that can be applied to every graph type, \textit{-- all --} is used. Graph types, to which the metric has been applied in literature, are listed in black. Other types, for which it could be calculated after simple modifications, in gray. 

In the column \textit{Scope}, the following abbreviations apply: \textit{local liability}: \textit{l}, \textit{local impact}: \textit{i} and \textit{global}: \textit{g}. The entry \textit{l \& i / g} indicates that a local metric version, which simultaneously assesses aspects about the  \textit{liability} and \textit{impact} of an entity, and a \textit{global} version exist and \textit{-- both --} that both local and global aspects are treated by one measure.
In \textit{Static}, the features are displayed as follows: \textit{static}: \textit{s} and \textit{dynamic}: \textit{d}; in the latter case, if only applicable to \textit{worst-case} scenarios: \textit{w}; or \textit{failures: f}.
Next, $\circ$ indicates that the codomain is independent of network parameters as $v$ and bounded; else, they are listed and unbounded codomains are marked by $\infty$. The entry \textit{N/A} indicates that the codomain could not be found in the literature, and '\textit{-}' that this column does not apply. Metrics that directly measure one or several of the \textit{Robustness Aspects}, \textit{Disconnection: d, Transmission: t, Traffic: o}, or evaluate the \textit{Backbone: b}, show the respective letters. 
If they are just an indirect \textit{indicator} for these aspects, the letters are put into brackets. The column \textit{Comp.} is checked if the metric is efficiently computable, and \textit{NP-h.} or \textit{NP-c.} if the computation is known to be NP-hard or NP-complete. If an exact or approximating algorithm is provided in the literature, its calculation order is displayed instead.\\

\begin{small}	
	\begin{longtable}
		{@{\hspace{-1mm}}l@{\hspace{4pt}}l@{\hspace{4pt}}r@{\hspace{3pt}}c@{\hspace{1pt}}cr@{\hspace{2pt}}c@{\hspace{2pt}}l@{\hspace{1pt}}r@{\hspace{0pt}}c@{\hspace{1pt}}
			cc@{\hspace{3pt}}c@{\hspace{3pt}}c@{\hspace{3pt}}c@{\hspace{3pt}}c} 
		
		\multicolumn{2}{l}{\textbf{Section \& Metric}} &\multicolumn{3}{c}{\textbf{Graphs}}&\multicolumn{3}{c}{\textbf{Scope}}&\multicolumn{2}{l}{\textbf{Static}}&\textbf{Codo.}&\multicolumn{4}{c}{\textbf{Robustness}}&\textbf{Comp.} \\
		\ref{sec:con:nod}&Node Degree / Degree-Freq. Distr.		&\, s&&[d]&\,l\hspace{2pt}\&\hspace{2pt}i&/&g&\;\; s&&$v$ / $\circ$ &[d]&[t]&&[b]& \checkmark \\	
		\ref{sec:con:str}&Strength / Strength Distribution				&&w& [d]&i&/&g&s&&$\infty$ /  $\circ$&&[t]&[o]&[b]&\checkmark \\	
		\ref{sec:con:ent}&Entropy 			&s&\textcolor{lightgray}{[w]}&\textcolor{lightgray}{[d]}&&&g&s&&$v$&[d]&[t]&&	 &\checkmark \\
\ref{sec:con:ske}&Skewness 			&s&\textcolor{lightgray}{[w]}&\textcolor{lightgray}{[d]}&&&g&s&&$\circ$&[d]&[t]&&	 &\checkmark \\
\ref{sec:con:vul}&Vulnerability Function		&s&&&&&g&s&&$\circ$&[d]&[t]&&	 &\checkmark \\
\ref{sec:con:ass}&Assortative Coefficient		&s&&&&&g&s&&$\circ$&[d]&&&	 &\checkmark \\
\ref{sec:con:ave}&Average Neighbor Connectivity &s& [w]&\textcolor{lightgray}{[d]}&&&g&s&&$v$ &[d]&&&[b]	&\checkmark \\ 
		\ref{sec:con:ric}&Rich-Club Connectivity &\multicolumn{3}{c}{-- all --}
		&&&g&s&&$\infty$&&[t]&[o]&[b]	&\checkmark \\
		\ref{sec:clu:clu}&Clustering Coefficient&\multicolumn{3}{c}{-- all --}&\textcolor{white}{l\hspace{2pt}\&\hspace{2pt}}i&/&g&s&&$\circ$&[d]&[t]&[o]&&\checkmark\\
		\ref{sec:clu:edg}&Edge Clustering Coefficient&s&\textcolor{lightgray}{[w]}&\textcolor{lightgray}{[d]}&\textcolor{white}{l\hspace{2pt}\&\hspace{2pt}}i&&&s&&$k_{i}$&[d]&[t]&&&\checkmark\\
		\ref{sec:clu:mod}&Modularity&s&\textcolor{lightgray}{[w]}&&&&g&s&&$\circ$&&&&&\checkmark\\
		\ref{sec:clu:mod:mod}&Modularity Matrix&s&&&
		\multicolumn{3}{c}{-- both --}
		&s&&-&d&&&&\checkmark\\
		\ref{sec:clu:mod:edg}&Edge Betweenness Part. Algorithm&s&&&\multicolumn{3}{c}{-- both --}&s&&-&d&&&&$<\mathcal{O}(v^3)$\\
		\ref{sec:clu:zsc}&Z-Score of Within Module-Degree&s&\textcolor{lightgray}{[w]}&\textcolor{lightgray}{[d]}&\textcolor{white}{l\hspace{2pt}\&\hspace{2pt}}i&&&s&&$\infty$&[d]&[t]&&&\checkmark\\
		\ref{sec:clu:par}&Participation Coefficient&s&\textcolor{lightgray}{[w]}&&l \& i&&&s&&$\circ$&[d]&[t]&&&\checkmark\\
		\ref{sec:min:con}&Vertex-, Edge- \& Cond. Connectivity&s&&&&&g&&w&$v$&&&&&NP-c.\\
		\ref{sec:min:spa}&Sparsity&s&&&&&g&&w&$v$&d&&&&\checkmark\\
		\ref{sec:min:che}&Cheeger Constant&s&&&&&g&&w&$v$&d&&&&NP-c.\\

		\ref{sec:min:minm}&Minimum m-Degree&s&&&&&g&&w&$v$&d&&&&$\mathcal{O}((\substack{v\\m})e)$\\
		\ref{sec:min:net}&Network Partitioning Algorithm&s&&&\multicolumn{3}{c}{-- both --}&&w&-&d&&&&$<\mathcal{O}(e^2)$\\
			\ref{sec:min:rat}&Ratio of Disruption&s&&&&&g&&w&$v$&d&&&&NP-c.\\
		\ref{sec:min:res}&Local Decay Resilience&s&&[d]&l\textcolor{white}{\hspace{2pt}\&\hspace{2pt}i}&/&g&&w&$v(h)$&d&t&&&\checkmark\\
		\ref{sec:min:toughness}&Toughness / Integrity / Scatt. No.&s&&&&&g&&w&N/A&d&&&&NP-c.\\
					\ref{sec:min:ten}&Tenacity / Edge-T. / Mixed-T.&s&&&&&g&&w&N/A&d&&&&NP-c.\\
		\ref{sec:min:per}&Percolation Threshold&s&&&&&g&&f&$\circ$&d&&&&\checkmark\\
		\ref{sec:min:rel}&Reliability Polynomial&s&&&&&g&&f&$\circ$&d&&&b&\checkmark\\
\ref{sec:min:resfac}&Partition Resilience Factor&s&&&&&g&&f&$\circ$&d&&&b&\checkmark\\		\ref{sec:min:ana}&Total Number of Isolated Components&s&&&&&g&&d&$v$&d&&&&\checkmark\\
		&Frac. of Nodes in Largest Component&s&&&&&g&&d&$\circ$&d&&&&\checkmark\\
		&Average Size of Isolated Components&s&&&&&g&&d&$v$&&&&&\checkmark\\
		&Distr. of Component Class Frequency&s&&&&&g&&d&$v$&d&&&&\checkmark\\
		&Distr. of Rel. No. of Nodes per Class&s&&&&&g&&d&$\circ$&d&&&&\checkmark\\
		&Reachability&s&&&&&g&&d&$\circ$&d&&&&\checkmark\\
		
		\ref{sec:dis:ave}&ASPL&s&\textcolor{lightgray}{[w]}&\textcolor{lightgray}{[d]}&\textcolor{white}{l\hspace{2pt}\&\hspace{2pt}}i&/&g&s&&$v$&&t&&&\checkmark \\ 
		&Diameter-Inverse-K&s&\textcolor{lightgray}{[w]}&\textcolor{lightgray}{[d]}&&&g&s&&$v$&&t&&&\checkmark\\
		&Diameter&s&\textcolor{lightgray}{[w]}&\textcolor{lightgray}{[d]}&&&g&s&&$\infty$ / $v$&&t&&&\checkmark\\
		\ref{sec:dis:glo}&Global Network Efficiency&s&\textcolor{lightgray}{[w]}&\textcolor{lightgray}{[d]}&\textcolor{white}{l\hspace{2pt}\&\hspace{2pt}}i&/&g&s&&$\circ$&&t&&b&\checkmark\\
		&Harm. Mean of Geodesic Distances&s&&&&&g&s&&$\infty$&&t&&&\checkmark\\
		\ref{sec:dis:loc}&Local Network Efficiency&s&\textcolor{lightgray}{[w]}&\textcolor{lightgray}{[d]}&l \& i&/&g&s&&$\circ$&d&t&[o]&&\checkmark\\
		&Cyclic Coefficient&s&\textcolor{lightgray}{[w]}&\textcolor{lightgray}{[d]}&l \& i&/&g&s&&$\circ$&d&t&[o]&&\checkmark\\
		\ref{sec:dis:cha}&Characteristic Path Length&l&\textcolor{lightgray}{[w]}&\textcolor{lightgray}{[d]}&\textcolor{white}{l\hspace{2pt}\&\hspace{2pt}}i&&&s&&$\infty$ / $v$&&t&&[b]&\checkmark\\
		\ref{sec:dis:exp}&Expansion&s&\textcolor{lightgray}{[w]}&[d]&\textcolor{white}{l\hspace{2pt}\&\hspace{2pt}}i&/&g&s&&$\circ$&[d]&t&&b&\checkmark\\
		\ref{sec:dis:eff}& Effective Eccentricity / Eff. Diameter&s&\textcolor{lightgray}{[w]}&\textcolor{lightgray}{[d]}&\textcolor{white}{l\hspace{2pt}\&\hspace{2pt}}i&/&g&s&&$v$&[d]&t&&b&\checkmark\\
		
		\ref{sec:thr:bet}&Betweenness Centrality, node / link&s&[w]&\textcolor{lightgray}{[d]}&\textcolor{white}{l\hspace{2pt}\&\hspace{2pt}}i&&&s&&
		$v$&[d]&[t]&o&	&\checkmark\\
		&AS Hegemony&s&&&\textcolor{white}{l\hspace{2pt}\&\hspace{2pt}}i&&&s&&
		N/A&[d]&[t]&o&	&\checkmark\\		&Edge Degree&s&\textcolor{lightgray}{[w]}&\textcolor{lightgray}{[d]}&\textcolor{white}{l\hspace{2pt}\&\hspace{2pt}}i&&&s&&$k_{max}$&&&&&\checkmark\\
		\ref{sec:thr:cen}&Central Point Dominance&s&\textcolor{lightgray}{[w]}&\textcolor{lightgray}{[d]}&&&g&s&&$\circ$&[d]&[t]&&&\checkmark\\
		\ref{sec:thr:eff}&Effective Load &s&\textcolor{lightgray}{[w]}&\textcolor{lightgray}{[d]}&l \& i&&&&d&$v$&&&o&&\checkmark\\
		\ref{sec:thr:per}&Performance&$\tilde{w}$&\textcolor{lightgray}{[w]}&\textcolor{lightgray}{[d]}&&&g&s&&$\infty$&&&o&&\checkmark\\
		\ref{sec:thr:ela}			&Elasticity&s&\textcolor{lightgray}{[w]}&\textcolor{lightgray}{[d]}&&&g&&d&$\circ$&&&o&&\checkmark\\
\ref{sec:thr:vuln}	&Vulnerability Impact Factors&l&w&\textcolor{lightgray}{[d]}&l \& i&/&g&&d&$\infty$ / $\circ$&&&o&&\checkmark\\
\ref{survfail}	&Survivability Function, Failures&&w&d&&&g&&f&$\circ$&&&o&b&\checkmark\\
		\ref{sec:con:sig}&Eigenvector Centrality&s&\textcolor{lightgray}{[w]}&\textcolor{lightgray}{[d]}&l \& i&&&s&&$\circ$&&&&&\checkmark\\
		\ref{sec:spe:sym}&Symmetry Ratio&s&&&&&g&s&&$v$&&&[o]&&\checkmark\\
		\ref{sec:spe:spe}&Spectral Cluster Identification&s&&[d]&\multicolumn{3}{c}{-- both --}&s&&-&d&[t]&[o]&&\checkmark\\
		\ref{sec:spe:alg}&Algebraic Connectivity&s&&&&&g&s&&$v,k_{min}$&d&&&&\checkmark\\
		\ref{sec:spe:goo}&Good Expansion&s&[w]&&&&g&s&&$\circ$&d&[t]&[o]&&NP-h.\\
		\ref{sec:spe:num}&Number of Spanning Trees&s&&&&&g&s&&$k_i$&d&&&&\checkmark\\
		\ref{sec:spe:nat}& Natural Connectivity&s&&&&&g&s&&$v$&d&&&&\checkmark\\	
		\ref{sec:spe:rwaspl}& Random Walk ASPL&s&[w]&\textcolor{lightgray}{[d]}&i&/&g&s&&$\infty$&&t&&&\checkmark\\	
		\ref{sec:spe:curcl}& Current-Flow Closeness&s&[w]&\textcolor{lightgray}{[d]}&i&/&g&s&&N/A&&t&&&\checkmark\\
		\ref{sec:spe:rwbe}& Random Walk Betweenness&\multicolumn{3}{c}{-- all --}&i&&&s&&$v,P(k)$&[d]&[t]&o&	&\checkmark\\
		\ref{sec:spe:curbe}& Current-Flow Betweenness&s&&&&&g&s&&$v$&[d]&[t]&o&	&\checkmark\\
		\ref{sec:spe:net}& Network Criticality&\multicolumn{3}{c}{-- all --}&&&g&s&&$v, a(G)$&[d]&t&o&&\checkmark\\	
		\ref{sec:oth:out}&Distance Strength &g&&\textcolor{lightgray}{[d]}&\textcolor{white}{l\hspace{2pt}\&\hspace{2pt}}i&&&s&&$\infty$&d&t&&&\checkmark\\
		\ref{survgeo}&Survivability Function, Geographical &g&&&&g&&&d&$\circ$&d&&&&\checkmark\\
		 &Outreach&g&w&\textcolor{lightgray}{[d]}&\textcolor{white}{l\hspace{2pt}\&\hspace{2pt}}i&&&s&&$\infty$&d&t&o&&\checkmark\\
		\ref{sec:dis:poi}&Pointwise Vulnerability&g&\textcolor{lightgray}{[w]}&\textcolor{lightgray}{[d]}&\textcolor{white}{l\hspace{2pt}\&\hspace{2pt}}i&&&&d&$\circ$&&t&[o]&&\checkmark\\
		&Global Vulnerability&g&\textcolor{lightgray}{[w]}&\textcolor{lightgray}{[d]}&&&g&&d&$\circ$&&t&&&\checkmark\\
		&Rel. Variance of Pointwise Vuln.&g&\textcolor{lightgray}{[w]}&\textcolor{lightgray}{[d]}&&&g&&d&$\infty$&&t&&&\checkmark\\
		\ref{sec:oth:eff}&Eff. / Total Geograph. Path Diversity&g&&\textcolor{lightgray}{[d]}&l\textcolor{white}{\hspace{2pt}\&\hspace{2pt}i}&/&g&s&&$\circ$&d&&&&\checkmark\\	
		
	\end{longtable}
\end{small}

\newpage



\bibliographystyle{acm}

\bibliography{references}

\begin{thebibliography}{100}

\bibitem{Aggarwal:2014}
{\sc Aggarwal, C., and Subbian, K.}
\newblock Evolutionary network analysis: A survey.
\newblock {\em ACM Computing Surveys 47}, 1 (2014), 10:1--10:36.

\bibitem{Albert2002}
{\sc Albert, R., and Barab{\'a}si, A.-L.}
\newblock Statistical mechanics of complex networks.
\newblock {\em {Reviews of Modern Physics} 74}, 1 (2002), 47.

\bibitem{albert2000barabasi}
{\sc Albert, R., Jeong, H., and Barab{\'a}si, A.-L.}
\newblock Error and attack tolerance of complex networks.
\newblock {\em Nature 406}, 6794 (2000), 378--382.

\bibitem{Barabasi:2016}
{\sc Barab{\'a}si, A.-L.}
\newblock {\em Network Science}.
\newblock Cambridge University Press, 2016.

\bibitem{barabasi1999emergence}
{\sc Barab{\'a}si, A.-L., and Albert, R.}
\newblock Emergence of scaling in random networks.
\newblock {\em Science 286}, 5439 (1999), 509--512.

\bibitem{barefoot1987vulnerability}
{\sc Barefoot, C.~A., Entringer, R., and Swart, H.}
\newblock Vulnerability in graphs -- a comparative survey.
\newblock {\em J. Combin. Math. Combin. Comput 1}, 38 (1987), 13--22.

\bibitem{Barrat2004}
{\sc Barrat, A., Barthelemy, M., Pastor-Satorras, R., and Vespignani, A.}
\newblock The architecture of complex weighted networks.
\newblock {\em Proceedings of the National Academy of Sciences (PNAS) 101}, 11
  (2004), 3747--3752.

\bibitem{Barthelemy2004}
{\sc Barth{\'e}lemy, M., Barrat, A., Pastor-Satorras, R., and Vespignani, A.}
\newblock Characterization and modeling of weighted networks.
\newblock {\em Physica A: Statistical Mechanics and its Applications 346}, 1-2
  (2005), 34--43.

\bibitem{baumann2014robust}
{\sc Baumann, A., and Fabian, B.}
\newblock How robust is the internet? -- insights from graph analysis.
\newblock In {\em International Conference on Risks and Security of Internet
  and Systems (CRiSIS 2014): Risks and Security of Internet and Systems},
  vol.~8924 of {\em LNCS}. Springer, 2014, pp.~247--254.

\bibitem{Baumann2013}
{\sc Baumann, A., and Fabian, B.}
\newblock Vulnerability against internet disruptions -- a graph-based
  perspective.
\newblock In {\em International Conference on Critical Information
  Infrastructures Security}, vol.~9578 of {\em LNCS}. Springer, 2015,
  pp.~120--131.

\bibitem{boccaletti2007multiscale}
{\sc Boccaletti, S., Buld{\'u}, J., Criado, R., Flores, J., Latora, V., Pello,
  J., and Romance, M.}
\newblock Multiscale vulnerability of complex networks.
\newblock {\em Chaos: An Interdisciplinary Journal of Nonlinear Science 17}, 4
  (2007), 043110.

\bibitem{boesch1970graphs}
{\sc Boesch, F., and Thomas, R.}
\newblock On graphs of invulnerable communication nets.
\newblock {\em IEEE Transactions on Circuit Theory 17}, 2 (1970), 183--192.

\bibitem{bonacich1972factoring}
{\sc Bonacich, P.}
\newblock Factoring and weighting approaches to status scores and clique
  identification.
\newblock {\em Journal of Mathematical Sociology 2}, 1 (1972), 113--120.

\bibitem{brandes2005centrality}
{\sc Brandes, U., and Fleischer, D.}
\newblock Centrality measures based on current flow.
\newblock In {\em Annual symposium on theoretical aspects of computer
  science\/} (2005), Springer, pp.~533--544.

\bibitem{bui1992finding}
{\sc Bui, T.~N., and Jones, C.}
\newblock Finding good approximate vertex and edge partitions is np-hard.
\newblock {\em Information Processing Letters 42}, 3 (1992), 153--159.

\bibitem{ccetinkaya2015multilevel}
{\sc {\c{C}}etinkaya, E.~K., Alenazi, M.~J., Peck, A.~M., Rohrer, J.~P., and
  Sterbenz, J.~P.}
\newblock Multilevel resilience analysis of transportation and communication
  networks.
\newblock {\em Telecommunication Systems 60}, 4 (2015), 515--537.

\bibitem{cetinkaya2013flow}
{\sc {\c{C}}etinkaya, E.~K., Peck, A.~M., and Sterbenz, J.~P.}
\newblock Flow robustness of multilevel networks.
\newblock In {\em 9th International Conference on the Design of Reliable
  Communication Networks (DRCN)\/} (2013), IEEE, pp.~274--281.

\bibitem{chartrand1996graphs}
{\sc Chartrand, G., and Lesniak, L.}
\newblock Graphs and digraphs.
\newblock {\em Wadsworth \& Brooks/Cole, Monterey, CA\/} (1996).

\bibitem{Cheng2013}
{\sc Cheng, Y., Li, J., and Sterbenz, J. P.~G.}
\newblock Path geo-diversification: design and analysis.
\newblock 46--53.

\bibitem{CHVATAL1973215}
{\sc Chv{\'a}tal, V.}
\newblock Tough graphs and hamiltonian circuits.
\newblock {\em Discrete Mathematics 5}, 3 (1973), 215 -- 228.

\bibitem{cohen2000resilience}
{\sc Cohen, R., Erez, K., Ben-Avraham, D., and Havlin, S.}
\newblock Resilience of the internet to random breakdowns.
\newblock {\em Physical Review Letters 85}, 21 (2000), 4626.

\bibitem{Colizza2006}
{\sc Colizza, V., Flammini, A., Serrano, M.~A., and Vespignani, A.}
\newblock Detecting rich-club ordering in complex networks.
\newblock {\em Nature Physics 2}, 2 (2006), 110--115.

\bibitem{Costa2007}
{\sc Costa, L. d.~F., Rodrigues, F.~A., Travieso, G., and Villas~Boas, P.~R.}
\newblock Characterization of complex networks: A survey of measurements.
\newblock {\em Advances in Physics 56}, 1 (2007), 167--242.

\bibitem{Criado2005}
{\sc Criado, R., Flores, J., Hern{\'a}ndez-Bermejo, B., Pello, J., and Romance,
  M.}
\newblock Effective measurement of network vulnerability under random and
  intentional attacks.
\newblock {\em Journal of Mathematical Modelling and Algorithms 4}, 3 (2005),
  307--316.

\bibitem{Crucitti2003}
{\sc Crucitti, P., Latora, V., Marchiori, M., and Rapisarda, A.}
\newblock Efficiency of scale-free networks: Error and attack tolerance.
\newblock {\em Physica A: Statistical Mechanics and its Applications 320\/}
  (2003), 622--642.

\bibitem{Crucitti2006}
{\sc Crucitti, P., Latora, V., and Porta, S.}
\newblock {Centrality measures in spatial networks of urban streets}.
\newblock {\em Physical Review E - Statistical, Nonlinear, and Soft Matter
  Physics 73}, 3 (2006), 1--5.

\bibitem{DallAsta2006}
{\sc Dall'Asta, L., Barrat, A., Barth{\'e}lemy, M., and Vespignani, A.}
\newblock Vulnerability of weighted networks.
\newblock {\em Journal of Statistical Mechanics: Theory and Experiment 2006},
  04 (2006), P04006.

\bibitem{Dekker2005}
{\sc Dekker, A.~H., and Colbert, B.}
\newblock The symmetry ratio of a network.
\newblock In {\em Proceedings of the 2005 Australasian Symposium on Theory of
  Computing (CATS '05)\/} (January 2005), vol.~41.

\bibitem{Doerr:2014}
{\sc Doerr, C., and Kuipers, F.~A.}
\newblock All quiet on the internet front?
\newblock {\em IEEE Communications Magazine 52}, 10 (2014), 46--51.

\bibitem{dong2007understanding}
{\sc Dong, J., and Horvath, S.}
\newblock Understanding network concepts in modules.
\newblock {\em BMC Systems Biology 1\/} (2007), 24.

\bibitem{Doyle2005}
{\sc Doyle, J.~C., Alderson, D.~L., Li, L., Low, S., Roughan, M., Shalunov, S.,
  Tanaka, R., and Willinger, W.}
\newblock The "robust yet fragile" nature of the internet.
\newblock {\em Proceedings of the National Academy of Sciences (PNAS) 102}, 41
  (2005), 14497--14502.

\bibitem{Edwards2012}
{\sc Edwards, B., Hofmeyr, S., Stelle, G., and Forrest, S.}
\newblock Internet topology over time.
\newblock {\em arXiv:1202.3993\/} (2012).

\bibitem{erdios1959r}
{\sc Erd{\"o}s, P., and R{\'e}nyi, A.}
\newblock On random graphs, i.
\newblock {\em Publicationes Mathematicae (Debrecen) 6\/} (1959), 290--297.

\bibitem{estrada2000characterization}
{\sc Estrada, E.}
\newblock Characterization of 3d molecular structure.
\newblock {\em Chemical Physics Letters 319}, 5 (2000), 713--718.

\bibitem{Estrada2007}
{\sc Estrada, E.}
\newblock Spectral scaling and good expansion properties in complex networks.
\newblock {\em EPL (Europhysics Letters) 73}, 4 (2006), 649.

\bibitem{Fabian:2017}
{\sc Fabian, B., Baumann, A., Ehlert, M., Ververis, V., and Ermakova, T.}
\newblock Coria - analyzing internet connectivity risks using network graphs.
\newblock In {\em IEEE International Conference on Communications (IEEE ICC
  2017), Paris, France\/} (2017).

\bibitem{Fabian:2015}
{\sc Fabian, B., Baumann, A., and Lackner, J.}
\newblock Topological analysis of cloud service connectivity.
\newblock {\em Computers and Industrial Engineering 88\/} (October 2015),
  151--165.

\bibitem{Fabian:2017a}
{\sc Fabian, B., Tilch, G., and Ermakova, T.}
\newblock A multilayer graph model of the internet topology.
\newblock Tech. rep., DOI: 10.5281/zenodo.1038599, October 2017.

\bibitem{faloutsos1999power}
{\sc Faloutsos, M., Faloutsos, P., and Faloutsos, C.}
\newblock On power-law relationships of the internet topology.
\newblock In {\em ACM SIGCOMM Computer Communication Review\/} (1999), vol.~29,
  ACM, pp.~251--262.

\bibitem{fiduccia1988linear}
{\sc Fiduccia, C.~M., and Mattheyses, R.~M.}
\newblock A linear-time heuristic for improving network partitions.
\newblock In {\em Papers on twenty-five years of electronic design
  automation\/} (1988), ACM, pp.~241--247.

\bibitem{Fiedler1973}
{\sc Fiedler, M.}
\newblock Algebraic connectivity of graphs.
\newblock {\em Czechoslovak Mathematical Journal 23}, 2 (1973), 298--305.

\bibitem{Fontugne:2017}
{\sc Fontugne, R., Shah, A., and Aben, E.}
\newblock As hegemony: A robust metric for as centrality.
\newblock In {\em Proceedings of the SIGCOMM Posters and Demos\/} (New York,
  NY, USA, 2017), ACM, pp.~48--50.

\bibitem{freeman1977set}
{\sc Freeman, L.~C.}
\newblock A set of measures of centrality based on betweenness.
\newblock {\em Sociometry\/} (1977), 35--41.

\bibitem{npcomplete}
{\sc Garey, M.~R., and Johnson, D.~S.}
\newblock {\em Computers and Intractability: A Guide to the Theory of
  NP-Completeness}, vol.~58.
\newblock W. H. Freeman and Company, 1979.

\bibitem{Ghedini2011}
{\sc Ghedini, C.~G., and Ribeiro, C. H.~C.}
\newblock {Rethinking failure and attack tolerance assessment in complex
  networks}.
\newblock {\em Physica A: Statistical Mechanics and its Applications 390},
  23-24 (2011), 4684--4691.

\bibitem{Girvan2002abc}
{\sc Girvan, M., and Newman, M. E.~J.}
\newblock Community structure in social and biological networks.
\newblock {\em Proceedings of the National Academy of Sciences (PNAS) 99}, 12
  (2002), 7821--7826.

\bibitem{Gkantsidis2003a}
{\sc Gkantsidis, C., Mihail, M., and Zegura, E.}
\newblock Spectral analysis of internet topologies.
\newblock 364--374.

\bibitem{gol2004vulnerability}
{\sc Gol'dshtein, V., Koganov, G.~A., and Surdutovich, G.~I.}
\newblock Vulnerability and hierarchy of complex networks.
\newblock {\em arXiv:cond-mat/0409298\/} (2004).

\bibitem{golnari2018random}
{\sc Golnari, G., Zhang, Z.-L., and Boley, D.}
\newblock Random walk fundamental tensor and its applications to network
  analysis.
\newblock {\em arXiv:1801.08583\/} (2018).

\bibitem{guimera2005functional}
{\sc Guimera, R., and Amaral, L. A.~N.}
\newblock Functional cartography of complex metabolic networks.
\newblock {\em Nature 433}, 7028 (2005), 895--900.

\bibitem{hakimi1962realizability}
{\sc Hakimi, S.~L.}
\newblock On realizability of a set of integers as degrees of the vertices of a
  linear graph. i.
\newblock {\em Journal of the Society for Industrial and Applied Mathematics\/}
  (1962), 496--506.

\bibitem{harary1983conditional}
{\sc Harary, F.}
\newblock Conditional connectivity.
\newblock {\em Networks 13}, 3 (1983), 347--357.

\bibitem{hariri2003impact}
{\sc Hariri, S., Qu, G., Dharmagadda, T., Ramkishore, M., and Raghavendra,
  C.~S.}
\newblock Impact analysis of faults and attacks in large-scale networks.
\newblock {\em IEEE Security \& Privacy 99}, 5 (2003), 49--54.

\bibitem{ho2007growth}
{\sc Ho, S.-C., Kauffman, R.~J., and Liang, T.-P.}
\newblock A growth theory perspective on b2c e-commerce growth in europe: An
  exploratory study.
\newblock {\em Electronic Commerce Research and Applications 6\/} (2007),
  237--259.

\bibitem{holme2002vertex}
{\sc Holme, P., and Kim, B.~J.}
\newblock Vertex overload breakdown in evolving networks.
\newblock {\em Physical Review E - Statistical, Nonlinear, and Soft Matter
  Physics 65}, 6 (2002), 066109.

\bibitem{holme2002attack}
{\sc Holme, P., Kim, B.~J., Yoon, C.~N., and Han, S.~K.}
\newblock Attack vulnerability of complex networks.
\newblock {\em Physical Review E - Statistical, Nonlinear, and Soft Matter
  Physics 65}, 5 (2002), 056109.

\bibitem{jabbar2010framework}
{\sc Jabbar, A.}
\newblock {\em A framework to quantify network resilience and survivability}.
\newblock PhD thesis, University of Kansas, 2010.

\bibitem{Jamakovic2007}
{\sc Jamakovic-Kapic, A., and Uhlig, S.}
\newblock Influence of the network structure on robustness.
\newblock {\em Proceedings of the 15th IEEE International Conference on
  Networks (ICON 2007)\/} (2007), 278--283.

\bibitem{Jung1978}
{\sc Jung, H.~A.}
\newblock On a class of posets and the corresponding comparability graphs.
\newblock {\em Journal of Combinatorial Theory, Series B 24}, 2 (1978),
  125--133.

\bibitem{Kamisinski:2015}
{\sc Kamisinski, A., Cholda, P., and Jajszczyk, A.}
\newblock Assessing the structural complexity of computer and communication
  networks.
\newblock {\em ACM Computing Surveys 47}, 4 (2015), 66:1--66:36.

\bibitem{karypis1998fast}
{\sc Karypis, G., and Kumar, V.}
\newblock A fast and high quality multilevel scheme for partitioning irregular
  graphs.
\newblock {\em SIAM Journal on scientific Computing 20}, 1 (1998), 359--392.

\bibitem{Kim2005}
{\sc Kim, H.-J., and Kim, J.~M.}
\newblock Cyclic topology in complex networks.
\newblock {\em Phys. Rev. E 72\/} (Sep 2005), 036109.

\bibitem{klemm2002growing}
{\sc Klemm, K., and Egu{\'\i}luz, V.~M.}
\newblock Growing scale-free networks with small-world behavior.
\newblock {\em Physical Review E - Statistical, Nonlinear, and Soft Matter
  Physics 65}, 5 (2002), 057102.

\bibitem{latora2001efficient}
{\sc Latora, V., and Marchiori, M.}
\newblock Efficient behavior of small-world networks.
\newblock {\em Physical Review Letters 87}, 19 (2001), 198701.

\bibitem{Lee2005}
{\sc Lee, E.~J., Goh, K.-I., Kahng, B., and Kim, D.}
\newblock {Robustness of the avalanche dynamics in data-packet transport on
  scale-free networks}.
\newblock {\em Physical Review E - Statistical, Nonlinear, and Soft Matter
  Physics 71}, 5 (2005), 1--5.

\bibitem{li2004first}
{\sc Li, L., Alderson, D., Willinger, W., and Doyle, J.}
\newblock A first-principles approach to understanding the internet's
  router-level topology.
\newblock {\em ACM SIGCOMM Computer Communication Review 34}, 4 (2004), 3--14.

\bibitem{liew1992framework}
{\sc Liew, S.~C., and Lu, K.~W.}
\newblock A framework for network survivability characterization.
\newblock In {\em Communications, 1992. ICC'92, Conference record,
  SUPERCOMM/ICC'92, Discovering a New World of Communications., IEEE
  International Conference on\/} (1992), IEEE, pp.~405--410.

\bibitem{lipman1985toward}
{\sc Lipman, M., and Pippert, R.}
\newblock Toward a measure of vulnerability ii. the ratio of disruption.
\newblock In {\em Graph theory with applications to algorithms and computer
  science\/} (1985), John Wiley \& Sons, Inc., pp.~507--517.

\bibitem{liu2017comparative}
{\sc Liu, J., Zhou, M., Wang, S., and Liu, P.}
\newblock A comparative study of network robustness measures.
\newblock {\em Frontiers of Computer Science 11}, 4 (2017), 568--584.

\bibitem{Magnien:2011}
{\sc Magnien, C., Latapy, M., and Guillaume, J.-L.}
\newblock Impact of random failures and attacks on poisson and power-law random
  networks.
\newblock {\em ACM Computing Surveys 43}, 3 (Apr. 2011), 13:1--13:31.

\bibitem{Sacks2009}
{\sc Magoni, D.}
\newblock Tearing down the internet.
\newblock {\em IEEE J.Sel. A. Commun. 21}, 6 (Sept. 2006), 949--960.

\bibitem{Mahadevan2007}
{\sc Mahadevan, P., Hubble, C., Krioukov, D., Huffaker, B., and Vahdat, A.}
\newblock Orbis: Rescaling degree correlations to generate annotated internet
  topologies.
\newblock {\em ACM SIGCOMM Computer Communication Review 37}, 4 (2007),
  325--336.

\bibitem{Mahadevan2005}
{\sc Mahadevan, P., Krioukov, D., Fomenkov, M., Dimitropoulos, X., Claffy, K.,
  and Vahdat, A.}
\newblock The internet as-level topology: three data sources and one definitive
  metric.
\newblock {\em ACM SIGCOMM Computer Communication Review 36}, 1 (Jan. 2006),
  17--26.

\bibitem{Manzano2011}
{\sc Manzano, M., Calle, E., and Harle, D.}
\newblock Quantitative and qualitative network robustness analysis under
  different multiple failure scenarios.
\newblock In {\em 3rd International Congress on Ultra Modern Telecommunications
  and Control Systems and Workshops (ICUMT)\/} (2011), IEEE, pp.~1--7.

\bibitem{medhi2017network}
{\sc Medhi, D., and Ramasamy, K.}
\newblock {\em Network Routing: Algorithms, Protocols, and Architectures},
  2nd~ed.
\newblock Morgan Kaufmann, 2017.

\bibitem{Merris1994}
{\sc Merris, R.}
\newblock Laplacian matrices of graphs: A survey.
\newblock {\em Linear Algebra and its Applications 197-198\/} (1994), 143--176.

\bibitem{mihail2002eigenvalue}
{\sc Mihail, M., and Papadimitriou, C.}
\newblock On the eigenvalue power law.
\newblock In {\em Randomization and approximation techniques in computer
  science}. Springer, 2002, pp.~254--262.

\bibitem{Alon1997}
{\sc Moazzami, D., and Salehian, B.}
\newblock On the edge-tenacity of graphs.
\newblock 929--936.

\bibitem{Mohar1989}
{\sc Mohar, B.}
\newblock {Isoperimetric numbers of graphs}.
\newblock {\em Journal of Combinatorial Theory, Series B 47}, 3 (1989),
  274--291.

\bibitem{Mohar1991}
{\sc Mohar, B.}
\newblock {Some applications of Laplace eigenvalues of graphs}.
\newblock {\em Graph Symmetry: Algebraic Methods and Applications 497\/}
  (1991), 225--275.

\bibitem{molisz2004survivability}
{\sc Molisz, W.}
\newblock Survivability function-a measure of disaster-based routing
  performance.
\newblock {\em IEEE Journal on Selected Areas in Communications 22}, 9 (2004),
  1876--1883.

\bibitem{motter2002cascade}
{\sc Motter, A.~E., and Lai, Y.-C.}
\newblock Cascade-based attacks on complex networks.
\newblock {\em Physical Review E - Statistical, Nonlinear, and Soft Matter
  Physics 66}, 6 (2002), 065102.

\bibitem{Muro:2017}
{\sc Muro, M. A.~D., Valdez, L.~D., Rego, H. H.~A., Buldyrev, S.~V., Stanley,
  H.~E., and Braunstein, L.~A.}
\newblock Cascading failures in interdependent networks with multiple
  supply-demand links and functionality thresholds.
\newblock {\em Scientific Reports 7}, 1 (2017).

\bibitem{newman2001scientific}
{\sc Newman, M. E.~J.}
\newblock Scientific collaboration networks. i. network construction and
  fundamental results.
\newblock {\em Physical Review E - Statistical, Nonlinear, and Soft Matter
  Physics 64}, 1 (2001), 016131.

\bibitem{Newman2002}
{\sc Newman, M. E.~J.}
\newblock Assortative mixing in networks.
\newblock {\em Physical Review Letters 89}, 20 (2002), 208701.

\bibitem{Newman2004}
{\sc Newman, M. E.~J.}
\newblock Detecting community structure in networks.
\newblock {\em The European Physical Journal B-Condensed Matter and Complex
  Systems 38}, 2 (2004), 321--330.

\bibitem{newman2005measure}
{\sc Newman, M. E.~J.}
\newblock A measure of betweenness centrality based on random walks.
\newblock {\em Social Networks 27}, 1 (2005), 39--54.

\bibitem{Newman2006}
{\sc Newman, M. E.~J.}
\newblock Modularity and community structure in networks.
\newblock {\em Proceedings of the National Academy of Sciences (PNAS) 103}, 23
  (2006), 8577--8582.

\bibitem{Newman2004a}
{\sc Newman, M. E.~J., and Girvan, M.}
\newblock Finding and evaluating community structure in networks.
\newblock {\em Phys. Rev. E 69\/} (Feb 2004), 026113.

\bibitem{ng2006structural}
{\sc Ng, A. K.~S., and Efstathiou, J.}
\newblock Structural robustness of complex networks.
\newblock {\em Physical Review 3\/} (2006), 175--188.

\bibitem{Oliveira2007}
{\sc Oliveira, R.~V., Zhang, B., and Zhang, L.}
\newblock {Observing the evolution of Internet as topology}.
\newblock {\em ACM SIGCOMM Computer Communication Review 37}, 4 (2007), 313.

\bibitem{Onnela2005}
{\sc Onnela, J.~P., Saram\"{a}ki, J., Kert\'{e}sz, J., and Kaski, K.}
\newblock {Intensity and coherence of motifs in weighted complex networks}.
\newblock {\em Physical Review E - Statistical, Nonlinear, and Soft Matter
  Physics 71}, 6 (2005).

\bibitem{Opsahl2010}
{\sc Opsahl, T., Agneessens, F., and Skvoretz, J.}
\newblock Node centrality in weighted networks: Generalizing degree and
  shortest paths.
\newblock {\em Social Networks 32}, 3 (2010), 245--251.

\bibitem{Opsahl2008}
{\sc Opsahl, T., Colizza, V., Panzarasa, P., and Ramasco, J.~J.}
\newblock Prominence and control: The weighted rich-club effect.
\newblock {\em Physical Review Letters 101}, 16 (2008), 168702.

\bibitem{Opsahl2009}
{\sc Opsahl, T., and Panzarasa, P.}
\newblock Clustering in weighted networks.
\newblock {\em Social Networks 31}, 2 (2009), 155--163.

\bibitem{page1988practical}
{\sc Page, L.~B., and Perry, J.~E.}
\newblock A practical implementation of the factoring theorem for network
  reliability.
\newblock {\em IEEE Transactions on Reliability 37}, 3 (1988), 259--267.

\bibitem{page1994reliability}
{\sc Page, L.~B., and Perry, J.~E.}
\newblock Reliability polynomials and link importance in networks.
\newblock {\em IEEE Transactions on Reliability 43}, 1 (1994), 51--58.

\bibitem{Palmer2001}
{\sc Palmer, C.~R., Siganos, G., Faloutsos, M., Faloutsos, C., and Gibbons,
  P.~B.}
\newblock {The connectivity and fault-tolerance of the Internet topology}.
\newblock In {\em Workshop on Network Related Data Management (NRDM 2001)\/}
  (2001), pp.~1--9.

\bibitem{park2003static}
{\sc Park, S.-T., Khrabrov, A., Pennock, D.~M., Lawrence, S., Giles, C.~L., and
  Ungar, L.~H.}
\newblock Static and dynamic analysis of the internet's susceptibility to
  faults and attacks.
\newblock In {\em INFOCOM 2003. Twenty-Second Annual Joint Conference of the
  IEEE Computer and Communications. IEEE Societies\/} (2003), vol.~3, IEEE,
  pp.~2144--2154.

\bibitem{park2004comparing}
{\sc Park, S.-T., Pennock, D.~M., and Giles, C.~L.}
\newblock Comparing static and dynamic measurements and models of the
  internet's as topology.
\newblock In {\em Twenty-third Annual Joint Conference of the IEEE Computer and
  Communications Societies (INFOCOM 2004)\/} (2004), vol.~3, IEEE,
  pp.~1616--1627.

\bibitem{PastorSatorras:2004}
{\sc Pastor-Satorras, R., and Vespignani, A.}
\newblock {\em Evolution and Structure of the Internet: A Statistical Physics
  Approach}.
\newblock Cambridge University Press, 2004.

\bibitem{Piazza1995}
{\sc Piazza, B.~L., Robertst, F.~S., and Stueckle, S.~K.}
\newblock Edge-tenacious networks.
\newblock {\em Networks 25}, 1 (1995), 7--17.

\bibitem{radicchi2004defining}
{\sc Radicchi, F., Castellano, C., Cecconi, F., Loreto, V., and Parisi, D.}
\newblock Defining and identifying communities in networks.
\newblock {\em Proceedings of the National Academy of Sciences (PNAS) 101}, 9
  (2004), 2658--2663.

\bibitem{Ravasz2003}
{\sc Ravasz, E., and Barab\'{a}si, A.~L.}
\newblock {Hierarchical organization in complex networks}.
\newblock {\em Physical Review E 67}, 2 (2003).

\bibitem{Rochat2009}
{\sc Rochat, Y.}
\newblock Closeness centrality extended to unconnected graphs: The harmonic
  centrality index.
\newblock In {\em Applications of Social Network Analysis (ASNA 2009)\/}
  (2009).

\bibitem{salles2011strategies}
{\sc Salles, R.~M., and Marino~Jr, D.~A.}
\newblock Strategies and metric for resilience in computer networks.
\newblock {\em The Computer Journal 55}, 6 (2011), 728--739.

\bibitem{Sato2009}
{\sc Sato, Y., Ata, S., and Oka, I.}
\newblock {A strategic approach for re-organizing the Internet topology by
  applying social behavior dynamics}.
\newblock {\em Journal of Network and Systems Management 17}, 1-2 (2009),
  208--229.

\bibitem{Seary2002}
{\sc Seary, A.~J., and Richards, W.~D.}
\newblock Spectral methods for analyzing and visualizing networks: An
  introduction.
\newblock {\em Workshop on Dynamic Social Network Modeling and Analysis\/}
  (2002), 1--20.

\bibitem{Soffer2005}
{\sc Soffer, S.~N., and V\'{a}zquez, A.}
\newblock {Network clustering coefficient without degree-correlation biases}.
\newblock {\em Physical Review E - Statistical, Nonlinear, and Soft Matter
  Physics 71}, 5 (2005), 2--5.

\bibitem{Sreenivasan2007}
{\sc Sreenivasan, S., Cohen, R., L\'{o}pez, E., Toroczkai, Z., and Stanley,
  H.~E.}
\newblock {Structural bottlenecks for communication in networks}.
\newblock {\em Physical Review E - Statistical, Nonlinear, and Soft Matter
  Physics 75}, 3 (2007), 1--4.

\bibitem{sterbenz2010resilience}
{\sc Sterbenz, J.~P., Hutchison, D., {\c{C}}etinkaya, E.~K., Jabbar, A.,
  Rohrer, J.~P., Sch{\"o}ller, M., and Smith, P.}
\newblock Resilience and survivability in communication networks: Strategies,
  principles, and survey of disciplines.
\newblock {\em Computer Networks 54}, 8 (2010), 1245--1265.

\bibitem{Sun2007}
{\sc Sun, S., Liu, Z., Chen, Z., and Yuan, Z.}
\newblock {Error and attack tolerance of evolving networks with local
  preferential attachment}.
\newblock {\em Physica A: Statistical Mechanics and its Applications 373\/}
  (2007), 851--860.

\bibitem{Sydney2008}
{\sc Sydney, A., Scoglio, C., Schumm, P., and Kooij, R.~E.}
\newblock {ELASTICITY}: Topological characterization of robustness in complex
  networks.
\newblock In {\em Bionetics 2008, Hyogo, Japan\/} (2008), vol.~16, p.~26.

\bibitem{Sydney2010}
{\sc Sydney, A., Scoglio, C., Youssef, M., and Schumm, P.}
\newblock Characterising the robustness of complex networks.
\newblock {\em International Journal of Internet Technology and Secured
  Transactions 2}, 3-4 (2010), 291--320.

\bibitem{szabo2003geometry}
{\sc Szab{\'o}, G.~J., Alava, M., and Kert{\'e}sz, J.}
\newblock Geometry of minimum spanning trees on scale-free networks.
\newblock {\em Physica A: Statistical Mechanics and its Applications 330}, 1
  (2003), 31--36.

\bibitem{Tangmunarunkit2002}
{\sc Tangmunarunkit, H., Govindan, R., Jamin, S., Shenker, S., and Willinger,
  W.}
\newblock Network topology generators: Degree-based vs. structural.
\newblock {\em ACM SIGCOMM Computer Communication Review 32}, 4 (2002), 147.

\bibitem{tangmunarunkit2001impact}
{\sc Tangmunarunkit, H., Govindan, R., Shenker, S., and Estrin, D.}
\newblock The impact of routing policy on internet paths.
\newblock In {\em Twentieth Annual Joint Conference of the IEEE Computer and
  Communications Societies (INFOCOM 2001)\/} (2001), IEEE, pp.~736--742.

\bibitem{Tauro1995}
{\sc Tauro, S.~L., Palmer, C., Siganos, G., and Faloutsos, M.}
\newblock A simple conceptual model for the internet topology.
\newblock In {\em IEEE Global Telecommunications Conference (GLOBECOM'01)\/}
  (2001), vol.~3, IEEE, pp.~1667--1671.

\bibitem{tizghadam2011robust}
{\sc Tizghadam, A., and Leon-Garcia, A.}
\newblock Robust network planning in nonuniform traffic scenarios.
\newblock {\em Computer Communications 34}, 12 (2011), 1436--1449.

\bibitem{Trpevski2010}
{\sc Trpevski, D., Smilkov, D., Mishkovski, I., and Kocarev, L.}
\newblock {Vulnerability of labeled networks}.
\newblock {\em Physica A: Statistical Mechanics and its Applications 389}, 23
  (2010), 5538--5549.

\bibitem{tyler2003proceedings}
{\sc Tyler, J.~R., Wilkinson, D.~M., and Huberman, B.~A.}
\newblock Email as spectroscopy: Automated discovery of community structure
  within organizations.
\newblock In {\em Communities and Technologies - Proceedings of the First
  International Conference on Communities and Technologies (C\&T 2003)\/}
  (2003), Kluwer, pp.~81--96.

\bibitem{wang2006entropy}
{\sc Wang, B., Tang, H., Guo, C., and Xiu, Z.}
\newblock Entropy optimization of scale-free networks' robustness to random
  failures.
\newblock {\em Physica A: Statistical Mechanics and its Applications 363}, 2
  (2006), 591--596.

\bibitem{Wang2008a}
{\sc Wang, Y., Xiao, S., Xiao, G., Fu, X., and Cheng, T.~H.}
\newblock {Robustness of complex communication networks under link attacks}.
\newblock In {\em International Conference on Advanced Infocomm Technology
  (ICAIT '08)\/} (2008), pp.~1--7.

\bibitem{wasserman1994social}
{\sc Wasserman, S., and Faust, K.}
\newblock {\em Social Network Analysis: Methods and Applications}, vol.~8.
\newblock Cambridge University Press, 1994.

\bibitem{Watts1998}
{\sc Watts, D.~J., and Strogatz, S.~H.}
\newblock Collective dynamics of 'small-world' networks.
\newblock {\em Nature 393}, 6684 (1998), 440--442.

\bibitem{Wu2011}
{\sc Wu, J., Barahona, M., Tan, Y.-J., and Deng, H.-Z.}
\newblock Spectral measure of structural robustness in complex networks.
\newblock {\em IEEE Transactions on Systems, Man, and Cybernetics-Part A:
  Systems and Humans 41}, 6 (2011), 1244--1252.

\bibitem{Wu2008}
{\sc Wu, J., Tan, Y.-J., Deng, H.-Z., Li, Y., Liu, B., and Lv, X.}
\newblock Spectral measure of robustness in complex networks.
\newblock {\em arXiv:0802.2564\/} (2008).

\bibitem{Xia2008}
{\sc Xia, Y., and Hill, D.~J.}
\newblock {Attack vulnerability of complex communication networks}.
\newblock {\em IEEE Transactions on Circuits and Systems II: Express Briefs
  55}, 1 (2008), 65--69.

\bibitem{Xiao2010}
{\sc Xiao, S., Xiao, G., and Cheng, T.~H.}
\newblock {Tolerance of local information-based intentional attacks in complex
  networks}.
\newblock {\em Journal of Physics A: Mathematical and Theoretical 43}, 33
  (2010), 335101.

\bibitem{yang2011transportation}
{\sc Yang, H.-X., Wang, W.-X., Xie, Y.-B., Lai, Y.-C., and Wang, B.-H.}
\newblock Transportation dynamics on networks of mobile agents.
\newblock {\em Physical Review E 83}, 1 (2011), 016102.

\bibitem{yazdani2010note}
{\sc Yazdani, A., and Jeffrey, P.}
\newblock A note on measurement of network vulnerability under random and
  intentional attacks.
\newblock {\em arXiv:1006.2791\/} (2010).

\bibitem{paperw}
{\sc Zhang, S.}
\newblock {\em {Cycles in Weighted Graphs and Related Topics}}.
\newblock Twente University Press, 2002.

\bibitem{Zhang2003}
{\sc Zhang, Y., Roughan, M., Lund, C., and Donoho, D.}
\newblock {An information-theoretic approach to traffic matrix estimation}.
\newblock In {\em Conference on Applications, Technologies, Architectures, and
  Protocols for Computer Communications (SIGCOMM '03)\/} (2003), pp.~301--312.

\bibitem{zhang2010enumeration}
{\sc Zhang, Z., Liu, H., Wu, B., and Zhou, S.}
\newblock Enumeration of spanning trees in a pseudofractal scale-free web.
\newblock {\em EPL (Europhysics Letters) 90}, 6 (2010), 68002.

\end{thebibliography}

\end{document}